\documentclass{amsart}
\pdfoutput=1 % To make arXiv use pdflatex

\usepackage{amssymb}

\usepackage[backend = biber,
    style = authoryear,
    giveninits = true,  % Shorten all first names.
    dashed = false,     % Prevent authors names from being replaced with dash.
    uniquename = false,  % This is just so that "Moritz, H." and "Moritz, Helmut" should not render extra initial to distinguish "different" authors.
    maxcitenames = 2  % To make three author list being cited as et al.
]{biblatex}
\DeclareCiteCommand{\cite}[\mkbibemph]
  {\usebibmacro{prenote}}%
  {\usebibmacro{citeindex}%
   \usebibmacro{cite}}
  {\multicitedelim}
  {\usebibmacro{postnote}}

\usepackage{tabularx,ragged2e} % For automatically determining the column widths.
\usepackage{hyperref}  % URL in references and internal links.
\usepackage{graphicx}  % Floats and resizing table and symbols.
\usepackage{caption}   %
\usepackage{mathtools} % For adjusted matrices.
\usepackage{longtable} % For tables spanning more than one page.

\newbox\VWbox \setbox\VWbox\hbox{v\kern-.21em v\kern-.21em v} 

\DeclareMathOperator{\atantwo}{atan2}

\DeclareMathOperator*{\argmin}{arg\,min}
\DeclareMathOperator{\ulp}{ulp}

\newcommand{\ellipse}{{\mathpalette\scaledCirc\relax}}
\newcommand{\scaledCirc}[2]{\scalebox{1.5}[1]{$#1\circ$}}

\newcommand\impltablewidth{0.43}

\newcommand{\sinixlimit}{I}
\newcommand{\tanixlimit}{J}
\newcommand{\mulixlimit}{L}
\newcommand{\fourierixlimit}{N}
\newcommand{\taylorixlimit}{M}
\usepackage{xparse}
\NewDocumentCommand{\lataddcorr}{ O{\fourierixlimit} O{\taylorixlimit} O{u} O{v}}{\omega_{#1,#2}(#3,#4)}
\NewDocumentCommand{\lataddcorrex}{ O{u} O{v}}{\omega(#1,#2)}
\newcommand{\lataddcorrexpr}{\lataddcorrex[h_c][\sin^2(\phi_c)]}
\NewDocumentCommand{\altaddcorr}{ O{\fourierixlimit} O{\taylorixlimit} O{u} O{v}}{\mu_{#1,#2}(#3,#4)}
\NewDocumentCommand{\altaddcorrex}{ O{u} O{v}}{\mu(#1,#2)}
\newcommand{\altaddcorrexpr}{\altaddcorrex[h_c][\sin^2(\phi_c)]}
\NewDocumentCommand{\sinlatmulcorr}{ O{\mulixlimit} O{v} O{w}}{\eta_{#1}(#2,#3)}

\NewDocumentCommand{\coslatmulcorr}{ O{\mulixlimit} O{v} O{w}}{\rho_{#1}(#2,#3)}

\NewDocumentCommand{\triglatmulevencorr}{ O{\mulixlimit} O{v} O{w}}{\sigma_{#1}(#2,#3)}
\NewDocumentCommand{\triglatmulevencorrTwo}{ O{\mulixlimit} O{\delta}}{\sigma_{#1}(#2)}
\NewDocumentCommand{\triglatmuloddcorr}{ O{\mulixlimit} O{v} O{w}}{\tau_{#1}(#2,#3)}
\NewDocumentCommand{\triglatmuloddcorrTwo}{ O{\mulixlimit}  O{w} O{\delta}}{\tau_{#1}(#2,#3)}
\NewDocumentCommand{\arcsinapprox}{ O{\sinixlimit}}{\chi_{#1}}
\NewDocumentCommand{\atanapprox}{ O{\sinixlimit}}{\psi_{#1}}
\NewDocumentCommand{\atantwoapprox}{ O{\tanixlimit}}{\xi_{#1}}

\title[Minimax polynomial ECEF to geodetic coordinate transformations]{Minimax polynomial ECEF to geodetic coordinate transformation approximations}
\author{John-Olof Nilsson}%\\{\scriptsize A\lowercase{vioniq} S\lowercase{weden} AB, \lowercase{jnil02@kth.se}}}
\address{The author is with Avioniq Awareness Sweden AB, e-mail: jnil02@kth.se}
\date{}

\usepackage[foot]{amsaddr} % To place address in footnote.

\begin{document}

\begin{abstract}
Minimax polynomial ECEF to geodetic coordinate transformation approximations are presented, including often preferable n-vector versions and pseudo-code implementations. The approximations provide high accuracy-to-computational-cost tunability and an unprecedented low latency down to an accuracy of $\sim10^{-5}$m, which is demonstrated in an extensive benchmark. This sets a new standard for coarse and fast ECEF to geodetic coordinate transformations and opens up a new realm of further improvement opportunities and extensions to other geodetic quantities.
\end{abstract}

\maketitle

\vspace{-8mm}

\section{Introduction}
Cartesian earth-centered earth-fixed (ECEF) coordinate $\{x,y,z\}$ to geodetic coordinates $\{\phi,\lambda,h\}$ (\underline{l}atitude, \underline{l}ongitude and \underline{a}ltitude, \emph{lla}) or $\{\bar{n},h\}$ (\underline{n}-\underline{v}ector and \underline{a}ltitude, \emph{nva}) transformations are integral to geodesy and related modeling and simulations and, for many applications, constitute a significant computational cost. The importance of the problem is demonstrated by over 230 publications listed in~\cite{Nilsson2024}, spanning more than 6 decades. However, with its apparent simplicity and many publications, it could be expected to be a closed chapter, but, somewhat surprisingly, transformation methods based on minimax polynomials, the cornerstone of \emph{approximation theory}, are completely lacking.
This work fills in this blank and demonstrates how the \emph{complete} transformations can be implemented with minimax polynomials, including often preferable n-vector versions. (See~\cite{Gade2010} for the benefits of the n-vector representation.) This results in high accuracy-to-computational-cost tunability, straight-forward arithmetic-only separable implementations and unprecedented low latency down to an accuracy of at least $\sim10^{-5}$m, at which the Bowring's method is equally good. %Note. the double precision numerical limit is $\sim10^{-9}$m.
The performance is demonstrated in an extensive benchmark of more than 40 methods. This sets a new standard for coarse (relative to the 64-bit numeric limit of $\sim10^{-9}$m) and fast ECEF to geodetic coordinate transformations (and evaluations) and opens up a new realm of further improvement and extension opportunities.

The improvement opportunities primarily come from the resemblance of the presented approximations to elementary function approximations, for which a vast literature and tool sets exist. For a starting point, see~\cite{Muller2006}. See also the \emph{potential improvement} in the discussion Section~\ref{sec:discussion}.
The extension opportunities primarily comes from the presented approximations being trivially extendable to other (in addition to latitude and altitude) ellipsoidal $z$-axis rotation symmetric and $xy$-plan reflection symmetric or anti-symmetric quantities.
The main caveat to it all is that the approximations are only valid for a \emph{preselected} altitude range. The range can be large enough to cover the vast majority of applications, with reasonable performance demonstrated for up to an altitude range of $~\sim500\;000$m. Significantly beyond that, one has to resort to range splitting, which, on the other hand, is normal for minimax approximations. %and possibly one should preferably use splitting long before that.
%
%Further, as suggested in~\cite{Claessens2019}, the method may also be of significance for a wider computational mathematics community for handling elliptical conic sections.
%
Further, the presented benchmark is valuable in itself. It is the most extensive benchmark to date and the first benchmark including \emph{nva} transformations, which arguably provides a superior view of the performance of different methods. The results indicate that, in contrast to frequent claims of the opposite; in terms of computational efficiency and for altitude ranges and conversion accuracies reasonable for most applications; there have only been a few marginal improvements beyond~\cite{Bowring1976} +~\cite{Bowring1985} method, with \cite{Lin1995}, \cite{Ohlson1996}, \cite{Wu2003}, \cite{Fukushima2006} and~\cite{Shu2010} being noteworthy. Similar conclusions have also been reached in a smaller benchmark~\cite{Claessens2019}.
%
%Finally, the transformation approximations are probably particularly interesting for physics simulation applications for which geopotential related properties, e.g. gravity and air density, need to be computed a great number of times but with moderate accuracy.

If you just want a fast polynomial approximation, skip to Table~\ref{tab:impl} of Section~\ref{sec:implementation} for pseudo-code implementations, then skip to the benchmark results in Figure~\ref{fig:lat_and_alt_err} of Section~\ref{sec:benchmark} to select polynomial orders and, finally, get the polynomials from Appendix~\ref{appendix:labeled_approx}.
Or take the long route and start in Section~\ref{sec:history} for some extended background and notes about novelty and continue with Section~\ref{sec:additive} where the core range reductions, series expansions and range reconstructions are presented.
Section~\ref{sec:partial} and~\ref{sec:range} follows with polynomial approximations of the transformation components and the complete transformations.
Section~\ref{sec:implementation} deals with the implementation, Section~\ref{sec:benchmark} with the benchmark and Section~\ref{sec:benchmark_discussion} discusses the benchmark results. Finally, general discussions, caveats and potential improvements are provided in Section~\ref{sec:discussion} followed by final recommendations in Section~\ref{sec:final_rec}. Many details are referred to the Appendices~\ref{appendix:derivation}-\ref{appendix:labeled_approx}.

\section{Background and Novelty}\label{sec:history}
As for most minimax polynomial approximations, there are four fundamental components of the presented transformation approximations; \emph{range reduction}, \emph{series expansion}, \emph{polynomial approximation} and \emph{range reconstruction}; for which some background and further comments about novelty of this work are provided here.
First a general comment, there is a significant (former) Soviet Bloc literature on the coordinate transformation subject. This literature is hard to access and has largely been ignored but references are listed in \cite{Nilsson2024}. However, there is no indication that it contains any transformation methods similar to the one presented here.

The \emph{range reduction} aims at reducing the essentially unbounded ECEF coordinate values to some bounded intervals. For polynomial methods, this essentially entail transforming the ECEF coordinates to the variables of the polynomial approximation.
Similar steps are present in most iterative and close-form methods but seldom explicitly recognized as such.
Here, the range reduction is split into a commonly appearing, but here implicit, geocentric latitude part, for the approximation derivations, and an explicit but significantly less expensive coordinate normalization, for the actual approximations.

\emph{Series expansions} are not strictly necessary for polynomial approximation but often helpful in that it provides structure for the later approximation. Polynomial approximation, and particularly series expansions, of ECEF (geocentric latitude) to geodetic coordinate transformations is nothing new. Rather it appears to have been the normal way to transform coordinates in the pre-digital-computers era. Expansions can be found on p40-43 in the standard reference~\cite{Helmert1880}. % \emph{Die mathematischen und physikalischen Theorieen der h{\"o}heren Geod{\"a}sie}~\cite{Helmert1880}. 
Helmert, in turn, cite even older literature. Further, an extensive treatment of series expansions can be found in \cite{Adams1921}. Moreover, the~\cite{Almanac1931}, see p718, even contains first order altitude corrections in the conversion from geocentric to geodetic coordinates but, although simple in this case, there is no explicit treatment of how to handle unknown altitude. (The corrections are late provided with some more explanation in~\cite{ExtraAlmanac1961}, see p58.)
%, as well as spherial astronomy literature, e.g. Smart's 1931 book~\cite{Smart1931}. Tabulated values for practical calculations can be found in the 1929 publication~\cite{Noaa1929}.% and cocurrent ephemeris include tabulated linear transformations, see p718 of the 1931 \emph{Nautical Almanac}~\cite{Cowell1929}. Dedicated publication with conversion tables can also be found~\cite{Lambert1935} %And series expansion has been used for the geodetic problems since at least the beginning of the 19th century~\cite{Karney2011}.
With the launch of Sputnik 1 on the 4th of October 1957 and the heightened interest and requirements of the space race, numerous publications of altitude dependent series expansions can be found from 1960 and up to 1976~\cite{Berger1960,Hirvonen1960,Morrison1961,Gersten1961,Hirvonen1963,
Pick1967,Mikhailov1967,Pavlov1968,Getchell1972,Long1974,Long1975,Deprit1975,Sunkel1976}.
%combined with reports of actual applications~\cite{Christ1965}.
%
However, after the seminal~\cite{Bowring1976} iterative method, they appear to wane in popularity. Some notable exceptions are~\cite{Eissfeller1985,Ohlson1996,Turner2013,Guo2014,Alberto2019}. There has also recently been some work on non-altitude dependent expansions~\cite{Orihuela2013,Li2022,Karney2023}.
However, the series expansions of the literature typically have three fundamental short-comings: 1) They partially rely on inefficient (local) derivative based series expansions. 2) The series variables are often not directly available. 3) The series are only for latitude, whereas altitude is computed with an exact latitude based formula and the approximated latitude. Here novel general series expansions are presented, directly in ECEF coordinate norm and normalized coordinates for latitude, altitude and the n-vector. The series expansions contains no derivatives.

The \emph{polynomial approximation} of the literature methods are, without exceptions, done by truncating the series expansions. This is simple but it is well known in \emph{approximation theory} that this is typically suboptimal. Rather minimax approximations should normally be used. The Remez algorithm for computing such polynomials has been around since~\cite{Remez1934} and the field can be described as well established since the 60s~\cite{Fraser1965}. Hence, it is somewhat surprising that minimax is nowhere mentioned within the over 230 references listed in~\cite{Nilsson2024}. However, there is a simple reason minimax approximations has not been used, so far; the coordinate transformation is not a 1-dimensional problem, making the techniques not directly applicable. On the other hand, closely related Chebyshew polynomials are typically used to describe ephemeris data~\cite{Newhall1988} and it is occasionally encountered in map projections and terrain model, so it is not completely unknown to the geodetic community. Nonetheless, here it is shown how minimax approximations can be used for ECEF to geodetic coordinate transformations. Only (Chebyshev) series which can be truncated with close to minimax performance are truncated. In all other places, optimal minimax approximations are used.

Finally, the \emph{range reconstruction} aim at transforming some intermediate approximated quantities to the final quantities of interest. For most approximation methods this means transforming some quantity, proportional to trigonometric function of the latitude, to the final latitude and altitude, i.e. the range reconstruction come in the form of inverse trigonometric functions. These inverse trigonometric functions are just as much a part of the coordinate transformation as any other part and should be approximated with suitable accuracy. Most likely, this is occasionally done in practice but with some rare exceptions~\cite{Toms1996,Toms1998,Zanevicius2010} this is not discussed, despite the range reconstruction making up half the computational cost (as will be shown) and suitable approximations being readily available since at least the 50s, see~\cite{Hastings1955}. Further, in many cases, n-vector representation is preferable~\cite{Gade2010}, and they can be though of as just a different range reconstruction. Most iterative and closed form methods are straight-forward to convert. However, most polynomial methods are not, in a sensible way. Here matched inverse trigonometric function approximations are provided and dedicated (minimax) approximations for the n-vector representation.

Altogether, the mentioned (novel) components make polynomial approximations competitive again. Further, the computer development have probably contributed here too. Modern super-scalar vector CPUs are very good at evaluating polynomials, giving polynomial methods and edge over iterative and closed-form methods.

Note that iterative methods appear to have been around for a long time as well. The classical method by Hirvonen was first published in~\cite{Hirvonen1959}\footnote{I have not yet managed to access the frequently cited 1958 publication by K. Rinner, or found any description of its results. Hence, it is not cited just yet.}. \cite{Morrison1961} states \emph{``Various procedures for obtaining a solution of these equations by iteration techniques exist.''} without any further references. In turn, (published) exact solutions have been around since at least~\cite{Ecker1967,Sugai1967} with more following soon after~\cite{Tomelleri1970,Paul1973,Benning1974,Hedgley1976}. (Also~\cite{Ecker1967} states that the transformation is normally carried out by iterative methods.) However,~\cite{Berger1960} refers to solutions by \emph{``Lagrange multiplier''} and \emph{``high-degree algebraic equation''}. Further, \cite{Vincenty1985} claims the earliest closed form solution can be found in~\cite{Dorrie1948} but, as pointed out by~\cite{Bajorek2014}, the latitude equation can be found e.g. on p49 in the book \emph{Analytical Conics}~\cite{Sommerville1924}. Finally, methods for solving quartic equations are known since Ferrari's 1545 solution, so exact solutions have probably been applied before Ecker's and Sugai's work.

%\section{Problem transformation}\label{sec:additive}
\section{Reductions, expansions and reconstructions}\label{sec:additive}
The multivariate and essentially unbounded nature of the coordinate transformation makes polynomial approximation techniques not directly applicable. The unboundedness is handled by the \emph{range reduction} comprised of the ECEF to geocentric coordinates transformations
\begin{equation}\label{eq:spherical}
\begin{matrix}
\phi_c\\
\lambda_c\\
h_c
\end{matrix}
\;
\begin{matrix}
=\\=\\=
\end{matrix}
\;
\begin{matrix*}[l]
\sin^{-1}(z/p) \\
\atantwo(y,x) \\
p - h_0 
\end{matrix*}
\quad\quad\text{and}\quad\quad
\bar{n}_c=
\begin{bmatrix}
x/p\\y/p\\z/p
\end{bmatrix}
=
\begin{bmatrix}
r\\s\\t
\end{bmatrix}
\end{equation}
where $p=\sqrt{x^2+y^2+z^2}$ and where $\phi_c$, $\lambda_c$, $h_c$, $h_0$ and $\bar{n}_c$ are the geocentric latitude, longitude, altitude, reference radius and n-vector, respectively. Note that
\begin{equation}\label{eq:n}
\bar{n}=
\begin{bmatrix}
\cos(\phi)\cos(\lambda)\\
\cos(\phi)\sin(\lambda)\\
\sin(\phi)
\end{bmatrix}
\quad\text{and}\quad
\bar{n}_c=
\begin{bmatrix}
\cos(\phi_c)\cos(\lambda_c)\\
\cos(\phi_c)\sin(\lambda_c)\\
\sin(\phi_c)
\end{bmatrix}
\end{equation}
and that $\lambda_c=\lambda$. Also note, later on, $h_0$ will be taken to be 0 but $h_c$ and $p$ are still kept separate since a non-zero $h_0$ could be used, or even some other expression for $h_c$.
Clearly, $\phi_c$, $\lambda_c$, $r$, $s$ and $t$ are bounded and $h_c$ is bounded in the sense that an interval of interest can typically be set.
In turn, the multivariate nature is handled by the following key \emph{series expansions} derived in Appendix~\ref{appendix:derivation}
\begin{equation}\label{eq:corrections}
\begin{alignedat}{3}
\phi &= \phi_c + t\sqrt{1 - t^2}\lataddcorrex[h_c][t^2]\quad\quad& \sin(\phi) &=\sin(\phi_c)\sinlatmulcorr[][t^2][\lataddcorrex[h_c][t^2]]\\
h &= \altaddcorrex[h_c][t^2] &\cos(\phi) &= \cos(\phi_c)\coslatmulcorr[][t^2][\lataddcorrex[h_c][t^2]]
\end{alignedat}
\end{equation}
where,  with the customary (abuse of) notation $\sum^\infty_{n=0}(\cdot)$ meaning $\lim_{N\to\infty}\sum^N_{n=0}(\cdot)$ and $\sum'$ meaning that the zeroth term is to be halved,
\begin{equation*}
\begin{split}
\lataddcorrex &= \sum^\infty_{n=1}\sum_{k=0}^{n-1}(-1)^k\binom{2n}{2k+1}v^k(1-v)^{n-k-1}b_n(u)\\
\altaddcorrex &= u + \mathop{\smash{\sideset{}{'}\sum}\vphantom{\sum}}\limits^{\infty}_{n=0}\sum_{k=0}^{n}(-1)^k\binom{2n}{2k}v^k(1-v)^{n-k}c_n(u)\\
\sinlatmulcorr[]&=\triglatmulevencorrTwo[][\delta(v,w)]+(1-v)\triglatmuloddcorrTwo[][w][\delta(v,w)] \\
\coslatmulcorr[]&=\triglatmulevencorrTwo[][\delta(v,w)]-v\triglatmuloddcorrTwo[][w][\delta(v,w)]
\end{split}
\end{equation*}
where $b_n(\cdot)$ and $c_n(\cdot)$ are altitude dependent Fourier coefficients, encapsulating the \emph{reference ellipsoid} properties, and
\begin{equation*}
\delta(v,w)=v(1-v)w^2
\end{equation*}
\begin{equation*}
\triglatmulevencorrTwo[]=\sum_{l\in\mathbb{N}_{e}}(-1)^{l/2}\frac{1}{l!}\delta^{l/2}
\quad\text{and}\quad
\triglatmuloddcorrTwo[]=w\sum_{l\in\mathbb{N}_{o}}(-1)^{(l-1)/2}\frac{1}{l!}\delta^{(l-1)/2}
\end{equation*}
where $\mathbb{N}_{e}$ and $\mathbb{N}_{o}$ are the sets of all positive even and odd natural numbers. Note, $\delta(t^2,\lataddcorrex[h_c][t^2])=(\phi - \phi_c)^2$, i.e. $\triglatmulevencorrTwo[]$ and $\triglatmuloddcorrTwo[]$ are polynomials in $(\phi - \phi_c)^2$. Further, note, the only assumption on $\phi$ and $h$ in the derivation is that they are rotation symmetric with respect to the z-axis and that they are reflection anti-symmetric and symmetric, respectively, with respect to the $xy$-plane, i.e. the expressions are valid for any such quantities. The quantity specific attributes are carried by $b_n(\cdot)$ and $c_n(\cdot)$.

For $h$ no \emph{range reconstruction} is required whilst, from~\eqref{eq:n}-\eqref{eq:corrections}, it follows for $n$ as
\begin{equation}\label{eq:n_rr}
\bar{n}\approx
\begin{bmatrix}
r\,\coslatmulcorr[][t^2][\lataddcorr[][][h_c][t^2]]\\
s\,\coslatmulcorr[][t^2][\lataddcorr[][][h_c][t^2]]\\
t\,\sinlatmulcorr[][t^2][\lataddcorr[][][h_c][t^2]]
\end{bmatrix}
\end{equation}
and for $\phi$ there are six reconstruction alternatives 

\hspace{-6mm}
\begin{minipage}{.53\textwidth}
\begin{equation}\label{eq:inverse}
\begin{split}
\hspace{-4mm}\phi & = \sin^{-1}\left(t\,\sinlatmulcorr[][t^2][\lataddcorrex[h_c][t^2]]\right)\\
\hspace{-4mm}\phi & = \sin^{-1}(t) + t\sqrt{1-t^2}\lataddcorrex[h_c][t^2]\\
\hspace{-4mm}\phi & = \text{sgn}(z)\cos^{-1}\!\left(q/p\,\coslatmulcorr[][t^2][\lataddcorrex[h_c][t^2]]\right)
\end{split}
\end{equation}
\end{minipage}
\hspace{-6mm}
\begin{minipage}{.47\textwidth}
\begin{equation*}
\begin{split}
\phi & = \text{sgn}(z)\cos^{-1}\!\left(q/p\right)\!+\!t\sqrt{1\!-\!t^2}\lataddcorrex[h_c][t^2] \\
\phi & = \tan^{-1}(z/q)\!+\!t\sqrt{1-t^2}\lataddcorrex[h_c][t^2]\\
\phi & = \tan^{-1}\left(\frac{z\,\sinlatmulcorr[][t^2][\lataddcorrex[h_c][t^2]]}{q\coslatmulcorr[][t^2][\lataddcorrex[h_c][t^2]]}\right)
\end{split}
\end{equation*}
\end{minipage}

\noindent
where $q=\sqrt{x^2+y^2}$. $\sin^{-1}$ is the least expensive to approximate but is ill-conditioned around the poles (the derivative goes to infinity) and using it anywhere near the poles requires a square root argument reduction. $\cos^{-1}$ requires a square root and a divide and is similarly ill-conditioned around the equator. $\tan^{-1}$ is well conditioned for all $\phi_c$ but requires a square root, a divide and some more arithmetic operations compared to $\sin^{-1}$ and $\cos^{-1}$. Similarly, the additive correction is ill-conditioned around the poles (the derivative of $t\sqrt{1-t^2}$ goes to infinity) whereas the multiplicative correction is well conditioned for all $\phi_c$. See Appendix~\ref{sec:selection} for some figures of what this ill-conditioning means.

As will be shown, for the least accurate approximations, $\sin^{-1}$-expressions are suitably used. Combining the $\sin^{-1}$ and $\cos^{-1}$ expressions for values around the poles gives potentially good performance but is a hard combination to tune, among other things to avoid discontinuous $\phi$ errors, and gives undesirable conditional code, and is not further investigated here. See Section~\ref{sec:discussion} for some more discussion on it. For more accurate approximations, the $\tan^{-1}$ expressions are suitably used.

\section{Partial polynomial approximations}\label{sec:partial}
The problem transformation of the previous sections reduces the transformation approximation problem to that of $\lataddcorrex[u][v]$, $\altaddcorrex[u][v]$, $\triglatmulevencorrTwo[]$, $\triglatmuloddcorrTwo[]$, $\sin^{-1}$, $\tan^{-1}$ and $\atantwo$.
Naturally, minimax approximation are sought.
The challenge is the two dimensional nature of $\lataddcorrex[u][v]$ and $\altaddcorrex[u][v]$. Fortunately, $\altaddcorrex[u][v]$, being a Fourier cosine series, is a Chebyshev series with respect to $v$ and truncated Chebyshev series give close to minimax approximations~\cite{Boyd2000}.
Similarly, truncation of $\lataddcorrex[u][v]$ gives a close to minimax optimal approximation with respect to the weighting $v\sqrt{1-v^2}$.
In contrast, for $b_n(h_c)$, $c_n(h_c)$ and $\triglatmulevencorrTwo[]$ and $\triglatmuloddcorrTwo[]$, explicit minimax approximations are used. Let
\begin{equation*}
\begin{split}
\{b^{(M)}_{n,m}:m\in[0,M]\} &= \argmin_{\{b_m:m\in[0,M]\}}\max_{h_c\in[{h_c}_{\min},{h_c}_{\max}]}\left| b_n(h_c) -  \sum_{m=0}^Mb_mh_c^m\right|
\\
\{c^{(M)}_{n,m}:m\in[0,M]\} &= \argmin_{\{c_m:m\in[0,M]\}}\max_{h_c\in[{h_c}_{\min},{h_c}_{\max}]}\left| c_n(h_c) -  \sum_{m=0}^Mc_mh_c^m\right|\\
\{\varsigma^{(L)}_{l}:l\in[0,\lfloor{L/2}\rfloor]\} &= \argmin_{\{\varsigma_l:l\in[0,\lfloor{L/2}\rfloor]\}}\max_{\delta\in[0,\delta_{\max}]}\left| \sigma(\delta) -  \sum_{l=0}^{\lfloor{L/2}\rfloor}\varsigma_l\delta^l\right|\\
\{\vartheta^{(L)}_{l}:l\in[0,\lfloor{(L-1)/2}\rfloor]\} &= \argmin_{\{\vartheta_l:l\in[0,\lfloor{(L-1)/2}\rfloor]\}}\max_{\delta\in[0,\delta_{\max}]}\left| \tau(1,\delta) -  \sum_{l=0}^{\lfloor{(L-1)/2}\rfloor}\vartheta_l\delta^l\right|
\end{split}
\end{equation*}
where $b^{(M)}_{n,m}$ and $c^{(M)}_{n,m}$ are the $m^{th}$ coefficients of the minimax approximation of degree $M$ with respect to $h_c$ of the $n^{th}$ Fourier coefficients. Similarly, $\varsigma^{(L)}_{l}$ and $\vartheta^{(L)}_{l}$ are the $l^{th}$ coefficients of the minimax approximation of degree $\lfloor{L/2}\rfloor$ and $\lfloor{(L-1)/2}\rfloor$, respectively, i.e. a total degree of the multiplicative corrections of $L$.
The ranges $[{h_c}_{\min},{h_c}_{\max}]$ and $[0,\delta_{\max}]$ are the ranges over which the polynomial approximations are minimax optimal. See Section~\ref{sec:implementation} for a discussions on $[{h_c}_{\min},{h_c}_{\max}]$. In turn, $\delta_{\max}$ is defined by $[{h_c}_{\min},{h_c}_{\max}]$ via
\begin{equation*}
\begin{split}
\delta_{\max} &= \max_{\substack{h_c\in[{h_c}_{\min},{h_c}_{\max}]\\\phi_c\in[-\pi/2,\pi/2]}}\delta(\sin^2(\phi_c),\lataddcorrexpr)\\
&=\max_{\substack{h_c\in[{h_c}_{\min},{h_c}_{\max}]\\\phi_c\in[0,\pi/2]}}(\phi-\phi_c)^2
%&=\max_{\substack{h_c\in[{h_c}_{\min},{h_c}_{\max}]\\\phi_c\in[0,\pi/2]}}\delta(\sin^2(\phi_c),(\phi-\phi_c)/(\sin(\phi_c)\sqrt{1-\sin^2(\phi_c)}))
\end{split}
\end{equation*}
%Note that, just truncating Taylor series of $b_n(h_c)$ and $c_n(h_c)$ actually only gives marginally worse performance than the minimax approximations.
%
Altogether, this gives the (essentially) minimax power series approximations
\begin{equation}\label{eq:partial}
\begin{alignedat}{3}
\lataddcorrex[u][v]&\approx\lataddcorr[\fourierixlimit][\taylorixlimit][u][v]
&\triglatmulevencorrTwo[]&\approx\triglatmulevencorrTwo\\
\altaddcorrex[u][v]&\approx\altaddcorr[\fourierixlimit][\taylorixlimit][u][v]
&\quad\quad\quad\triglatmuloddcorrTwo[]&\approx\triglatmuloddcorrTwo
\end{alignedat}
\end{equation}
where
%where and $\altaddcorrex[u][v]$ are the (truncated) power series
\begin{equation*}
\begin{split}
\lataddcorr &= \sum^N_{n=1}\sum_{k=0}^{n-1}(-1)^k\binom{2n}{2k+1}v^k(1-v)^{n-k-1}\sum_{m=0}^Mb^{(N)}_{n,m}u^m\\
\altaddcorr &= u + \mathop{\smash{\sideset{}{'}\sum}\vphantom{\sum}}\limits^N_{n=0}\sum_{k=0}^{n}(-1)^k\binom{2n}{2k}v^k(1-v)^{n-k}\sum_{m=0}^Mc^{(M)}_{n,m}u^m\\
\triglatmulevencorrTwo&=\sum_{l=0}^{\lfloor{L/2}\rfloor}\varsigma^{(L)}_l\delta^{l}\quad\text{and}\quad
%\triglatmulevencorrTwo&=\sum_{l\in\mathbb{N}_{e}\land l\leq L}(-1)^{l/2}\frac{1}{l!}\delta^{l/2}\\
%\triglatmulevencorr&=\sum_{l\in\mathbb{N}_e\land l\leq L}(-1)^{l/2}\frac{1}{l!}v^{l/2}(1-v)^{l/2}w^l\\
%
\triglatmuloddcorrTwo=w\sum_{l=0}^{\lfloor{(L-1)/2}\rfloor}\vartheta^{(L)}_l\delta^{l}
%\triglatmuloddcorrTwo&=w\sum_{l\in\mathbb{N}_{o}\land l\leq L}(-1)^{(l-1)/2}\frac{1}{l!}\delta^{(l-1)/2}
%\triglatmuloddcorr&=w\sum_{l\in\mathbb{N}_e\land l\leq L-1}(-1)^{l/2}\frac{1}{(l+1)!}v^{l/2}(1-v)^{l/2}w^{l}
%\triglatmuloddcorr&=w\sum_{l\in\mathbb{N}_o\land l\leq L}(-1)^{(l-1)/2}\frac{1}{l!}v^{(l-1)/2}(1-v)^{(l-1)/2}w^{(l-1)}
\end{split}
\end{equation*}

Minimax approximations are also naturally used for $\sin^{-1}$, $\tan^{-1}$ and $\atantwo$; math library functions are typically minimax approximations. Many different approximations are conceivable. \emph{The focus here is not these approximations} for which a vast literature and tool sets exist, see e.g.~\cite{Muller2006,Muller2020,Darulova2019,Brunie2015}. However, to exemplify the performance and to make this work self-contained, basic approximations are used based on the sub-range ($[0,1]$ and $[-1,1]$) approximations
\begin{equation*}
\begin{split}
\{\alpha^{(I)}_{i}:i\in[0,I]\} &= \argmin_{\{\alpha_i:i\in[0,I]\}}\max_{x\in[0,1]}\left| \sin^{-1}(x) -  \tfrac{\pi}{2}+\sqrt{1-x}\sum_{i=0}^I\alpha_ix^i\right|
%\{\alpha^{(I)}_{i}:i\in[1,I]\} &= \argmin_{\{\alpha_i:i\in[0,I]\}}\max_{x\in[0,1]}\left| \sin^{-1}(x) -  \tfrac{\pi}{2}+\sqrt{1-x}\left(\tfrac{\pi}{2}+\sum_{i=1}^I\alpha_ix^i\right)\right|
\\
\{\beta^{(J)}_{j}:j\in[0,J]\} &= \argmin_{\{\beta_j:j\in[0,J]\}}\max_{x\in[-1,1]}\left| \tan^{-1}(x) -  \sum_{j=0}^J\beta_{j}x^{2j+1}\right| %\\
%\{\gamma^{(I)}_{i}:i\in[0,I]\} &= \argmin_{\{\gamma_i:i\in[0,I]\}}\max_{x\in[0,1]}\left| \sin^{-1}(x) -  \tfrac{\pi}{2}+\sqrt{1-x}\sum_{i=0}^I\alpha_ix^i\right|
\end{split}
\end{equation*}
and the resulting polynomials
\begin{equation*}
\arcsinapprox'(x)=\sum^{I}_{i=1}\alpha^{(I)}_ix^i
%\arcsinapprox'(x)=\tfrac{\pi}{2}+\sum^{I}_{i=1}\alpha^{(I)}_ix^i
\quad\text{and}\quad
\atantwoapprox'(x)=\sum^{J}_{j=0}\beta^{(J)}_{j}x^{2j+1}
\end{equation*}
%Note, forcing the zeroth term of $\arcsinapprox'(x)$ to $\tfrac{\pi}{2}$ is a simple way of ensures a continuous approximation at the cost of a factor two in accuracy.
%
%The mapping of these approximations over the partial ranges to the full ranges of $\sin^{-1}$, $\tan^{-1}$ and $\atantwo$ can be implemented in many ways. Particularly, 
These approximations are mapped to the full ranges by the following conditional-free bit-twiddling transformations
\begin{equation}\label{eq:trig}
\begin{split}
\sin^{-1}(x)&\approx\arcsinapprox(x)=\texttt{copysign}\left(\tfrac{\pi}{2}-\sqrt{1-|x|}\arcsinapprox'(|x|),x\right)\\
\tan^{-1}(y,x)&\approx\atanapprox(y,x)=\texttt{copysign}\left(\tfrac{\pi}{4}+\atantwoapprox'\left(\tfrac{|y|-x}{|y|+x}\right),y\right)\quad\forall\quad x\geq 0\\
\atantwo(y,x)&\approx\atantwoapprox(y,x)=\pi d(y,x) + \texttt{copysign}\left(\tfrac{\pi}{4}+\atantwoapprox'\left(\tfrac{|y|-|x|}{|y|+|x|}\right),x\oplus y\right)
\end{split}
\end{equation}
where $d(y,x)=(((x\land y) \ggg (\texttt{sizeof}(x)\cdot 8 -2)) \land -2)\lor(x\gg(\texttt{sizeof}(x)\cdot 8 -1))$, and -2 within it, are Two's complement integers of the size of $x$ and $y$ implicitly converted to real in the multiplication with $\pi$ and where $\texttt{copysign}(\cdot,\cdot)$ and $\texttt{sizeof}(\cdot)$ have the obvious (C/C++) meanings. Further, all logical operators ($\oplus$ is xor) are bit-wise and $\ggg$ and $\gg$ are bit shift with and without sign extension, respectively.
Note, $\atanapprox(0,0)$ and $\atantwoapprox(0,0)$ contain $0/0$. For $\phi$ and $\atanapprox(y,x)$ this is acceptable since it corresponds to the middle of the earth for which $\phi$ is not defined. For $\lambda$ and $\atantwoapprox(0,0)$ this is acceptable since it corresponds to the poles for which $\lambda$ is not defined.
Note the two argument version of arctangent $\tan^{-1}(y,x)$. In contrast to $\atantwo(y,x)$, the second argument is limited and the output is in the range $[-\pi/2,\pi/2]$. The fact that the $\tan^{-1}$ argument can be split in a dividend and a \emph{positive} divisor is instead used to make the mapping from $[-1,1]$ conditional free at the cost of two extra additions. Further, note that the bit manipulations of the real values, which are typically implementation defined for most programming languages, only rely on the real value representation having an initial leading sign bit, so it is most likely possible to implement on almost all platforms but may require some extra care.

%Continuous!? To avoid large relative errors.

\section{Complete polynomial approximations}\label{sec:range}
Combining~\eqref{eq:n},~\eqref{eq:corrections},~\eqref{eq:inverse},  \eqref{eq:partial}, \eqref{eq:n_rr} and~\eqref{eq:trig} gives the four latitude approximations
\begin{align}
\label{eq:lat_approx1}\phi&\approx\arcsinapprox(t) + t\sqrt{1-t^2}\,\lataddcorr[\fourierixlimit_\phi][\taylorixlimit_\phi][h_c][t^2]\\
\label{eq:lat_approx2}\phi&\approx\atanapprox(z,q) + t\sqrt{1-t^2}\,\lataddcorr[\fourierixlimit_\phi][\taylorixlimit_\phi][h_c][t^2]\\
\label{eq:lat_approx3}\phi&\approx\arcsinapprox(t\,\sinlatmulcorr[\mulixlimit_{\phi}][t^2][\lataddcorr[\fourierixlimit_\phi][\taylorixlimit_\phi][h_c][t^2]])\\
\label{eq:lat_approx4}\phi&\approx\atanapprox\left(z\,\sinlatmulcorr[\mulixlimit_{\phi}][t^2][\lataddcorr[\fourierixlimit_\phi][\taylorixlimit_\phi][h_c][t^2]],q\coslatmulcorr[\mulixlimit_{\phi}][t^2][\lataddcorr[\fourierixlimit_\phi][\taylorixlimit_\phi][h_c][t^2]]\right)
\end{align}
and the longitude, n-vector and altitude approximations
\begin{align}
\label{eq:lon_approx}\lambda&\approx\atantwoapprox(y,x)\\
\label{eq:n_approx}\bar{n}&\approx
\begin{bmatrix}
r\,\coslatmulcorr[\mulixlimit_n][t^2][\lataddcorr[\fourierixlimit_n][\taylorixlimit_n][h_c][t^2]]\\
s\,\coslatmulcorr[\mulixlimit_n][t^2][\lataddcorr[\fourierixlimit_n][\taylorixlimit_n][h_c][t^2]]\\
t\,\sinlatmulcorr[\mulixlimit_n][t^2][\lataddcorr[\fourierixlimit_n][\taylorixlimit_n][h_c][t^2]]
\end{bmatrix}\\
\label{eq:alt_approx}h &\approx \altaddcorr[\fourierixlimit_h][\taylorixlimit_h][h_c][t^2]
\end{align}
From these follow the four \emph{lla} ($\{\phi,\lambda,h\}$) transformation approximations
\begin{equation}\label{eq:lat_lon_alt_approx}
\begin{split}
f_{\sin\!+,\{N_\phi,M_\phi,N_h,M_h,I,J\}}(\{x,y,z\})&=\left\{
\eqref{eq:lat_approx1},\eqref{eq:lon_approx},\eqref{eq:alt_approx}
\right\}\\
f_{\tan\!+,\{N_\phi,M_\phi,N_h,M_h,I,J\}}(\{x,y,z\})&=\left\{
\eqref{eq:lat_approx2},\eqref{eq:lon_approx},\eqref{eq:alt_approx}
\right\}\\
f_{\sin\!\times,\{L_\phi,N_\phi,M_\phi,N_h,M_h,I,J\}}(\{x,y,z\})&=\left\{
\eqref{eq:lat_approx3},\eqref{eq:lon_approx},\eqref{eq:alt_approx}
\right\}\\
f_{\tan\!\times,\{L_\phi,N_\phi,M_\phi,N_h,M_h,I,J\}}(\{x,y,z\})&=\left\{
\eqref{eq:lat_approx4},\eqref{eq:lon_approx},\eqref{eq:alt_approx}
\right\}
\end{split}
\end{equation}
and the \emph{nva} ($\{\bar{n},h\}$) transformation approximation
\begin{equation}\label{eq:nh_approx}
f_{\bar{n},\{L_n,N_n,M_n,N_h,M_h\}}(\{x,y,z\})=\left\{
\eqref{eq:n_approx},\eqref{eq:alt_approx}
\right\}
\end{equation}
for finite index limits $I$, $J$, $\mulixlimit_{\phi}$, $\mulixlimit_{n}$, $\fourierixlimit_{\phi}$, $\taylorixlimit_{\phi}$, $\fourierixlimit_{n}$, $\taylorixlimit_{n}$, $\fourierixlimit_{h}$ and $\taylorixlimit_{h}$.

\section{Implementation}\label{sec:implementation}
Implementing~\eqref{eq:lat_lon_alt_approx} and~\eqref{eq:nh_approx} entail computation of the coefficients, suitable polynomial transformations %polynomial evaluation schemas,
 and arithmetic sequences for the polynomials and auxiliary quantities. For an application, a specific approximation with specific index limits, giving the best performance for a given platform for the desirable accuracy or computational cost design constraints, also have to be selected, i.e. a benchmark as exemplified in the next section should be carried out. (Many application will probably get away with just selecting an approximation from Appendix~\ref{appendix:labeled_approx}.)

Further, the reference radius $h_0$ and the range $[{h_c}_{\min},{h_c}_{\max}]$ are parameters of the approximations. Given minimax approximations, $h_0$ is of less importance and can be chosen as $h_0=0$, which eliminates one addition for each transformation. (If other polynomial approximations, i.e. Taylor, Padé, etc., are used, a non-zero $h_0$ will be of importance.) $[{h_c}_{\min},{h_c}_{\max}]$ should cover all $h_c$ values of an application. For most applications, it is easier to set an altitude range $[h_{\min},h_{\max}]$ which  gives $[{h_c}_{\min},{h_c}_{\max}] = [h_{\min} + b - h_0, h_{\max} + a - h_0]$, where $a$ and $b$ are the reference ellipsoid semi-major and semi-minor axes. Note, even a zero altitude range still requires a geocentric altitude range of $a-b\approx 21385$m. For the benchmark, the limits $h_{\min}=-5000$m and $h_{\max}=100\,000$m are primarily used but results for larger and smaller ranges are also provided in Appendix~\ref{appendix:more_results}.
%
%Given minimax approximations, $h_0$ is of less importance and can be chosen as $h_0=0$, which eliminates an addition for each transformation.

The computations of the coefficients $\varsigma^{(L)}_l$, $\vartheta^{(L)}_l$, $\alpha^{(I)}_{i}$ and $\beta^{(J)}_{i}$ only depend on arithmetic and elementary functions. Hence, they were easily computed with the Remez exchange algorithm implementation of \texttt{Sollya}~8.0~\cite{Chevillard2010}. In contrast, the computations of the coefficients $b^{(M)}_{n,m}$ and $c^{(M)}_{n,m}$ were surprisingly hard. They appear numerically ill-conditioned and include numerical solutions of integrals making corresponding proper interval arithmetic challenging to implement. Instead, they were computed with a straight-forward basic Remez exchange algorithm implementation. However, to ensure that the linear system of the algorithm iterations were non-singular for the higher polynomial degrees and higher order Fourier coefficients, it had to be implemented with arbitrary precision arithmetic. Consequently, the integrals of $b_n(h_c)$ and $c_n(h_c)$ had to be solved numerically with arbitrary precision. This in turn requires reference transformations implemented in arbitrary precision. For the arbitrary precision arithmetic, the GMP~\cite{Granlund2012} (linked against), the MPFR~\cite{Mpfr2007} (basic arbitrary precision arithmetic) and the MPLAPACK~\cite{Nakata2022} (overloaded operators and arbitrary precision linear algebra) libraries were used. For the reference transformation, the closed-form transformation by~\cite{Vermeille2004} was used. For the integrals, the Composite Simpson's 1/3 rule was used with 128 sub-intervals. (In principle something more suitable such as Clenshaw-Curtis integration should be used but, in practice, the integrals are benign enough to use Simpson's.) All computations were done with 200 bits of precision. The Remez algorithm termination criteria was set to ratio of the maximum and minimum error being less than $1+10^{-5}$.
This gave coefficient values with $\sim$20 decimal digits of precision, giving some more precision than the $\leq$17 digits of double precision to handle final polynomial coefficients computations of $\lataddcorr[\fourierixlimit][\taylorixlimit][u][v]$ and $\altaddcorr[\fourierixlimit][\taylorixlimit][u][v]$, which were done with~\cite{Ginac}.

Beyond the coefficients, the polynomial evaluation has many degrees of freedom and even if arithmetic trees of minimal size could be found, in practice, they will typically be suboptimal due to instruction parallelism of modern superscalar processors and, in general, a large number of permutations has to be benchmarked for each polynomial order to find the best evaluation, a daunting task. See~\cite{Ewart2020} for a glimpse of what throughout optimization entails. And even an optimal evaluation, put in the overall arithmetic sequence, may not be optimal anymore. That said, Estrin's schema has been used as a middle-of-the-road approach. Only marginal differences has been observed in comparison with other general evaluation schema such as direct evaluation or Horner-2. The power factors of the schema were constructed by starting from the smallest power and solving the minimal \emph{change-making} problem with respect to the already computed powers.
%
% a plain balanced evaluation with respect to the variable of the highest degree typically gives close-to-optimal performance and has been used throughout. % for a wide range of general purpose processors.
%Further, as seen in Fig.~\ref{fig:ix_limits}, in general, $M<L$ and, hence, for exemplifying preformance, Horner-2 with respect to $v$, and secondarily with respect to $u$, has been used throughout. %The polynomials, on Horner-2 form, of selected approximations can be found in Appendix~\ref{appendix:labeled_approx}. %Relative to a Horner-1 evaluation, the performance improvement is a few percent.

With the polynomials at hand, a reasonable overall arithmetic tree and an overall arithmetic sequence (traversal of the overall arithmetic tree/graph) of the polynomials and the auxiliary quantities is straight-forward to find. Pseudo-code of the approximations, defining such sequences are found in Table~\ref{tab:impl}.
The approximations of Table~\ref{tab:impl}, as well as the literature methods, have been implemented in C++, using IEEE-754 double precision. %(Note, an ulp of IEEE-754 single precision at sea level is roughly 0.5m making it insufficient for many applications.)
Polynomial coefficients were pre-computed and other derived constants computed at compile time. The implementations were compiled with gcc 13.2 with \texttt{-O3 -march=native}. \texttt{-O3} enables auto-vectorization and \texttt{-march=native} enables AVX instructions for the benchmark platform, which is beneficial for the polynomial evaluations. Results without AVX instructions are also provided in Appendix~\ref{appendix:more_results}. %, giving a reduction in latency 0 to $\sim$30\% for the tested methods but does not change the overall picture.

\begin{table}[p]
\begin{tabular}[t]{|l|l|}
\hline
$f_{\sin\!+,\{N_\phi,M_\phi,N_h,M_h,I,J\}}(\{x,y,z\})$ & $f_{\tan\!+,\{N_\phi,M_\phi,N_h,M_h,I,J\}}(\{x,y,z\})$\\
\hspace{2mm}
\begin{minipage}{\impltablewidth\textwidth}
\setlength{\jot}{0mm}% Decrease align vertical spacing.
\vspace{-3mm}
$\begin{aligned}
p&=\sqrt{x^2+y^2+z^2}\\
t&=z/p\\
t^2&=tt\\[-1mm]
\phi&=\arcsinapprox(t) + t\sqrt{1-t^2}\,\lataddcorr[\fourierixlimit_\phi][\taylorixlimit_\phi][p][t^2]\\
\lambda&=\atantwoapprox(y,x)\\
h&=\altaddcorr[\fourierixlimit_h][\taylorixlimit_h][p][t^2]\\[1mm]
&\hspace{-2.5mm}\text{return}\,\{\phi,\lambda,h\}
\end{aligned}$
\end{minipage}
&
\hspace{2mm}
\begin{minipage}{\impltablewidth\textwidth}
\setlength{\jot}{0mm}% Decrease align vertical spacing.
\vspace{1mm}
$\begin{aligned}
p&=\sqrt{x^2+y^2+z^2}\\[-1mm]
q&=\sqrt{x^2+y^2}\\
t&=z/p\\[-0.5mm]
t^2&=tt\\[-1mm]
\phi&=\atanapprox(z,q) + t\sqrt{1-t^2}\,\lataddcorr[\fourierixlimit_\phi][\taylorixlimit_\phi][p][t^2]\\
\lambda&=\atantwoapprox(y,x)\\
h&=\altaddcorr[\fourierixlimit_h][\taylorixlimit_h][p][t^2]\\[1mm]
&\hspace{-2.5mm}\text{return}\,\{\phi,\lambda,h\}
\end{aligned}$
\end{minipage}
\hspace{2mm}
\\ \hline
$f_{\sin\!\times,\{L_\phi,N_\phi,M_\phi,N_h,M_h,I,J\}}(\{x,y,z\})$ &
$f_{\tan\!\times,\{L_\phi,N_\phi,M_\phi,N_h,M_h,I,J\}}(\{x,y,z\})$\\
\hspace{2mm}
\begin{minipage}{\impltablewidth\textwidth}
\setlength{\jot}{0mm}% Decrease align vertical spacing.
$\begin{aligned}
p&=\sqrt{x^2+y^2+z^2}\\
t&=z/p\\
t^2&=tt\\[-0.5mm]
\omega &= \lataddcorr[\fourierixlimit_\phi][\taylorixlimit_\phi][p][t^2]\\
\delta &= t^2 (1 - t^2) w^2\\
\sigma &=\triglatmulevencorrTwo[L_\phi][\delta]\\
\tau &=\triglatmuloddcorrTwo[L_\phi][\omega][\delta]\\
\phi&=\arcsinapprox(t(\sigma+(1-t^2)\tau))\\
\lambda&=\atantwoapprox(y,x)\\
h&=\altaddcorr[\fourierixlimit_h][\taylorixlimit_h][p][t^2]\\[1mm]
&\hspace{-2.5mm}\text{return}\,\{\phi,\lambda,h\}
\end{aligned}$
\end{minipage}
&
\hspace{2mm}
\begin{minipage}{\impltablewidth\textwidth}
\setlength{\jot}{0mm}% Decrease align vertical spacing.
\vspace{1mm}
$\begin{aligned}
p&=\sqrt{x^2+y^2+z^2}\\[-1mm]	
q&=\sqrt{x^2+y^2}\\
t^2&=(z/p)^2\\
%t^2&=tt\\[-1.5mm]
\omega &= \lataddcorr[\fourierixlimit_\phi][\taylorixlimit_\phi][p][t^2]\\
\delta &= t^2 (1 - t^2) w^2\\
\sigma &=\triglatmulevencorrTwo[L_\phi][\delta]\\
\tau &=\triglatmuloddcorrTwo[L_\phi][\omega][\delta]\\
%\eta&=\sigma+(1-t^2)\tau\\
%\rho&=\sigma-t^2\tau\\
\phi&=\atanapprox\left(z(\sigma+(1-t^2)\tau),q(\sigma-t^2\tau)\right)\\
\lambda&=\atantwoapprox(y,x)\\
h&=\altaddcorr[\fourierixlimit_h][\taylorixlimit_h][p][t^2]\\[1mm]
&\hspace{-2.5mm}\text{return}\,\{\phi,\lambda,h\}
\end{aligned}$
\end{minipage}
\\ \hline $f_{\bar{n},\{L_n,N_n,M_n,N_h,M_h\}}(\{x,y,z\})$ &\\
\hspace{2mm}
\begin{minipage}{\impltablewidth\textwidth}
\setlength{\jot}{0mm}% Decrease align vertical spacing.
\vspace{1mm}
$\begin{aligned}
p&=\sqrt{x^2+y^2+z^2}\\
t&=z/p\\
t^2&=tt\\[-0.5mm]
\omega &= \lataddcorr[\fourierixlimit_n][\taylorixlimit_n][p][t^2]\\
\delta &= t^2 (1 - t^2) w^2\\
\sigma &=\triglatmulevencorrTwo[L_n][\delta]\\
\tau &=\triglatmuloddcorrTwo[L_n][\omega][\delta]\\
\rho'&=(\sigma-t^2\tau)/p\\
\bar{n}&=\left[x\rho',y\rho',t(\sigma+(1-t^2)\tau)\right]\\
h&=\altaddcorr[\fourierixlimit_h][\taylorixlimit_h][p][t^2]\\[1mm]
&\hspace{-2.5mm}\text{return}\,\{\bar{n},h\}
\end{aligned}$
\end{minipage}
&\\
\hline
\end{tabular}
\caption{\label{tab:impl}\footnotesize Pseudo-code implementation  for the five different ECEF ($\{x,y,z\}$) to geodetic ($\{\phi,\lambda,h\}$ or $\{\bar{n},h\}$) coordinate transformations. Equality sign imply assignment. Powers are assumed implemented inline unless available as a left hand side variable. Note that $h_0=0$ and has been eliminated. A selection of polynomials can be found in Appendix~\ref{appendix:labeled_approx}.
}
\end{table}

\section{Benchmark}\label{sec:benchmark}
The performance of the coordinate transformation methods are quantified with pairs of computational cost and error measurements. The desirable way to measure error may vary significantly for different applications. %Overall, the target here is worst case performance for general modern CPUs.
Hence, the results of multiple error measures are presented.
In contrast, for most applications, the most interesting measure of computational cost, and the only one presented, is latency.

Normally, measuring anything but mean latency for algorithms is difficult, especially for the current algorithms since the latencies are in many cases data dependent. Hence, the mean has to be with respect to data. To separate the algorithm from the system influence, the minimum mean latency of multiple tests is used.
The latency results were obtained with Nanobench 4.3~\cite{Ankerl2022} using the minimum of the mean latency per sample of sets of 1000 samples repeated $10\,000$ times. This gave stable results for all methods. %Note that the values for the n-vector versions also include the computation of altitude.
For further details of the latency measurements, see Appendix~\ref{sec:appendix_benchmark}.

Error measurements were separately computed with $5\cdot 10^8$ uniformly distributed ECEF samples (ignoring elliptical effects) $\{x^\ellipse_k, y^\ellipse_k, z^\ellipse_k\}$ and geodetic references $\{\phi_k,\lambda_k,h_k\}$ and $\bar{n}_k$ over the volume of the respective altitude ranges, see~\cite{Rosca2010} for details.
The surface area of the earth is $\sim 5\cdot 10^8\text{km}^2$ so it means one sample per $\text{km}^2$.
Extra samples were added around the poles, where methods frequently have sharp error peaks. % TODO desribe better.
%
%See Appendix~\ref{appendix:a} for details of how the uniform samples are obtained.
%
All error computations were done in the Intel 80-bit extended precision format (\texttt{long double} on Linux), ensuring that the limits of the IEEE-754 double precision can be observed.
Let $f_{lla/nva}(x,y,z)$ denote the transformation to be evaluated and the corresponding \emph{lla} and \emph{nva} transformation samples
\begin{equation*}
\{\tilde{\phi}_k,\tilde{\lambda}_k,\tilde{h}_k\}
= f_{lla}(x^\ellipse_k, y^\ellipse_k, z^\ellipse_k)\quad\textrm{and}\quad
\{\tilde{n}_k,\tilde{h}_k\}= f_{nva}(x^\ellipse_k, y^\ellipse_k, z^\ellipse_k)
\end{equation*}
respectively. Further, for the \emph{nva}, latitude and longitude values are computed with
\begin{equation*}
\begin{split}
\tilde{\phi}_k&=\atantwo({n_k}_z, \sqrt{{n_k}_x^2 + {n_k}_y^2})\\
\tilde{\lambda}_k & =  \atantwo({n_k}_y, {n_k}_x)
\end{split}
\end{equation*}
with extended precision, where $\tilde{n}_k=[{n_k}_x;{n_k}_y;{n_k}_z]$.
Let $e_\alpha(\phi_k,\tilde{\phi}_k,\lambda_k,\tilde{\lambda}_k,h_k,\tilde{h}_k)$ and $e_\alpha(\phi_k,\tilde{\phi}_k,\lambda_k,\tilde{\lambda}_k,h_k,\tilde{h}_k, \bar{n}_k, \tilde{n}_k)$ denote $\alpha$ error measures for \emph{lla} and \emph{nva} transformations, respectively. See Appendix~\ref{appendix:b} for the definitions of the errors measures. The accuracy of $f_{lla/nva}(x,y,z)$ is then quantified with the maximum error
\begin{equation*}
\begin{split}
&\max_k(|e_\alpha(\phi_k,\tilde{\phi}_k,\lambda_k,\tilde{\lambda}_k,h_k,\tilde{h}_k)|)\\
&\max_k(|e_\alpha(\phi_k,\tilde{\phi}_k,\lambda_k,\tilde{\lambda}_k,h_k,\tilde{h}_k, \bar{n}_k, \tilde{n}_k)|)
\end{split}
\end{equation*}
respectively. (However, with the uniform samples, any statistics of the errors could be computed.)

For the polynomial approximations, the number of possible index limit combinations are too great for them to be exhaustively benchmarked.
Initial narrowing of index limits is discussed in Appendix~\ref{sec:selection}. This still gives a test set of $\sim3000$ index limit combination. Therefore, only results for selected approximations are labeled.
The \emph{nva} approximations have distinct error levels corresponding to different $N$. Consequently, the selection process is rather straight-forward. In contrast, the \emph{lla} approximations demonstrates an almost continuous range of errors. Consequently, a simple selection of \emph{lla} approximations matching the selected \emph{nva} approximations is used. The selected, in the charts labeled and in Appendix~\ref{appendix:labeled_approx} provided (corresponding polynomials) approximations are:

%\begin{table}[ht]
%\resizebox{\textwidth}{!}{%
{\footnotesize
\begin{longtable}{p{.11\textwidth} p{.84\textwidth}} 
%\begin{tabularx}{\textwidth}{@{}>{\bfseries}l@{\hspace{.5em}}X@{}}
$f_{\bar{n},\{0,0,0,0,0\}}$ & The simplest possible \emph{nva} approximation corresponding to a spherical earth. The maximum Euclidean, horizontal and altitude errors are 
$2.1\cdot10^4$m, 
$2.1\cdot10^4$m and 
$1.1\cdot10^4$m, respectively. \\
$f_{\bar{n},\{1,1,0,1,0\}}$ & $N=1$ approximation with the maximum Euclidean, horizontal and altitude errors
$85$m, 
$85$m and 
$13$m, respectively. Note that there is a significant n-vector magnitude error of  $2.8\cdot10^{-6}$. This gives an equivalent gravity error which is order of magnitude the same as typical gravity anomalies.\\
%https://en.wikipedia.org/wiki/Gravity_anomaly
$f_{\bar{n},\{2,2,1,2,0\}}$ & $N=2$ approximation with the maximum Euclidean, horizontal and altitude errors
$0.44$m, 
$0.44$m and 
$0.10$m, respectively. The magnitude error is $9.9\cdot10^{-12}$. Note that there are faster approximations with roughly the same Euclidean error but with a significant magnitude error. The selected approximation gives the lowest relative gravity error of $6.8\cdot10^{-8}$ for the range, well below typical gravity anomalies. \\
$f_{\bar{n},\{3,3,2,3,2\}}$ & $N=3$ approximation with the maximum Euclidean, horizontal and altitude errors
$1.2\cdot10^{-3}$m, 
$1.2\cdot10^{-3}$ and 
$4.8\cdot10^{-4}$m, respectively. The magnitude error is $6.7\cdot10^{-13}$. \\
$f_{\bar{n},\{3,4,3,4,2\}}$ & $N=4$ approximation with the maximum Euclidean, horizontal and altitude errors
$4.7\cdot10^{-6}$m, 
$4.7\cdot10^{-6}$m and 
$1.2\cdot10^{-6}$m, respectively. The magnitude error is again $6.7\cdot10^{-13}$.\\
$f_{\bar{n},\{4,5,4,5,3\}}$ & $N=5$ approximation with the maximum Euclidean, horizontal and altitude errors
$2.1\cdot10^{-8}$m, 
$2.1\cdot10^{-8}$m and 
$4.1\cdot10^{-9}$m, respectively. The magnitude error is at the numerical limit at $2.9\cdot10^{-16}$. Note, $L=3$ gives a marginally faster approximation with essentially the same Euclidean error but the magnitude error become significant giving worse $g$ approximation. (However, as noted, such $g$ errors are below typical gravity anomalies.) \\
$f_{\bar{n},\{4,6,4,5,3\}}$ & $N=6$ approximation with the maximum Euclidean, horizontal and altitude errors just above the numerical limits at 
$5.5\cdot10^{-9}$m, 
$5.3\cdot10^{-8}$m and 
$4.1\cdot10^{-9}$m, respectively. The magnitude error is at the numerical limit at $2.9\cdot10^{-16}$. Note that the altitude approximation only has $N=5$.

%\end{tabularx}
\end{longtable}
}
%}%
%\vspace{2mm}
%\caption{\label{tab:lit_methods}\footnotesize Numbering in the plots and comments to implemented methods form the literature.
%}
%\end{table}

%\begin{table}[ht]
%\resizebox{\textwidth}{!}{%
{\footnotesize
\begin{longtable}{p{.2\textwidth} p{.75\textwidth}} 
%\begin{tabularx}{\textwidth}{@{}>{\bfseries}l@{\hspace{.5em}}X@{}}
$f_{\sin\!\times,\{0,0,0,0,0,2,3\}}$ & The simplest possible \emph{lla} approximation. The Euclidean, horizontal and altitude errors are 
$2.0\cdot10^4$m,
$2.0\cdot10^4$m and
$1.1\cdot10^4$m, respectively. \\
$f_{\sin\!+,\{1,0,1,0,4,5\}}$ & $N=1$ approximation with Euclidean, horizontal and altitude errors of
$114$m, 
$113$m and 
$13$m, respectively. \\
$f_{\sin\!+,\{2,1,2,0,7,8\}}$ & $N=2$ approximation with Euclidean, horizontal and altitude errors of
$0.41$m, 
$0.40$m and 
$0.10$m, respectively. \\
$f_{\tan\!+,\{3,2,3,1,12,12\}}$ & $N=3$ approximation with Euclidean, horizontal and altitude errors of 
$1.2\cdot10^{-3}$m, 
$1.2\cdot10^{-3}$m and 
$4.8\cdot10^{-4}$m, respectively.\\
$f_{\tan\!\times,\{3,4,3,4,2,14,14\}}$ & $N=4$ approximation with Euclidean, horizontal and altitude errors of 
$7.9\cdot10^{-6}$m, 
$7.9\cdot10^{-6}$m and 
$1.2\cdot10^{-6}$m, respectively. \\
$f_{\tan\!\times,\{4,5,4,5,3,17,17\}}$ & $N=5$ approximation with Euclidean, horizontal and altitude errors of 
$2.5\cdot10^{-8}$, 
$2.5\cdot10^{-8}$ and 
$4.1\cdot10^{-9}$, respectively. \\
$f_{\tan\!\times,\{4,6,4,5,3,18,18\}}$ & $N=6$ approximation with Euclidean, horizontal and altitude error of 
$8.2\cdot10^{-9}$, 
$7.7\cdot10^{-9}$ and 
$4.1\cdot10^{-9}$, respectively. Note that the altitude approximation only has $N=5$.
%\end{tabularx}
\end{longtable}
}
%}%
%\vspace{2mm}
%\caption{\label{tab:lit_methods}\footnotesize Numbering in the plots and comments to implemented methods form the literature.
%}
%\end{table}

The results of the literature methods, against which the approximations are benchmarked, are labeled with numbers. These numbers and the corresponding methods are listed in chronological order below, together with short comments about the implementation and the results.
Note that \emph{all iterative methods have been implemented without trigonometric function evaluations in the iteration loop!} Methods of which trigonometric functions within the iteration loop has not been avoidable, e.g.~\cite{Jones2002}, simply have much longer latencies and have not been included in the benchmark.

%\begin{table}[ht]
%\resizebox{\textwidth}{!}{%
{\footnotesize
\begin{longtable}{p{.08\textwidth} p{.88\textwidth}} 
%\begin{tabularx}{\textwidth}{@{}>{\bfseries}l@{\hspace{.5em}}X@{}}
0 & Spherical earth approximation. Not including any elementary function approximations, i.e. direct implementation of~\eqref{eq:spherical}. This is in many ways the fastest (and least accurate) conceivable approximation and a special case of the presented method. \\
1-6 & Iterative method by~\cite{Hirvonen1959} with increasing number of iterations. The results start with zero iterations, using only the suggested initialization. Note that the method is often referred to as Hirvonen and Mortiz's since it appear in a publications bearing both names~\cite{Hirvonen1963}. Further, it is also often referred to as Heiksanen and Mortiz's since it appears in their 1967 book~\cite{Heiskanen1967}. Similarly, the same methods is also referred to as Torge's, e.g.~\cite{Voll1990}, since it appears in~\cite{Torge1975}. \\ %Implemented without trigonometric function evaluations in the iteration loop. \\
7 & Series expansion method by~\cite{Morrison1961}. All but the necessary arctan:s have been replaced with multi-angle formulas and small angle approximations. The altitude formula has been replaced by that of Bowring. \\
8 & Coarse method by~\cite{Gersten1961}. \\
9-11 & Method by~\cite{Baird1964}. It performs poorly around the poles and, therefore, the maximum errors levels out. Closer than $\sim6$cm from the poles, it returns NaN and this area has been replaced with a spherical approximation.\\
12 & Method by~\cite{Paul1973}. It performs poorly around the equator, hence, the poor performance. \\
13-14 & Original method by~\cite{Bowring1976} with one and two iterations. \\
15 & Closed form method by~\cite{Heikkinen1982}. \\% The method is also occasionally referred to as Kaplan's from his presentation in his 1996 book~\cite{Kaplan1996}. Heikkinen's original publication is in German.\\
16-18 & Method by~\cite{Bowring1976} with refined altitude calculations from~\cite{Bowring1985} and 1-3 iterations. \\
19 & Method by~\cite{Ozone1985}. Poor conditioning makes the method perform poorly around the poles, hence the poor results. Except for the poles, it achieves the numerical precision. \\
20-26 & Iterative method by~\cite{Wei1986} starting from zero iterations. \\
27-33 & Second iterative method by~\cite{Wei1986} starting from zero iterations. \\
34-37 & Method by Goad (1987) retrieved from~\cite{Voll1990} with increasing number of iterations starting from zero. \\
38 & Closed formula method by~\cite{Borkowski1987} excluding inside-evolute handling. \\
39 & Closed formula method by~\cite{Zhu1994}. \\
40-42 & Iterative method by~\cite{Lin1995} with increasing number of iterations. The results start with zero iterations, using only the suggested initialization.\\
43 & Approximation method by~\cite{Ohlson1996}. No n-vector method is implementable since the algorithm works directly on the latitude. \\ %Fast, partially due to usage of $\sin^{-1}$ and $\cos^{-1}$ rather than $\tan^{-1}$, with obvious drawbacks.\\
44 & Single iteration Bowrings's method optimized by~\cite{Toms1996}. The method is not suitable for n-vector implementation. \\
45-50 & Iterative method by~\cite{Sjoberg1999} with increasing number of iterations starting from zero. The suggested polar region iterations are not implemented but rather only the adapted altitude formula~\cite{Wahlberg2009}. \\
51-56 & Iterative method by~\cite{Sjoberg1999} with altitude computations according to~\cite{Bowring1985}. \\
57-62 & Iterative method by~\cite{Fukushima1999} with increasing number of iterations. \\
63 & Sofair's revised method~\cite{Sofair2000}. It does perform particularly poor at the poles but not particularly good anywhere. \\
64-67 & Iterative altitude first method by~\cite{Pollard2002} with increasing number of iterations. The results start with two iterations. Less iterations gives error outside the plots. \\
68-71 & Iterative latitude first method by~\cite{Pollard2002} with increasing number of iterations. The results start with zero iterations, using only the suggested initialization. \\
72-74 & Method by~\cite{Wu2003} with 0-2 iterations. One square root is avoided by reformulating the inverse cotangent to an inverse tangent and for the n-vector version, the altitude and n-vector normalization is done jointly, avoiding an extra square root.\\
75 & Closed formula method by~\cite{Vermeille2004} with n-vector version by~\cite{Gade2007}. \\
76 & Iterative method by~\cite{Fukushima2006}. Note, with more than one iteration it is possible to find points for which it diverges. It also been noted for high altitudes by Ward~\cite{Ward2020} but it may also happen for low altitudes. \\
77 & Close form method by~\cite{Sjoberg2008}. \\
78-83 & Iterative method \emph{a.2} by~\cite{Feltens2008} with increasing number of iterations starting from zero iterations. \\
84-89 & Iterative method \emph{b.1} by~\cite{Feltens2008} with increasing number of iterations starting from zero iterations. \\
90-91 & Method by~\cite{Shu2010}. \\
92 & Closed formula method by~\cite{Vermeille2011} excluding inside evolute handling. \\
93-95 & Method I by~\cite{Ligas2011} with one to three iterations. \\
96-97 & Method II by~\cite{Ligas2011} with one and two iterations. \\
98 & Method by~\cite{Karney2011} provided in \texttt{GeographicLib} but with special handling for non-WGS84 spheroids removed. Note, this is an adaptation of~\cite{Vermeille2004}. \\
99 & Closed formula method by\cite{Osen2017}. \\
100-102 & Perturbation method by~\cite{Hmam2018} with increasing ($3^{rd}$, $4^{th}$ and $5^{th}$) order of the approximation. \\
103,104 & Method by~\cite{Sampson1982} and modified version by Claessens. Implementations from~\cite{Claessens2019}. \\
105,106 & Method by~\cite{Uteshev2018} and modified version by Claessens. Implementations from~\cite{Claessens2019}. \\
107 & Bowring's method modified by~\cite{Claessens2019}. \\
108 & Spherical method by~\cite{Claessens2019}. \\
109-111 & Method by Dave Knopp found in the Peridetic code base~\cite{Knopp2021} with 1-3 iterations.
%60 & Method by Nautilya~\cite{Nautiyal1988}. \\
%61 & Lupash's 1985 iterative method~\cite{Lupash1985}. The method diverges for points close to the poles and hence it is not seen in the plots. Lupash is very similar to Baird's 1964 method~\cite{Churchyard1985}. \\% Compare with by Wu
%\end{tabularx}
\end{longtable}
}
%}%
%\vspace{2mm}
%\caption{\label{tab:lit_methods}\footnotesize Numbering in the plots and comments to implemented methods form the literature.
%}
%\end{table}

\begin{figure}[!p]
\begin{center}
\includegraphics[width=0.49\textwidth]{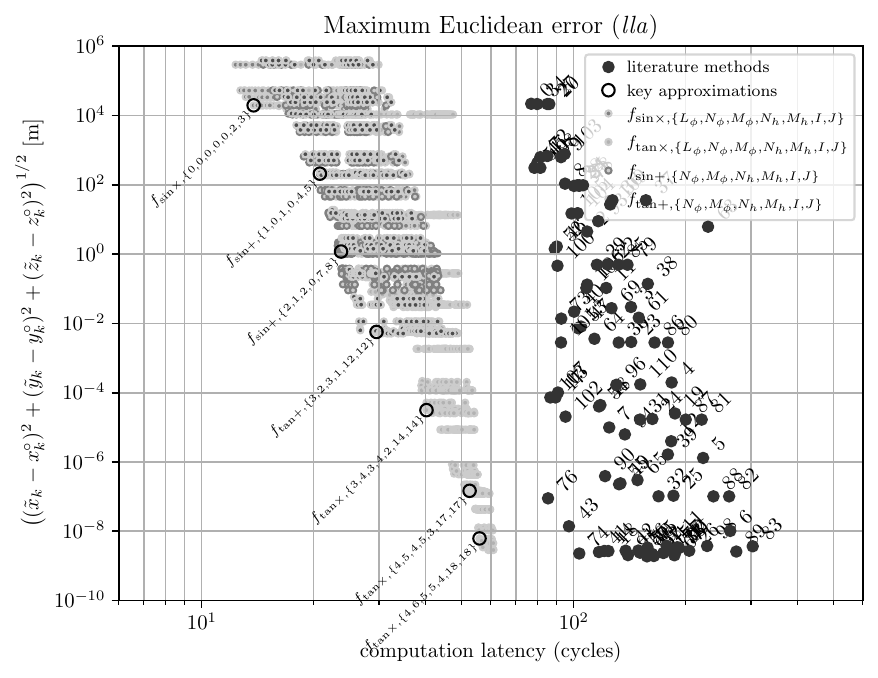}
\includegraphics[width=0.49\textwidth]{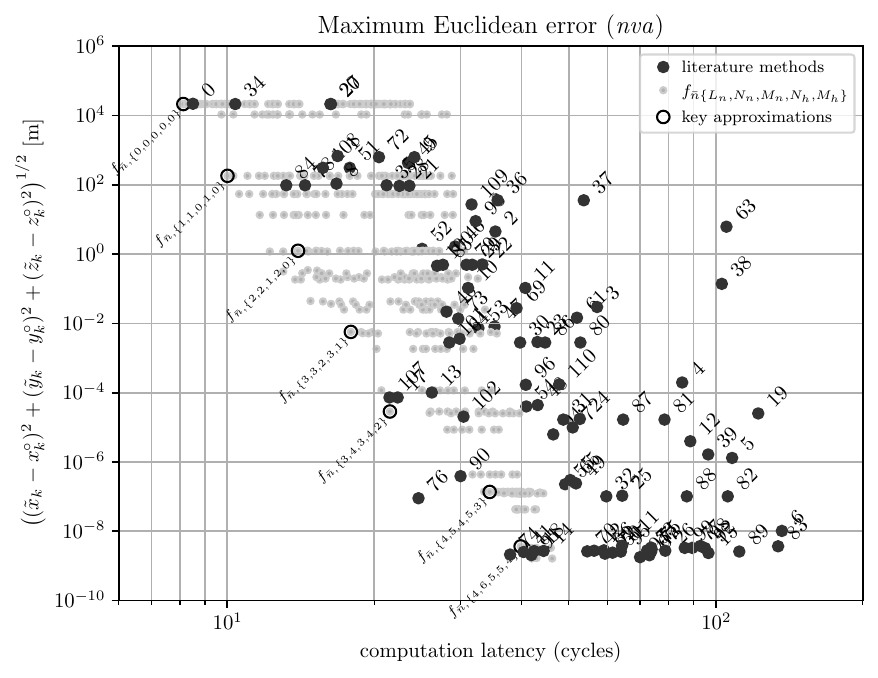}
\includegraphics[width=0.49\textwidth]{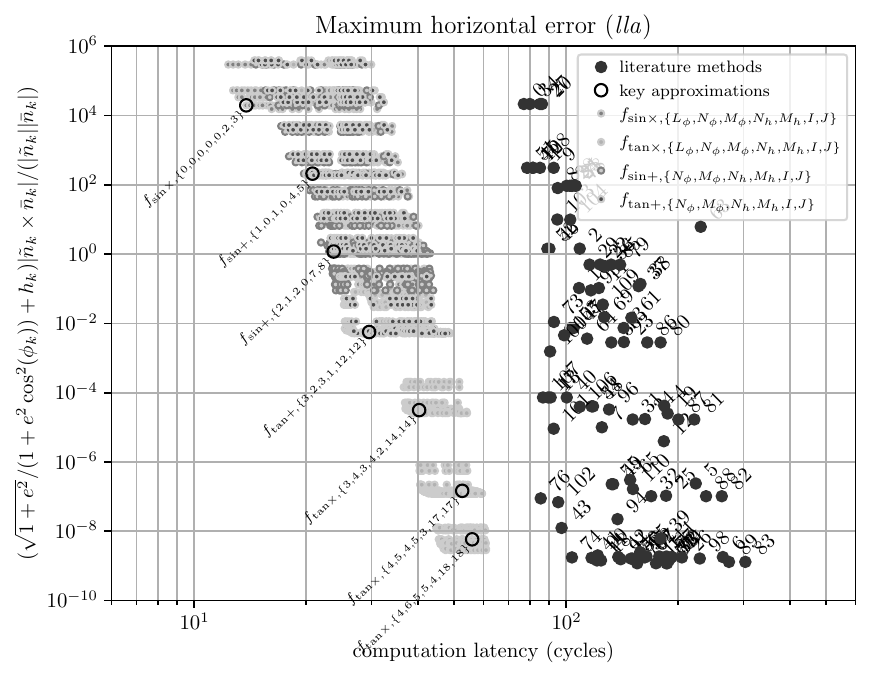}
\includegraphics[width=0.49\textwidth]{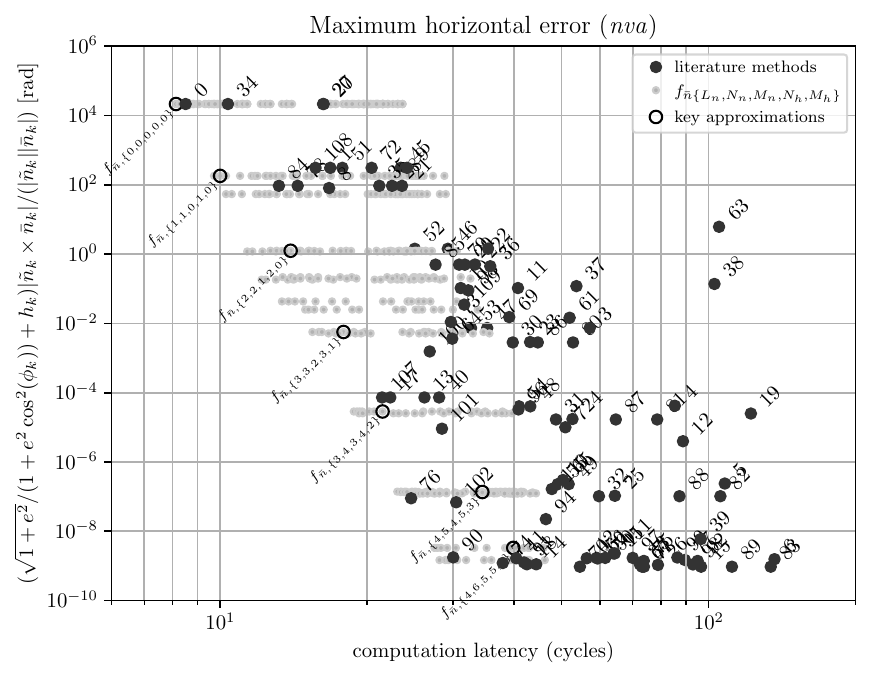}
\includegraphics[width=0.49\textwidth]{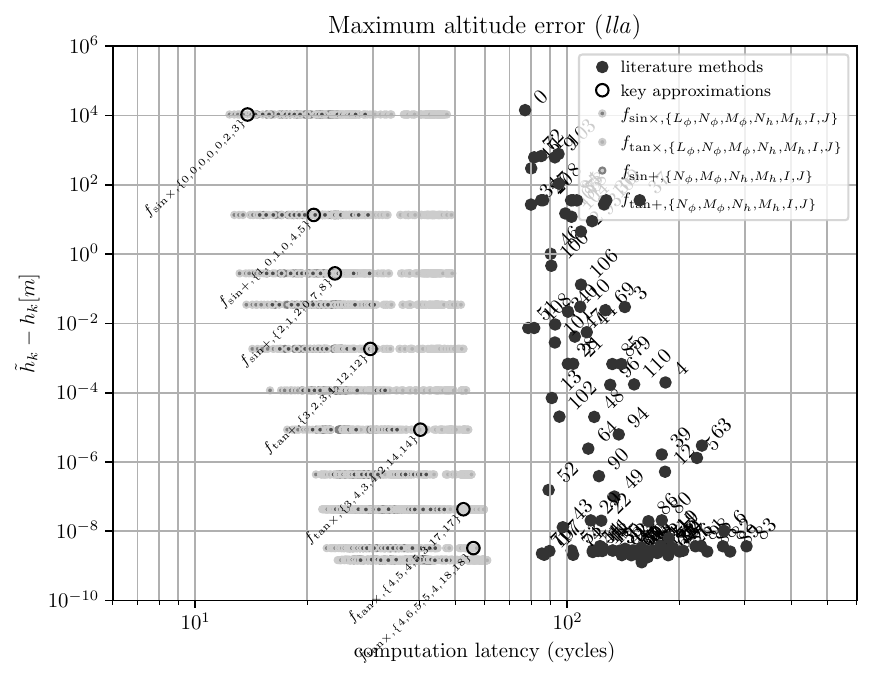}
\includegraphics[width=0.49\textwidth]{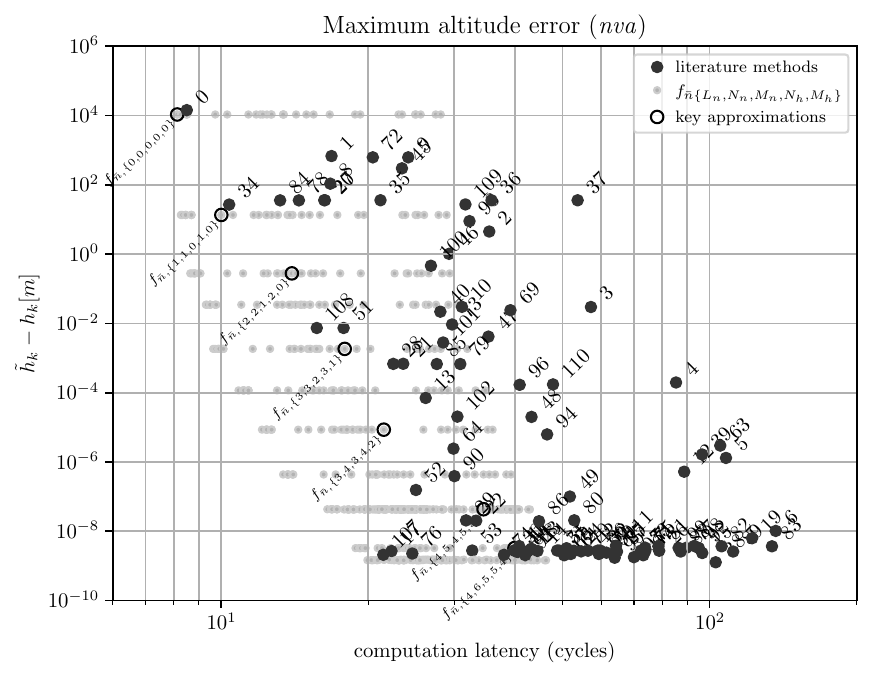}
\caption{\footnotesize Total (Euclidean distance), horizontal (spherical and small angle approximation), and vertical (altitude) maximum absolute error over the volume of the altitude interval $[-5000,100\,000]$. See Appendix~\ref{appendix:b} for detailed definitions. Grey points are the complete set of tested polynomial approximations. Circled and labeled points are approximation of particular interest. Black points are the indicated literature methods.}\label{fig:lat_and_alt_err}
\end{center}
\end{figure}

\begin{figure}[!p]
\begin{center}
\includegraphics[width=0.49\textwidth]{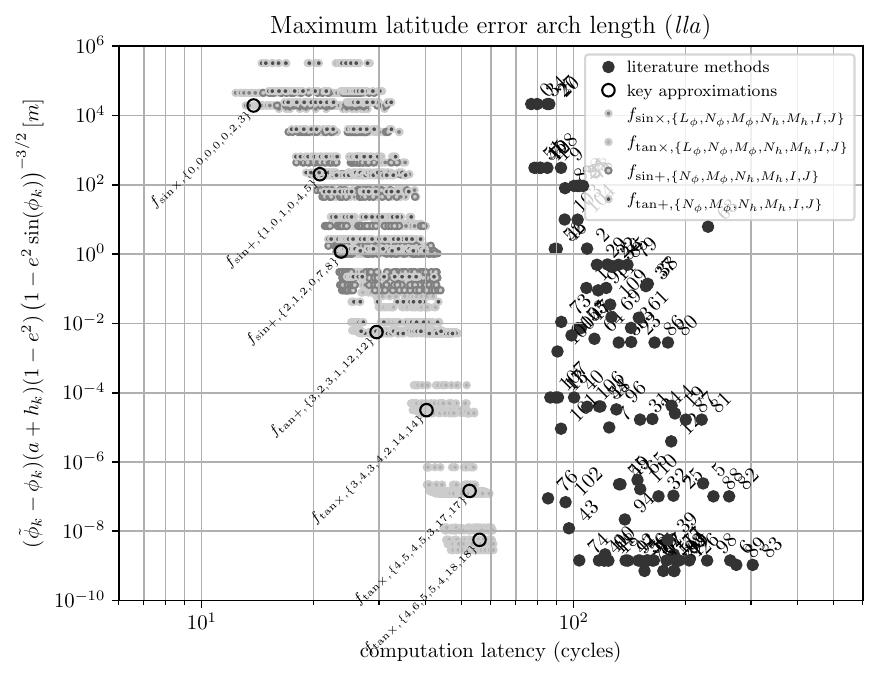}
\includegraphics[width=0.49\textwidth]{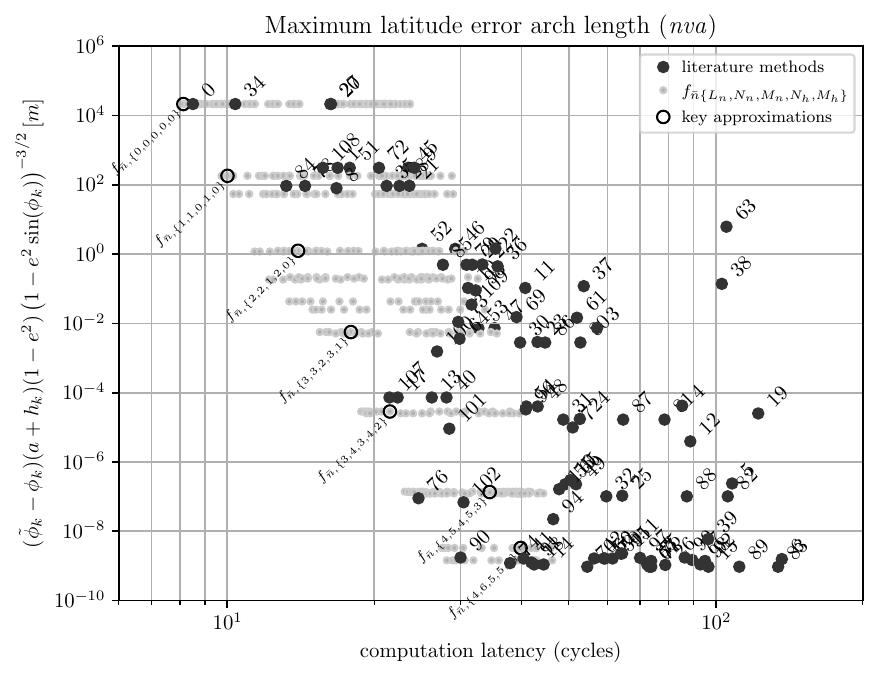}
\includegraphics[width=0.49\textwidth]{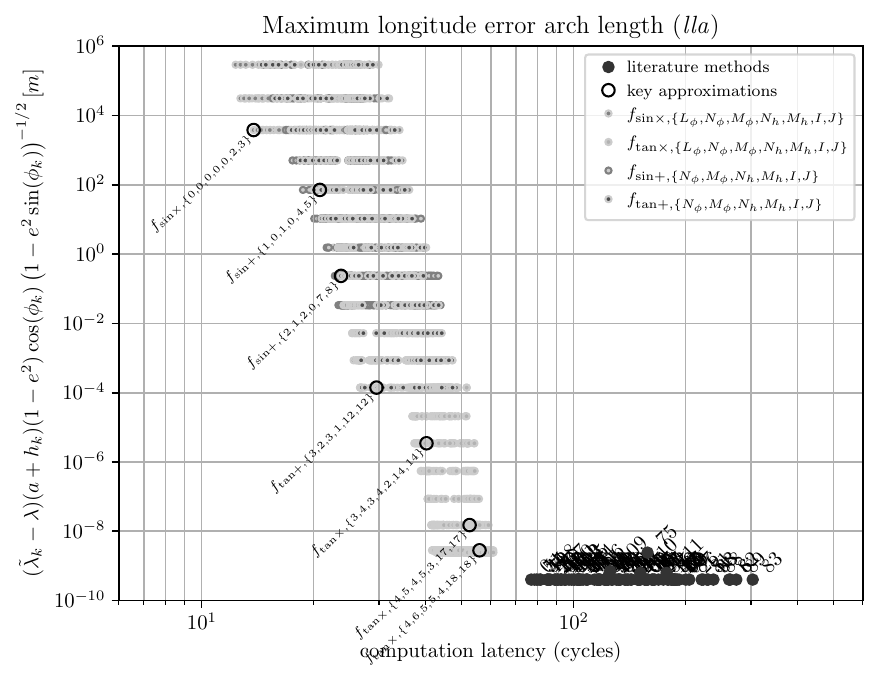}
\includegraphics[width=0.49\textwidth]{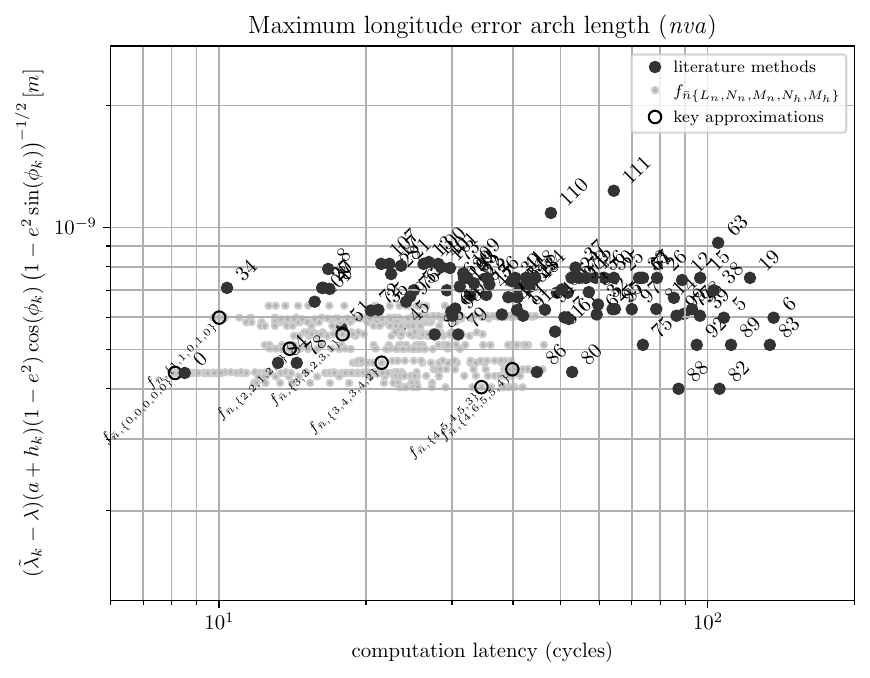}
\includegraphics[width=0.49\textwidth]{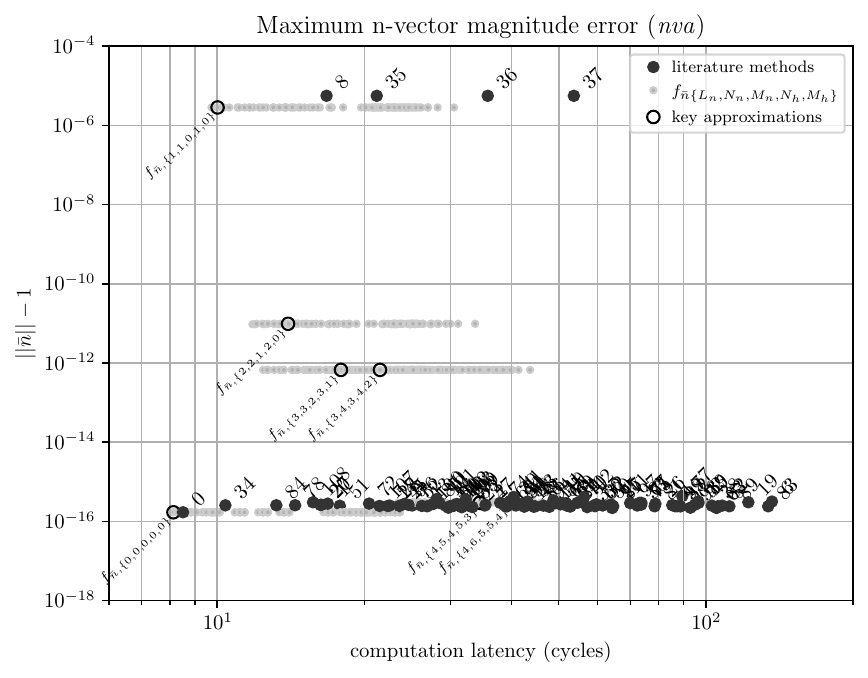}
\includegraphics[width=0.49\textwidth]{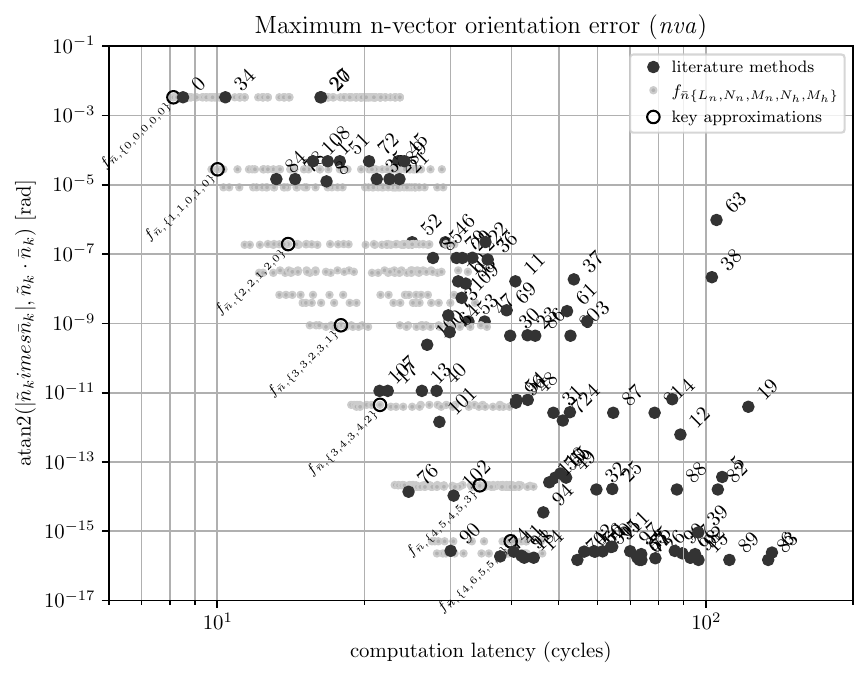}
\caption{\footnotesize Latitude and longitude (arch length) and n-vector (magnitude and direction) maximum absolute error over the volume of the altitude interval $[-5000,100\,000]$. See Appendix~\ref{appendix:b} for detailed definitions. Grey points are the complete set of polynomial approximations. Black points are the indicated literature methods.}\label{fig:lat_lon_and_n_err}
\end{center}
\end{figure}

\begin{figure}[!ht]
\begin{center}
\includegraphics[width=0.49\textwidth]{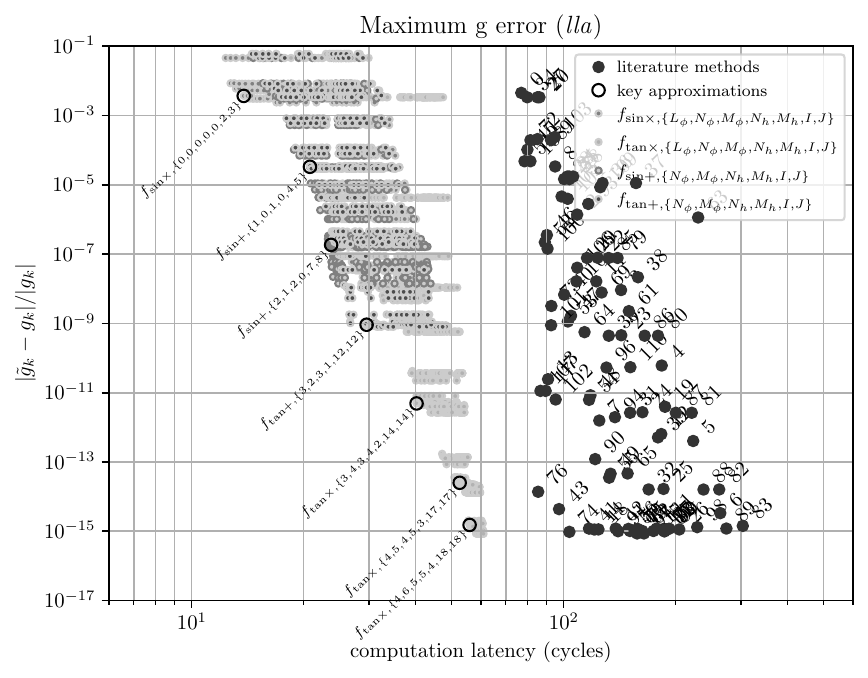}
\includegraphics[width=0.49\textwidth]{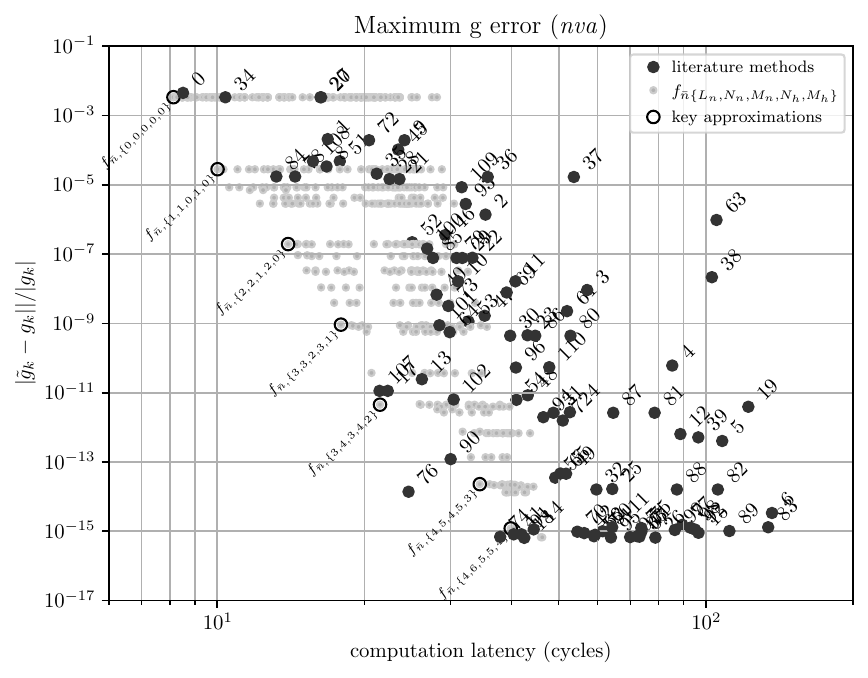}
\caption{\footnotesize Fractional gravity (Euclidean distance) maximum absolute error over the volume of the altitude interval $[-5000,100\,000]$. See Appendix~\ref{appendix:b} for detailed definitions. Grey points are the complete set polynomial approximations. Black points are the indicated literature methods. Clearly, even though the polynomial approximations introduces a magnitude error of the n-vector, they still improve on the achievable performance on the gravity computations.}\label{fig:g_err}
\end{center}
\end{figure}

In Figure~\ref{fig:lat_and_alt_err} to~\ref{fig:g_err}, the latency results versus the different error measures are shown for a nominal altitude range of $[-5000, 100\,000]$m.
Additional results for smaller and larger altitude ranges as well as result without AVX instructions are provided in Appendix~\ref{appendix:more_results}.

%In Fig. 1, the maximum Euclidean position errors, the maximum latitude error converted to an arch length and the maximum altitude error are shown relative to the latency for all combinations with repetitions for $L_{\phi/n},N_{\phi/h/n},M_{\phi/h/n}\in[0,3]$. $I\in[1,10]$ and $J\in[2,12]$ with only results shown for values giving an accuracy within a factor 10 below and above that of the inner approximation.
%
%In Fig. 2, the corresponding maximum gravity errors resulting from the coordinates and the n-vector and the WGS84 gravity model are shown. Finally, also in Fig. 2, the n-vector magnitude and orientation errors are shown. 

\section{Benchmark discussion}\label{sec:benchmark_discussion}
The most important aspect of the benchmark is naturally correctness. As discussed in Appendix~\ref{sec:appendix_benchmark}, there are many factors influencing the results which together warrant a level of doubt in any benchmark results of this kind. Therefore, to minimize this doubt, in Appendix~\ref{sec:cross_validation}, the benchmark machinery is cross-validated by reproducing results published elsewhere. It also contains some discussions about other benchmarks found in the literature, specifically note~\cite{Ward2020} and \cite{Voll1990}.

% TODO Make note about geodynamical effect in the order of 10^-7m setting a practical limits to required accuracy. Couldn't find any good source on the number. Mortiz sais 10^-7 in the forwards to his book.

About the results, the first thing to note are the outer bounds. The apparent lower numerical limits of the Euclidean, horizontal, altitude, latitude, longitude errors are comparable to the numerical precision (ulp) of IEEE-754 double precision of $\ulp_{64}(a)\approx9.3\cdot10^{-10}$. A further detailed numerical analysis is beyond the scope of this work. The upper limits are comparable to the difference between the major and minor axes, $a-b\approx2.14\cdot10^4$. The lowest latency measurements $\sim10$ clock cycles can be expected from a square root and a handful of arithmetic operations of the spherical transformation approximation. The upper values of $\sim200$ clock cycles can be expected from numerous square and cubic roots and a large number of arithmetic operations. Naturally, all these limits only changes marginally with varying altitude range.
%
%Note that, published error results below the ulp limit can be found in the literature, e.g.~\cite{Claessens2019}. These results are probably the result of incorrect intermediate rounding. Some other sub-ulp results rely on extended precision, e.g.~\cite{Ligas2011}.
With a glance at the results for greater altitude ranges in Appendix~\ref{appendix:more_results} another outher bound may be observed. Simply, the presented methods becomes inefficient for altitude ranges beyond $\sim10^6$. This sets a limit beyond which range splitting is required to use the presented methods. See discussion on improvements in the next section.

%About the individual results, the first
The second thing to note is that the \emph{nva} latencies are roughly half that of the \emph{lla} latencies, i.e. inverse trigonometric functions make up roughly half the cost of an \emph{lla} method. The vast majority of \emph{lla} methods contains the same inverse trigonometric functions, see~\cite{Ohlson1996} for an exception, and, hence, the most relevant latencies are those for the \emph{nva} methods. The differences between transformation methods are simply easier to observe when not diluted by common trigonometric components.
However, not all methods can be converted to \emph{nva} form and some applications requires \emph{lla} representation
%the methods by Feltens~\cite{Feltens2008}, Gade~\cite{Gade2007} and Knopp~\cite{Knopp2021} are actually described in \emph{nva} form but converting most of the methods is rather straight-forward but, as noted, not possible for all of them,
so \emph{lla} latencies are still relevant when comparing such methods, and for determining the computational cost for such applications.
Note that latencies for trigonometric functions may vary significantly from implementation to implementation, e.g. see CORE-MATH benchmark~\cite{COREMATH}, requiring care when making comparison of absolute \emph{lla} latency measurements from different sources.

From the \emph{lla} results, it may appear that the presented approximations provide superior performance over the full accuracy range. However, the matched trigonometric approximations may be introduced to any method and, considering the \emph{nva} result, a more balanced conclusion is rather that matched trigonometric approximations should be employed for any transformation method. Instead, the \emph{nva} results, suggests that the presented approximations provide significantly improved performance down to $\sim10^{-5}$m accuracy. Of course, the exact number depends on the considered altitude range and may vary from platform to platform.
For at hint on the variability of the results, see~\cite{Claessens2019}.

%The convergence rates of the suggested method is roughly ...

Note, for the current approximations, the pattern that 1) the most preferable combination have identical $N_\phi$ and $N_h$ values, 2) the Euclidean errors are normally almost identical to the horizontal errors and 3) the altitude errors are smaller than the horizontal errors, can be expected. From Fig.~\ref{fig:ix_limits}, for a given error, the altitude is less expensive to compute and, hence, for a given minimum maximum Euclidean error for a computational latency, the altitude error can be expected to be smaller. Further, for identical $N_\phi$ and $N_h$, the latitude and altitude errors are out-of-phase, which is preferable from an Euclidean error point of view. Combined with the fact that the altitude error can be expected to be smaller, the maximum Euclidean error can be expected to be dominated by the horizontal error.

Apart from the presented methods, below $10^{-5}$, the following set of methods are within expected platform variability in terms of computational cost:~\cite{Bowring1976} with modifications of~\cite{Bowring1985}, \cite{Ohlson1996} (only \emph{lla}),~\cite{Wu2003}, \cite{Fukushima2006} and~\cite{Shu2010}. Further, for very great altitude ranges~\cite{Hmam2018} curiously appear to provide a relatively inexpensive and essentially altitude independent error. Beyond that, the benchmark results speaks for themselves. The horizontal, vertical, latitude and longitude errors can be viewed as basic error components and the Euclidean and g-errors can be viewed as examples of common application specific combined error measures. The n-vector errors are special errors primarily relevant for the presented method (containing explicit n-vector approximations).

\section{Discussions}\label{sec:discussion}
A set of fundamentally new ECEF to geodetic coordinate transformation approximations based on minimax polynomials have been presented. The benchmark results show that these approximations improves the achievable computational cost down to an accuracy of $\sim10^{-5}$m. Corresponding improvements are also seen in other presented error measures and are large enough and shown in a convincingly large benchmark to be labeled \emph{significant} despite natural uncertainties associated with the latency measurements. The improvements comes with the caveats that
\begin{itemize}
\item The approximations are valid over a preselected altitude range $[h_{\min},h_{\max}]$ which has to be selected \emph{before} the transformation are implemented. One cannot just iterate until convergence. Note that, a zero-range still imply an approximation range of $a-b\approx 21385$m.
\item For the n-vector approximations, a magnitude error is also introduced.
\item The performance of the methods is relatively improved by vector (AVX) instructions, especially for high order polynomials. Non-AVX results are provided in Appendix~\ref{appendix:more_results}.
\end{itemize}
On the other hand, for the vast majority of applications, the exemplified ranges will likely suffice and the \emph{potential improvements} listing below contains several potential remedies for the problem. Moreover, even considering the magnitude errors, the approximation improves the achievable performance for e.g. gravity computations for which the n-vector magnitude enters directly. Finally, vector instructions have been prevalent in CPUs for more than a decade.
Further, in addition to the improved performance, the approximations are attractive in that
\begin{itemize}
\item Explicit \emph{nva} approximations are provided which is of particular importance for wide area physics simulations, e.g. flight and weather simulations, for which forces (gravity and drag) are computed many times but with moderate accuracy requirements.
\item They resemble traditional polynomial approximations of elementary functions and related tools and techniques may be used to improve implementations and even to implement fixed-point versions, see e.g.~\cite{Mouilleron2014}.
%\item Explicit n-vector versions are presented.
\item They are easy to specify and implement with straight-forward unbranching code only requiring regular arithmetic operations and the simplest elementary function, i.e. square-root, giving minimal dependencies on standard libraries. No special handling around the poles, the equator, etc. is required.
\item The latitude, longitude, altitude and n-vector approximations are naturally separable making it possible to obtain only individual values at a lower computational cost.
\item They are highly tunable in that a specific approximation may be selected to fulfill a computational cost or accuracy requirement. Especially, the horizontal and vertical accuracy can be set independently.
\item Although only exemplified with WGS84 latitude and altitude, the method is extendable to a wider set of rotation symmetric and equatorial plan reflection symmetric and antisymmetric properties and the underlying series expansions may have a wider applicability within other geodesy problems enabling similar results to be obtained for these.
\item The polynomial, separable and extendable nature of the components of the transformation approximation makes them amenable to vectorization (SIMD instructions) making them suitable for modern computational platforms.
\end{itemize}
%Some methods found in the literature also express some of these properties to varying degree but none has them all.
%
In contrast, the greatest drawback and limitations of the approximations are that
\begin{itemize}
%\item The approximations are only valid for a preselected altitude range. Changing this range requires all coefficient to be recomputed. %If the approximations are to be used for large ranges, some lookup table should probably be used.
%\item They become inefficient for higher accuracies and they cannot directly be extended to full numerical precision. On the other hand, they could be used to initialize, and hence speed-up, other iterative methods to achieve full numerical precision.
%\item The n-vector version introduces a magnitude error which makes the error picture more complicated.
\item A large number of coefficients need to be precomputed to arrive at an actual approximation implementation, making the approximations somewhat hard to use out-of-the-box. Changing the preselected altitude range requires all coefficients to be recomputed.
\item The convergence rate with respect to a greater number of coefficients indicates that the approximations becomes inefficient for accuracies beyond double precision, i.e. other methods have better error convergence rates.
\item The results for great altitude ranges indicates that the approximations becomes inefficient for ranges beyond $\sim10^6$m, i.e. handling such ranges requires range splitting.
\end{itemize}
However, coefficients covering most regular use-cases are provided in Appendix~\ref{appendix:labeled_approx} and applications requiring more than double precision, i.e. subatomic accuracy, and ranges beyond $\sim10^6$m, i.e. surface to orbital altitudes, are rare.
%
%Performance at higher altitudes. How fast does the error grow.
%Valid for all imput?
%
%The improved performance is significant for e.g. aerospace simulations, for which external forces very slowly with the geodetic coordinates and where computational cost may be a limiting factor; geographical information services/processing with data for which the underlying granularity is in the range of the performance improvement and for which a large number requests need to be serviced; and for initializing other iterative coordinate transformation methods providing a potential speed-up of existing methods.
%
%Used to initialize other methods.
%

Clearly, the presented approximations sets a new standard for coarse (above $\sim10^{-5}$m) and fast ECEF to geodetic coordinate transformations. Further, their polynomial nature opens up a new realm of further improvements. Specifically, note the following overall areas of potential improvement
\begin{itemize}
\item \emph{Lookup tables} It is well known that many elementary functions are often implemented with a combination of minimax polynomials and coefficient lookup tables rather than just a single polynomial. Diverging from clean arithmetic code, for many platforms, such strategies could probably be employed to further improve the approximation performance for the geodetic power series approximations. For the inverse trigonometric function approximations, this is definitely the case. Specifically, lookup tables with respect to $p$ would be required to handle altitude ranges beyond $\sim10^6$ but may be beneficial for much smaller altitude ranges as well. %Further, this could also possibly be used to construct methods which essentially reaches the numerical limits of this type of transformations~\cite{COREMATH}. %Also based on the results for Ohlson's methods~\cite{Ohlson1996}, combined $\sin^{-1}$ and $\cos^{-1}$ methods could also improve performance.
\item \emph{Evaluation optimization} The expression trees of the individual (multivariat) polynomials have numerous degrees of freedom and the naive employed Estrin's schema are probably suboptimal, both in a number of operations sense and in a latency sense for a given platform. Hence, it may be improved upon both in a general sense, see e.g.~\cite{Kupiers2013}, and in a platform specific sense, see e.g.~\cite{Ewart2020}. The fact that the $\lataddcorr[\fourierixlimit][\taylorixlimit][\cdot][\cdot]$ and $\altaddcorr[\fourierixlimit][\taylorixlimit][\cdot][\cdot]$ polynomials are Bernstein polynomials with respect to $t^2\in[0,1]$ could potentially be exploited, see e.g.~\cite{Chudy2021}. However, changing the individual evaluations schema may come in conflict with the following improvement option.
\item \emph{Cross polynomial evaluation optimization} The approximations are comprised of multiple polynomials with common factors and terms. This is only exploited through the polynomial arguments. However, more common components may be separated out and computed jointly in order to reduce the overall number of required arithmetic operations. For example, for specific $N$ and $M$, joint evaluation schema for $\altaddcorr$ and $\lataddcorr$ with less arithmetic operations than the Horner scheme can be constructed. Further, as the results with and without AVX show, vectorization is important for performance. For platforms with vector operations, any evaluation schema has to take this into account and manual vectorization could potentially improve the latency. Furthermore, since the presented method can be extended to other geodetic quantities, vectorized method computing numerous such quantities at no extra cost may be constructed.
\item \emph{Alternative polynomial approximations} In general, regular minimax approximations provide good accuracy-to-computational-cost performance. However, weighted (relative) minimax, rational minimax, Pad\'{e} or similar approximations may provide better performance. See~\cite{Turner2009} and \cite{Turner2013} for examples. Other approximations may also give a different error growth relative some nominal $h_0$ which may be preferable for some applications. Further, it is also conceivable to parameterize the polynomials in ${h_c}^{-1}$ rather than $h_c$. This could provide better properties for great altitude ranges.
\item \emph{Full coefficient optimization} The polynomial approximations are split over inner and outer polynomials, with one assuming no errors in the other. Further, latitude, longitude and altitude errors are only connected via the index limits selection. This is clearly suboptimal and coefficients could potentially be adjusted to provide better overall approximations, i.e. higher-dimensional minimax approximations. Potentially, a low hanging fruit would be to replace the minimax optimization over $h_c$ with the same over a first order altitude approximation.
% The truncation and the $\ell_2$ norms of the Fourier series, the multi-dimensionality and the minimax optimization with respect to subpolynomials means that the polynomial approximations are not optimal in the targeted $||\cdot||_\infty$ sense. Hence, the polynomial coefficients may be optimized for specific polynomial approximations to reduce the errors. This is in general a difficult optimization but the current approximations provides starting points for the optimization improving the chance of convergence considerably. Futher, cross terms of the joint multidimensional polynomials could potentially be trimmed.
% Relative absolute error rather than absolute error.
\item \emph{Combined sin-cos approximations} Polynomial \emph{lla} transformation approximations with accuracies below $\sim10^{-4}$m combining $\sin^{-1}$ and $\cos^{-1}$ approximation for different $t$ ranges with better performance than the presented approximations can easily be constructed. Note, in this case, neither the $\sin^{-1}$ nor the $\cos^{-1}$ approximations require a square root argument reduction. However, for a proper implementation, errors at the range extrema have to be matched to avoid error discontinuities. Further, approximation orders and ranges have to be optimized. This is, overall, a challenging but doable task.
%\item \emph{Other problem transformation} The enabling problem transformation of Section~\ref{sec:additive} shows that there is at least one transformation enabling polynomial approximations. However, this does not mean that it is unique and other, potentially better, transformations may exist.
%\item \emph{Changing minimax criteria} All coefficient are computed to minimize the \emph{absolute} errors. For multiplicative corrections, it may be more appropriate to minimize the \emph{relative} errors. Further, the $[h_{\min},h_{\max}]$ is with respect to $h_c$. Ideally, we would like to minimize the errors over a range of $h$.
%\item Alternatively, the structure and common component of the polynomials may be exploited to efficiently implement the polynomial evaluation. % structures Exploit structures of the polynomials. E.g. both contains $\omega$, Bernstein polynomials. 
%\item Prune terms which are insignificant.
%\item Fused multiply and add.
%\item \emph{Combined methods} Some iterative methods such as \cite{Lin1995}and~\cite{Bowring1976} appear to have a higher rate of convergence than the polynomial methods NO IT DOESN'T WORK BECAUSE THE BASE COMPUTATIONAL COST IS NOT IN THE INITIALIZATION BUT IN THE PROCESSING AFTER THE ITERATIONS.
\item \emph{Series expansions for coefficient calculations} The coefficients $b_n(u)$ and $c_n(u)$ are rather cumbersome to compute and, therefore, computing the corresponding minimax approximations is also cumbersome. The problem is that the integrals cannot be solved analytically and one has to resort to numerical integration. This could be circumvented by using series expansions for the differences. See for example~\cite{Long1974}. The series sum could then be moved out of the integral and each term solved analytically resulting in a new series which would be much easier to implement and approximate with plain truncation. In turn, this would enable use of standard tools for computing the minimax approximations.
\end{itemize}
However, these are all very different technical areas from the work presented here and are left for future improvements.

%Note that the $\lataddcorr[\fourierixlimit][\taylorixlimit][\cdot][\cdot]$, $\altaddcorr[\fourierixlimit][\taylorixlimit][\cdot][\cdot]$, $\coslatmulcorr[\mulixlimit][\cdot][\cdot]$ and $\sinlatmulcorr[\mulixlimit][\cdot][\cdot]$ polynomials are Bernstein polynomials with respect to $t^2\in[0,1]$. This may be of theoretical interest in the sense that the approximations may be interpreted as Bézier curves and of practical interest since there are specific algorithms for evaluating Bernstein polynomials. However, this is not further investigated here.

\section{Final recommendations}\label{sec:final_rec}
Before selecting a ECEF to geodetic coordinate transformation method, do review the actual application transformation requirements. Favor \emph{nva} representation. Compared to \emph{lla} representation and related methods, the computational cost is roughly half. If \emph{lla} representation is required, make sure to use matched trigonometric function approximations.
As for the actual transformation method: if the computational cost is important, the accuracy requirement is not below $~\sim10^{-5}m$ and the altitude range is below $10^6$, do use the presented polynomial approximations. In contrast
\begin{itemize}
\item If computational cost is not a significant concern, use a closed form method like~\cite{Vermeille2011} or~\cite{Karney2011}. Then you do not have to consider number of iterations, matched trigonometric functions, accuracy, etc.
\item If computational cost is important, the accuracy requirement is moderate (above $~\sim10^{-5}m$) but the altitude range cannot be bounded, instead consider methods by~\cite{Bowring1976} +~\cite{Bowring1985} or~\cite{Shu2010} and make a look-up table in $p$ for number of iterations until convergence. Be wary that the often suggested method by \cite{Fukushima2006} has numerical problems at very high altitudes. For astronomical altitude ranges, also consider~\cite{Hmam2018}.
\item If computational cost is important and the required accuracy is below $~\sim10^{-5}m$, also consider methods by~\cite{Bowring1976} +~\cite{Bowring1985},~\cite{Ohlson1996},~\cite{Wu2003}, \cite{Fukushima2006} and~\cite{Shu2010}. %Alternatively, especially if handling of points inside the earth's evolute is required, also consider closed-form methods by~\cite{Vermeille2011} and~\cite{Karney2011} (GeographicLib).
%\item If very large altitude ranges are to be handled, including points inside the earth's evolute, consider closed-form methods by~\cite{Vermeille2011} and (GeographicLib)~\cite{Karney2011}.
\end{itemize}
% Further, for \emph{lla} representation, if mean computational cost is important rather than worst-case computational cost, consider implementing a sin-cos splitted version like~\cite{Ohlson1996}.
%
%
Further, if you are implementing widely used library methods or computational resources are scarce and there is plenty of time and human resources, consider the areas of potential improvement of the previous section. Great altitude ranges can be handled and there is likely more performance to be found!

%\addresseshere

\section*{References}
References are sorted in alphabetical order. References which have not been accessed are labeled accordingly and citations to one or more references citing and providing information about the inaccessible source are provided. To the extent possible, DOI:s or URL:s are provided.

\vspace{5mm}

%\bibliographystyle{plain}
%\bibliography{ref_geo,ref_misc}\label{refs}
\printbibliography[heading=none]

\appendix

\section{Power series expansions}\label{appendix:derivation}
To handle the multivariate nature of the coordinate transformation, the approach is to express the difference between~\eqref{eq:spherical} and $\phi$, $h$ and $\bar{n}$ in Fourier (Chebyshev) series, with respect to $\phi_c$ and with $h_c$ dependent coefficients. This separates the dependencies and enable truncation of the series with close to minimax properties. %, and separate minimax approximation of the coefficients.
To start, relations between $\phi$ and $\phi_c$ and $h$ and $h_c$ are required. The actual analytical relations are of less importance and it is sufficient that they are numerically evaluable. Hence, let
\begin{equation*}
\begin{split}
\phi&=f(x,y,z)\\
h&=g(x,y,z)
\end{split}
\end{equation*}
be some reference ECEF-to-geodetic coordinate transformation of choice, e.g. see~\cite{Vermeille2004}. (Note, the opaque nature of the reference means that the following derivation may be carried out for any $x=y=0$ rotation symmetric and $z=0$ plane reflection symmetric and anti-symmetric properties.)
%Further, with $q=\sqrt{x^2+y^2}$, from simple geometry
%\begin{equation*}
%\begin{split}
%q &= (h_c + h_0)\cos(\phi_c)\\
%z &= (h_c + h_0)\sin(\phi_c)
%\end{split}
%\end{equation*}
%From the above relations and 
%Then, from simple geometry and from the symmetry of the earth ellipsoid, 
The geocentric to geodetic coordinate transformations follows as
\begin{equation*}
\begin{split}
\phi&=f_c(\phi_c,h_c)\\
    &=f((h_c + h_0)\cos(\phi_c),0,(h_c + h_0)\sin(\phi_c))\\
h&=g_c(\phi_c,h_c)\\
    &=g((h_c + h_0)\cos(\phi_c),0,(h_c + h_0)\sin(\phi_c))
\end{split}
\end{equation*}
%Hence, with the initial round-earth geodetic coordinate transformation the target for the polynomial
%
$f_c(\cdot)$ and $g_c(\cdot)$ are naturally defined for $\phi_c\in[-\pi/2,\pi/2]$ %and $f_c(\pm\pi/2,h_c)=\pm\pi/2$ and $g_c(\pm\pi/2,h_c)=h$. Hence,
and, hence, the differences between the geocentric and the geodetic coordinates, $f_c(\phi_c,h_c)-\phi_c=\phi - \phi_c$ and $g_c(\phi_c,h_c) - h_c=h - h_c$, are continuous and may be viewed as periodic with period $\pi$.
Obviously, $\phi - \phi_c$ is antisymmetric and $h-h_c$ is symmetric around $\phi_c=0$. Hence, they may be expressed as the Fourier $\sin$ and $\cos$ series
\begin{equation*}
\phi - \phi_c = \sum^\infty_{n=1}b_n(h_c)\sin\left(2n\phi_c\right)
\end{equation*}
where
\begin{equation*}
b_n(h_c) = \frac{2}{\pi}\int^{\pi/2}_{-\pi/2}(f_c(\phi_c,h_c) - \phi_c)\sin\left(2n\phi_c\right)\textrm{d}\phi_c
\end{equation*}
and
\begin{equation*}
h - h_c = \mathop{\smash{\sideset{}{'}\sum}\vphantom{\sum}}\limits^\infty_{n=0}c_n(h_c)\cos\left(2n\phi_c\right)
\end{equation*}
where
\begin{equation*}
c_n(h_c) = \frac{2}{\pi}\int^{\pi/2}_{-\pi/2}(g_c(\phi_c,h_c)-h_c)\cos\left(2n\phi_c\right)\textrm{d}\phi_c
\end{equation*}
%
%\begin{equation*}
%\phi - \phi_c = \frac{a_0(h_c)}{2} + \sum^\infty_{n=1}\left(a_n(h_c)\cos\left(2n\phi_c\right)+b_n(h_c)\sin\left(2n\phi_c\right)\right)
%\end{equation*}
%where
%\begin{equation*}
%a_n(h_c) = \frac{2}{\pi}\int^{\pi/2}_{-\pi/2}(f_c(\phi_c,h_c) - \phi_c)\cos\left(2n\phi_c\right)\textrm{d}\phi_c
%\end{equation*}
%\begin{equation*}
%b_n(h_c) = \frac{2}{\pi}\int^{\pi/2}_{-\pi/2}(f_c(\phi_c,h_c) - \phi_c)\sin\left(2n\phi_c\right)\textrm{d}\phi_c
%\end{equation*}
%and
%\begin{equation*}
%h - h_c = \frac{1}{2}c_0(h_c) + \sum^\infty_{n=1}\left(c_n(h_c)\cos\left(2n\phi_c\right)+d_n(h_c)\sin\left(2n\phi_c\right)\right)
%\end{equation*}
%where
%\begin{equation*}
%c_n(h_c) = \frac{2}{\pi}\int^{\pi/2}_{-\pi/2}(g_c(\phi_c,h_c)-h_c)\cos\left(2n\phi_c\right)\textrm{d}\phi_c
%\end{equation*}
%\begin{equation*}
%d_n(h_c) = \frac{2}{\pi}\int^{\pi/2}_{-\pi/2}(g_c(\phi_c,h_c)-h_c)\sin\left(2n\phi_c\right)\textrm{d}\phi_c
%\end{equation*}
%Obviously, $\phi - \phi_c$ is antisymmetric around $\phi_c=0$ and, hence, $a(h_c) = 0\forall n$.
%
%Similarly, $h-h_c$ is symmetric around $\phi_c=0$ and, hence, $d_n(h_c) = 0\forall n$.
%
Further, from the multi-angle formulas
% From https://mathworld.wolfram.com/Multiple-AngleFormulas.html
\begin{equation*}
\begin{split}
\sin(2n\phi_c)&=\sum_{k=0}^{n-1}(-1)^k\binom{2n}{2k+1}\sin^{2k+1}(\phi_c)\cos^{2n-2k-1}(\phi_c)\\
&=\sin(\phi_c)\sqrt{1 - \sin^2(\phi_c)}\sum_{k=0}^{n-1}(-1)^k\binom{2n}{2k+1}\sin^{2k}(\phi_c)(1-\sin^2(\phi_c))^{n-k}\\
\cos(2n\phi_c)&=\sum_{k=0}^{n}(-1)^k\binom{2n}{2k}\sin^{2k}(\phi_c)\cos^{2n-2k}(\phi_c)\\
&=\sum_{k=0}^{n}(-1)^k\binom{2n}{2k}\sin^{2k}(\phi_c)(1-\sin^2(\phi_c))^{n-k}
\end{split}
\end{equation*}
It may not be obvious why a Fourier series expansion would be suitable but the fact that the differences are smooth and periodic gives a hint that a few series terms could be sufficient to make a good approximation of the differences. Further, from the multi-angle formulas, it can be seen that the Fourier series can be expressed as a power series in $\sin(\phi_c)=z/p$, which can easily be computed.
%
%Further, to arrive at a polynomial form (power series) of the differences, express the $h_c$ dependence of $b_n(h_c)$ and $c_n(h_c)$ in a polynomial basis, i.e. as a generalized Fourier series. The exact basis and the inner product is of less importance since it will be replaced with finite polynomial approximation but, for the sake of argument, let us use an easily computed Taylor series around $h_0$
%\begin{equation*}
%b_n(h_c) = \sum_{m=0}^\infty b_n^{(m)}h_c^m\quad %c_n(h_c) = \sum_{m=0}^\infty c_n^{(m)}h_c^m
%\end{equation*}
%where
%\begin{equation*}
%b_n^{(m)}=\left.\frac{1}{m!}\frac{\partial^m}{\partial\delta^m}b_n(\delta)\right|_{h_0} \quad\textrm{and}\quad c_n^{(m)}=\left.\frac{1}{m!}\frac{\partial^m}{\partial\delta^m}c_n(\delta)\right|_{h_0}
%\end{equation*}
%
%Again, it may not be obvious why a Taylor series would be suitable here. In general, Taylor series does not provide efficient function approximations. However, the fact that the typical range of $h_c$ is tiny in comparison with $p$ gives a hint that a few series terms could be sufficient.
%Note that this essentially gives an expansion around $h_0$.
%
%\begin{equation*}
%b_n(h_c) = b_n(0) + \left.\frac{\partial}{\partial\delta}b_n(\delta)\right|_{0}h_c+\left.\frac{1}{2}\frac{\partial^2}{\partial\delta^2}b_n(\delta)\right|_{0}h_c^2+\dots
%\end{equation*}
%and
%\begin{equation*}
%c_n(h_c) = c_n(0) + \left.\frac{\partial}{\partial\delta}c_n(\delta)\right|_{0}h_c+\left.\frac{1}{2}\frac{\partial^2}{\partial\delta^2}c_n(\delta)\right|_{0}h_c^2+\dots
%\end{equation*}
%
Combining it all gives the expressions for the additive corrections to $\phi$ and $h$
\begin{equation}\label{eq:direct}
\begin{split}
\phi &= \phi_c + \sin(\phi_c)\sqrt{1 - \sin^2(\phi_c)}\lataddcorrexpr\\
     &= \phi_c + t\sqrt{1 - t^2}\lataddcorrex[h_c][t^2]\\
h &= \altaddcorrexpr\\
  &= \altaddcorrex[h_c][t^2]
\end{split}
\end{equation}
where $\lataddcorrex[\cdot][\cdot]$ and $\altaddcorrex[\cdot][\cdot]$ are the power series
\begin{equation*}
\begin{split}
\lataddcorrex &= \sum^\infty_{n=1}\sum_{k=0}^{n-1}(-1)^k\binom{2n}{2k+1}v^k(1-v)^{n-k-1}b_n(u)\\
\altaddcorrex &= u + \mathop{\smash{\sideset{}{'}\sum}\vphantom{\sum}}\limits^\infty_{n=0}\sum_{k=0}^{n}(-1)^k\binom{2n}{2k}v^k(1-v)^{n-k}c_n(u)
\end{split}
\end{equation*}
Further, 
%For the n-vector, additive corrections are of no good. From the additive $\phi_c$ correction
from \eqref{eq:direct} and the Taylor series of $\sin(\phi)$ and $\cos(\phi)$ around $\phi_c$
\begin{equation*}
\begin{split}
\frac{\sin(\phi)}{\sin(\phi_c)} \!\!\!&\,\,\,= \frac{\sin\left(\phi_c + \sin(\phi_c)\sqrt{1 - \sin^2(\phi_c)}\lataddcorrexpr\right)}{\sin(\phi_c)} \\
=&\sum^\infty_{l=0}\left.\frac{\partial^l}{\partial\varphi^l}\frac{\sin(\varphi)}{l!}\right|_{\phi_c}\frac{\left(\sin(\phi_c)\sqrt{1 - \sin^2(\phi_c)}\lataddcorrexpr\right)^l}{\sin(\phi_c)} \\
=&\sum_{l\in\mathbb{N}_e}(-1)^{l/2}\frac{1}{l!}\left(\sin(\phi_c)\sqrt{1 - \sin^2(\phi_c)}\lataddcorrexpr\right)^l \\
 &+\sum_{l\in\mathbb{N}_o}(-1)^{(l-1)/2}\frac{1}{l!}\frac{\cos(\phi_c)}{\sin(\phi_c)}\left(\sin(\phi_c)\sqrt{1 - \sin^2(\phi_c)}\lataddcorrexpr\right)^l\\
=&\sum_{l\in\mathbb{N}_e}(-1)^{l/2}\frac{1}{l!}(\sin^2(\phi_c))^{l/2}(1-\sin^2(\phi_c))^{l/2}\lataddcorrexpr^l \\
 &+\sum_{l\in\mathbb{N}_o}(-1)^{(l-1)/2}\frac{1}{l!}(\sin^2(\phi_c))^{(l-1)/2}(1-\sin^2(\phi_c))^{(l+1)/2}\lataddcorrexpr^l
\end{split}
\end{equation*}
\begin{equation*}
\begin{split}
\frac{\cos(\phi)}{\cos(\phi_c)} \!\!\!&\,\,\,= \frac{\cos\left(\phi_c + \sin(\phi_c)\sqrt{1 - \sin^2(\phi_c)}\lataddcorrexpr\right)}{\cos(\phi_c)} \\
=&\sum^\infty_{l=0}\left.\frac{\partial^l}{\partial\varphi^l}\frac{\cos(\varphi)}{l!}\right|_{\phi_c}\frac{\left(\sin(\phi_c)\sqrt{1 - \sin^2(\phi_c)}\lataddcorrexpr\right)^l}{\cos(\phi_c)} \\
=&\sum_{l\in\mathbb{N}_e}(-1)^{l/2}\frac{1}{l!}\left(\sin(\phi_c)\sqrt{1 - \sin^2(\phi_c)}\lataddcorrexpr\right)^l \\
 &+\sum_{l\in\mathbb{N}_o}(-1)^{(l+1)/2}\frac{1}{l!}\frac{\sin(\phi_c)}{\cos(\phi_c)}\left(\sin(\phi_c)\sqrt{1 - \sin^2(\phi_c)}\lataddcorrexpr\right)^l\\
=&\sum_{l\in\mathbb{N}_e}(-1)^{l/2}\frac{1}{l!}(\sin^2(\phi_c))^{l/2}(1-\sin^2(\phi_c))^{l/2}\lataddcorrexpr^l \\
 &+\sum_{l\in\mathbb{N}_o}(-1)^{(l+1)/2}\frac{1}{l!}(\sin^2(\phi_c))^{(l+1)/2}(1-\sin^2(\phi_c))^{(l-1)/2}\lataddcorrexpr^l
\end{split}
\end{equation*}
where $\mathbb{N}_{e}$ and $\mathbb{N}_{o}$ are the sets of all positive even and odd natural numbers.
Let
\begin{equation*}
\delta(v,w)=v(1-v)w^2
\end{equation*}
\begin{equation*}
\triglatmulevencorrTwo[]=\sum_{l\in\mathbb{N}_{e}}(-1)^{l/2}\frac{1}{l!}\delta^{l/2}
\quad\text{and}\quad
\triglatmuloddcorrTwo[]=w\sum_{l\in\mathbb{N}_{o}}(-1)^{(l-1)/2}\frac{1}{l!}\delta^{(l-1)/2}
\end{equation*}
and
\begin{equation*}
\sinlatmulcorr[]=\triglatmulevencorrTwo[][\delta(v,w)]+(1-v)\triglatmuloddcorrTwo[][w][\delta(v,w)]
\end{equation*}
\begin{equation*}
\coslatmulcorr[]=\triglatmulevencorrTwo[][\delta(v,w)]-v\triglatmuloddcorrTwo[][w][\delta(v,w)]
\end{equation*}
Then, multiplicative sin and cos corrections follow as
\begin{equation*}
\begin{split}
\sin(\phi) %&= \sin(\phi_c)\sinlatmulcorr[][\sin^2(\phi_c)][\lataddcorrexpr]\\
           &=\sin(\phi_c)\sinlatmulcorr[][t^2][\lataddcorrex[h_c][t^2]]\\
\cos(\phi) %&= \cos(\phi_c)\coslatmulcorr[][\sin^2(\phi_c)][\lataddcorrexpr]\\
           &= \cos(\phi_c)\coslatmulcorr[][t^2][\lataddcorrex[h_c][t^2]]
\end{split}
\end{equation*}

\section{Reasonable index limit combinations}\label{sec:selection}
\begin{figure}[ht]
\begin{center}
\includegraphics[width=0.49\textwidth]{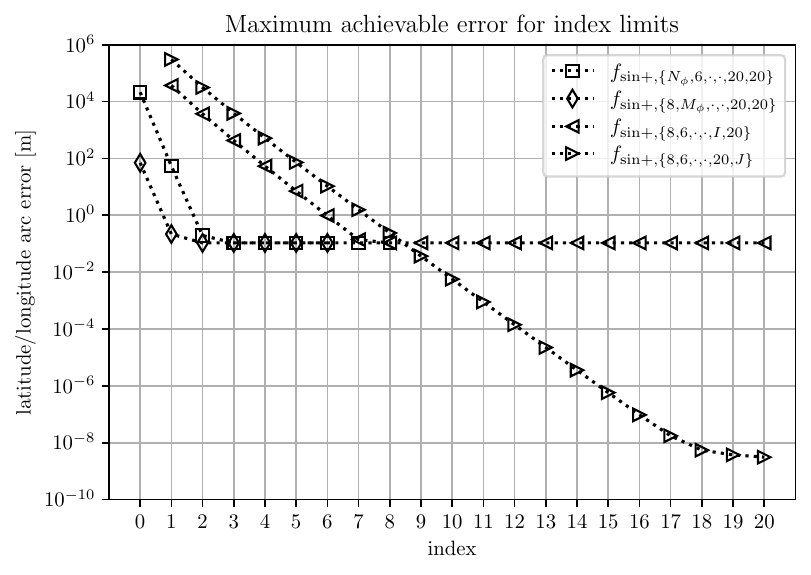}
\includegraphics[width=0.49\textwidth]{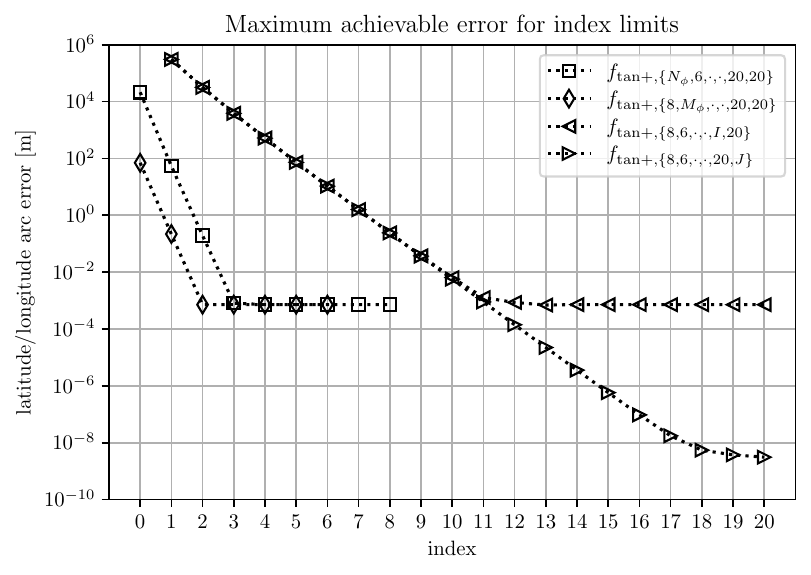}
\includegraphics[width=0.49\textwidth]{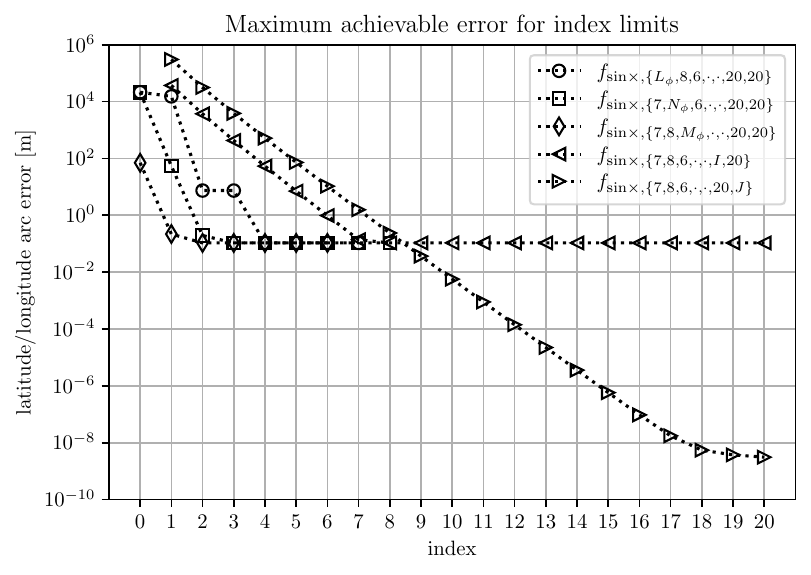}
\includegraphics[width=0.49\textwidth]{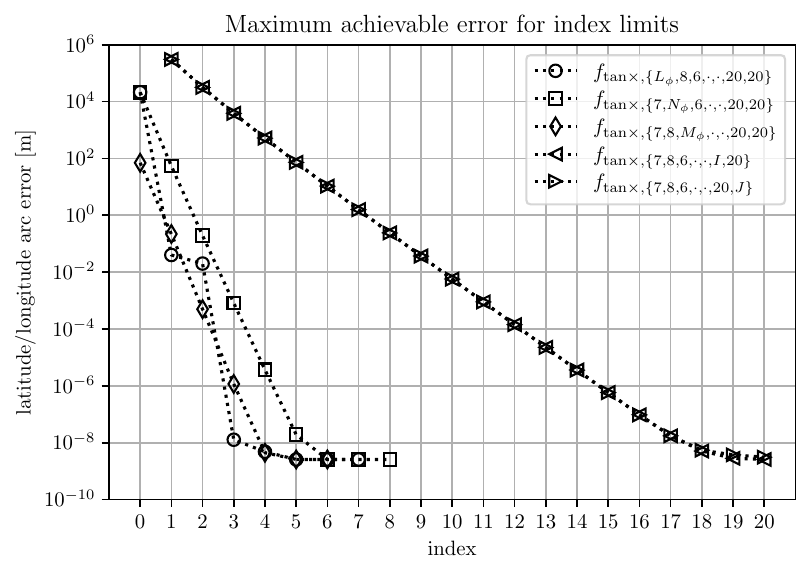}
\includegraphics[width=0.49\textwidth]{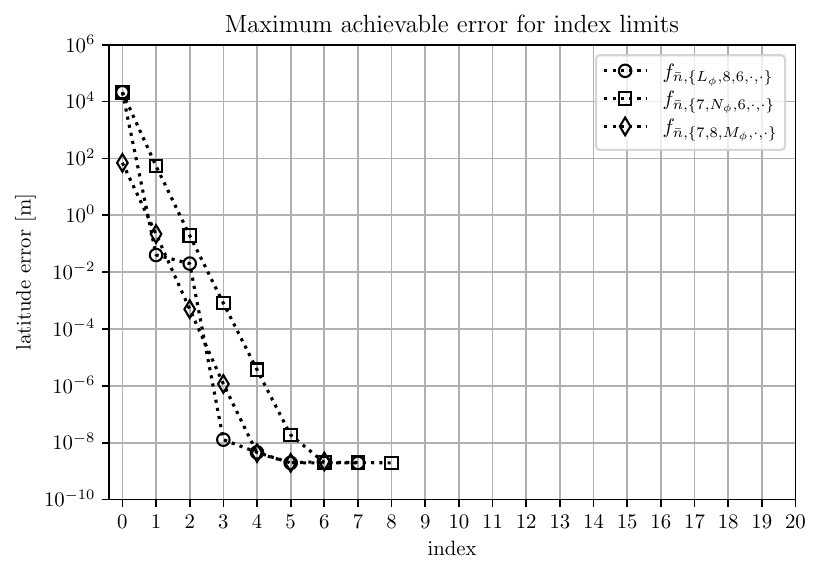}
\includegraphics[width=0.49\textwidth]{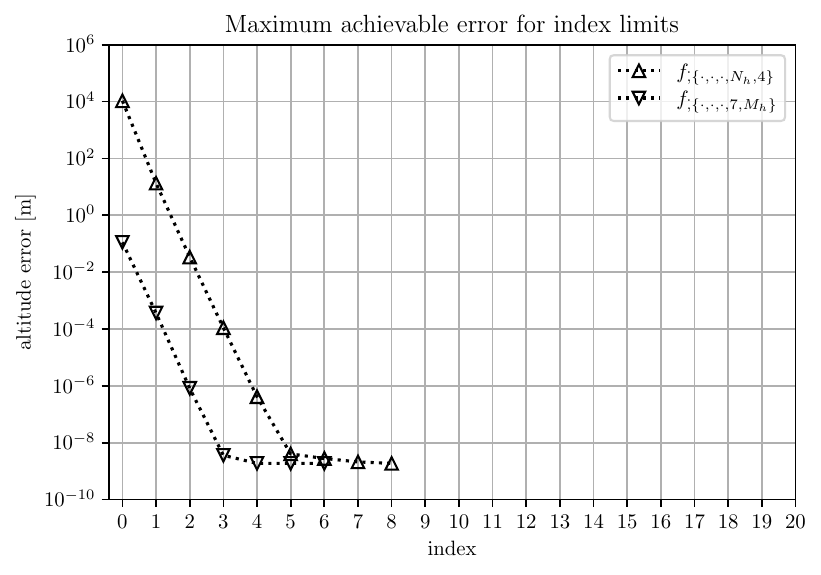}
\caption{\footnotesize Maximum error (latitude arc error, longitude arc error and altitude error) of the labeled approximation and index limit set to the indicated value and the remaining index limits set to $\sinixlimit=20$, $\tanixlimit=20$, $\mulixlimit_{\phi/n}=6$, $\fourierixlimit_{\phi/h/n}=8$, $\taylorixlimit_{\phi/h/n}=6$. This provides an approximate achievable lower maximum error limit given individual index limits. The leveling out of some of the curves above the numerical limit of $10^{-1}$ is caused by poor numerical conditioning and corresponding errors of $\sin^{-1}$, i.e. $\arcsinapprox$, and $t\sqrt{1-t^2}$ around the poles.}\label{fig:ix_limits}
\end{center}
\end{figure}

The polynomial orders determines the absolute errors and computational costs. Hence, they will, to some extent, be application and platform specific and will ultimately have to be set by benchmarking different combinations, which is done in Section~\ref{sec:benchmark}. Unfortunately, the combinations $\{\sinixlimit,\tanixlimit,\mulixlimit_{\phi},\fourierixlimit_{\phi},\taylorixlimit_{\phi},\fourierixlimit_h,\taylorixlimit_h\}\in\mathbb{N}^7$ or $\{\sinixlimit,\tanixlimit,\fourierixlimit_{\phi},\taylorixlimit_{\phi},\fourierixlimit_h,\taylorixlimit_h\}\in\mathbb{N}^6$, for the \emph{lla} approximations, and $\{\mulixlimit_n,\fourierixlimit_n,\taylorixlimit_n,\fourierixlimit_h,\taylorixlimit_h\}\in\mathbb{N}^5$, for the \emph{nva} approximations, are too many to be tested exhaustively. Therefore, the combinations have to be narrowed down based on their relative errors which is here made based on the three assumptions
\begin{itemize}
\item Applications typically have a horizontal accuracy requirement rather than separate latitude and longitude accuracy requirements. Hence, $\{\sinixlimit,\tanixlimit,\mulixlimit_{\phi/n},$ $\fourierixlimit_{\phi/n},\taylorixlimit_{\phi/n}\}$ are narrowed down jointly.
\item Applications may have vastly different requirements on horizontal and vertical accuracy. Hence, $\{\sinixlimit,\tanixlimit,\mulixlimit_{\phi/n},\fourierixlimit_{\phi/n},\taylorixlimit_{\phi/n}\}$ and $\{\fourierixlimit_h,\taylorixlimit_h\}$ are narrowed down independently.
\item A part from position error, $\mulixlimit_{n}$ also affect n-vector magnitude errors, which are hard to compare and significant for $\mulixlimit_{\phi}\leq3$. Hence, for $\mulixlimit_{\phi}\leq3$ the limitations set by $\mulixlimit_{\phi}$ are ignored.
\end{itemize}
Starting from sufficiently high limits, attaining horizontal (latitude and longitude) and altitude errors at the numerical limits of the IEEE-754 double precision implementation, and lowering one limit at at a time gives the maximum error shown in Fig.~\ref{fig:ix_limits}.
%The maximum horizontal (latitude, longitude) and altitude errors resulting from starting from index limits sufficient to provide the achievable errors, given IEEE-754 double precision implementation, for each approximation and lowering one index limit at a time are shown in Fig.~\ref{fig:ix_limits}.
%
Let $E^{f_i}_{K}(\cdot)$ denote the values shown in Fig.~\ref{fig:ix_limits} for the respective approximation $f_i$ and index limit $K$. Start with complete sets of index limit value combinations $\{L,N,...\}$. Split according to the assumptions.
Let $E^{f_i}_{\max}=\max_K(\{E^{f_i}_{K}(X) : X\in\{L,N,...\}\}$ be the maximum error for such a set.
Then reject the combinations if either
\begin{itemize}
\item $K>K_{\min}$ and $E^{f_i}_{K}(K-1)<E^{f_i}_{\max}$ and $E^{f_i}_{K}(K) < E^{f_i}_{\max} 0.25$
\item $K>K_{\min}$ and $E^{f_i}_{K}(K) < 0.95E^{f_i}_{K}(K-1)$
\end{itemize}
The first conditions handles the potential constructively or destructively inference between errors, meaning the combinations cannot be reduced to the index limits providing limiting errors closest to each other. The second condition handles increasing index limits not reducing the errors more. 0.25 and 0.95 are trade-offs between ensuring that all reasonable combinations are captured and limiting the total number of tested combination. 0.25 and 0.95 give $\sim3000$ combinations to test for the benchmark.

\section{Error definitions}\label{appendix:b}
The errors used in the comparison of different approximations and literature methods are defined as follows. Throughout, the \emph{Earth's prime-vertical radius of curvature} is $\tilde{N}_k=a^2(a^2\cos^2(\tilde{\phi}_k) + b^2\sin^2(\tilde{\phi}_k))^{-1/2}$ and \emph{Earth's meridional radius of curvature} $\tilde{M}_k=(1-e^2)/a^2\tilde{N}^3_k$. Note, for $\{x,y,z\}=\{0,0,0\}$ only altitude error is well defined but this is not repeated for every measure.
\begin{itemize}
\item Euclidean error in meter
\begin{equation*}
e_{\leftrightarrow}=\sqrt{(\tilde{x}_k-x^\ellipse_k)^2+(\tilde{y}_k-y^\ellipse_k)^2+(\tilde{z}_k-z^\ellipse_k)^2}
\end{equation*}
where
\begin{equation*}
\begin{split}
\tilde{x}_k &= \left(\tilde{N}_k + \tilde{h}_k\right)\cos(\tilde{\phi}_k)\cos(\tilde{\lambda}_k)\\
\tilde{y}_k &= \left(\tilde{N}_k + \tilde{h}_k\right)\cos(\tilde{\phi}_k)\sin(\tilde{\lambda}_k)\\
\tilde{z}_k &= \left(\tfrac{b^2}{a^2}\tilde{N}_k + \tilde{h}_k\right)\sin(\tilde{\phi}_k)
\end{split}
\end{equation*}
\item Geodetic horizontal error in meter. Computing the error is equivalent to the \emph{inverse geodetic problem}. However, since the errors (distances) can be assumed small, sufficient accuracy can be achieved with a local spherical approximation which can be further simplified
\begin{equation*}
\begin{split}
e_{\frown}&=(\sqrt{1+e^2}/(1+e^2\cos^2(\phi_k)) + h_k)\atantwo(|\tilde{n}_k\times\bar{n}_k|, \tilde{n}_k\cdot\bar{n}_k)\\
&\approx (\sqrt{1+e^2}/(1+e^2\cos^2(\phi_k)) + h_k)|\tilde{n}_k\times\bar{n}_k|/(|\tilde{n}_k||\bar{n}_k|)
\end{split}
\end{equation*}
where $\sqrt{1+e^2}/(1+e^2\cos^2(\phi_k)) + h_k$ is the local spherical curvature radius and $e^2=(1-b^2/a^2)$ is the Earth's eccentricity.
\item Latitude error arch length in meter. This in general requires a numerical solution of the \emph{meridian arc} but for small error $\tilde{\phi}_k-\phi_k$, relative to the change in curvature of the earth, it can be approximated with
\begin{equation*}
e_{\phi}=(\tilde{\phi}_k-\phi_k)(a + h_k)/a\tilde{M}_k
\end{equation*}
% https://en.wikipedia.org/wiki/Meridian_arc
\item Longitude error arch length in meter
\begin{equation*}
e_{\lambda}=
\begin{cases}
\phi_k=\pi/2 & 0 \\
\text{otherwise} & (\tilde{\lambda}_k-\lambda_k)\cos(\phi_k)(a + h_k)/a\tilde{N}_k
\end{cases}
\end{equation*}
Longitude is not defined for the polar axis and the errors there is always $0$.
\item Altitude error in meter
\begin{equation*}
e_{h}=\tilde{h}_k - h_k
\end{equation*}
\item Fractional gravity vector error based on
Somigliana's (WGS84) gravity formula with second order free-air anomaly correction~\cite{Torge2001}
\begin{equation*}
\quad\quad \textsl{g}(\phi,h) = \gamma_a\frac{1+\kappa_\textsl{g}\sin(\phi)}{(1-e^2\sin(\phi))^{-1/2}}(1 - (\kappa_1 - \kappa_2\sin(\phi))h + \kappa_3h^2)
\end{equation*}
where
\begin{itemize}
\item $\kappa_\textsl{g}=\frac{b\gamma_b}{a\gamma_a} - 1$
\item $\gamma_a = 9.780\,326\,771\,5\;\textrm{m/s}^2$ is the gravity at the equator
\item $\gamma_b = 9.832\,186\,368\,5\;\textrm{m/s}^2$ is the gravity the poles
\item $\kappa_1 = 2(1+f+m)/a=3.157\,04\cdot10^{-7}\textrm{m}^{-1}$
\item $\kappa_2 = 4f/a = 2.102\,69\cdot10^{-9}\textrm{m}^{-1}$
\item $\kappa_3/a^2 = 7.374\,52\cdot^{-14}\textrm{m}^{-1}$
\end{itemize}
giving the error
\begin{equation*}
e_{g}=||\textsl{g}(\tilde{\phi}_k,\tilde{h}_k)\tilde{n}_k - \textsl{g}(\phi_k,h_k)\bar{n}_k||/|\textsl{g}(\phi_k,h_k)|
\end{equation*}
\item N-vector magnitude error
\begin{equation*}
e_{|n|}=1 - ||\tilde{n}_k||
\end{equation*}
\item N-vector orientation error in radians
\begin{equation*}
e_{\arg(n)}=\atantwo(|\tilde{n}_k\times\bar{n}_k|, \tilde{n}_k\cdot\bar{n}_k)
\end{equation*}
This may be implemented as
\begin{equation*}
\begin{split}
e_{\arg(n)}&=\sin^{-1}(|\tilde{n}_k\times\bar{n}_k|/|\tilde{n}_k|\bar{n}_k|)\\
&\approx|\tilde{n}_k\times\bar{n}_k|/(|\tilde{n}_k||\bar{n}_k|)
\end{split}
\end{equation*}
for angles below $\pi/2$ and small angles, respectively. 
\end{itemize}

\section{More benchmark results}\label{appendix:more_results}
\begin{figure}[!ht]
\begin{center}
\includegraphics[width=0.49\textwidth]{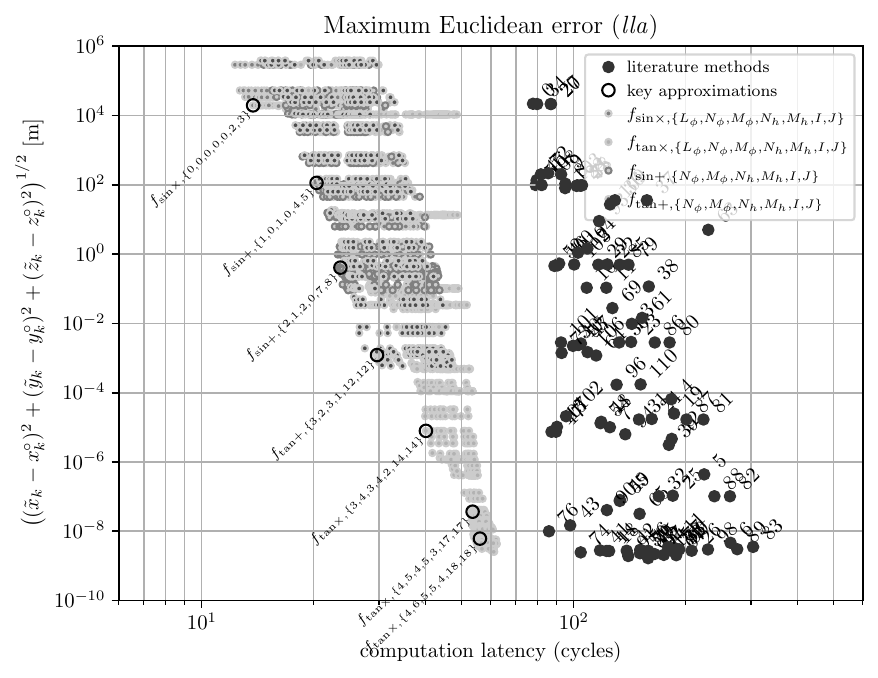}
\includegraphics[width=0.49\textwidth]{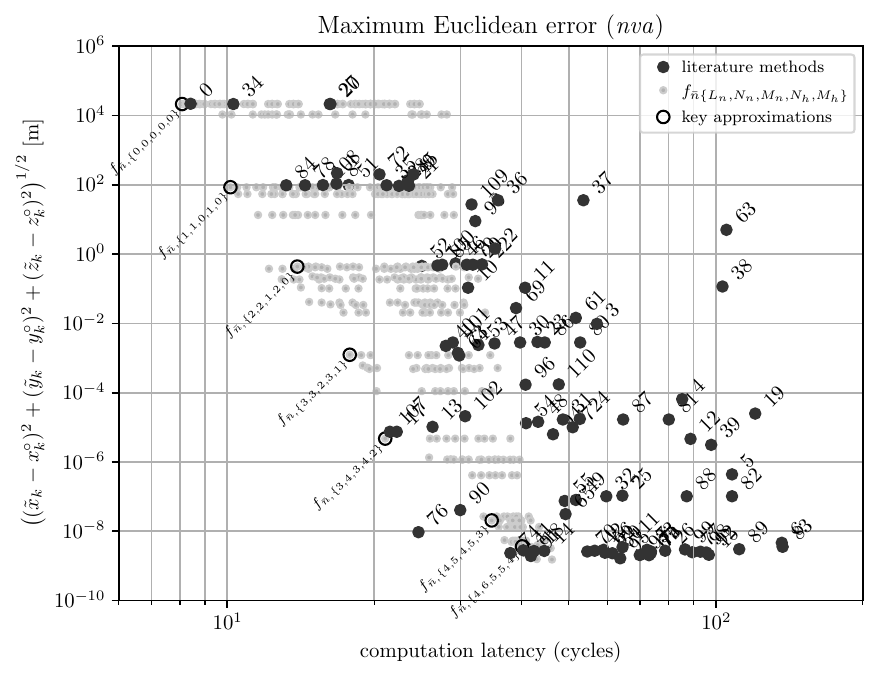}
\includegraphics[width=0.49\textwidth]{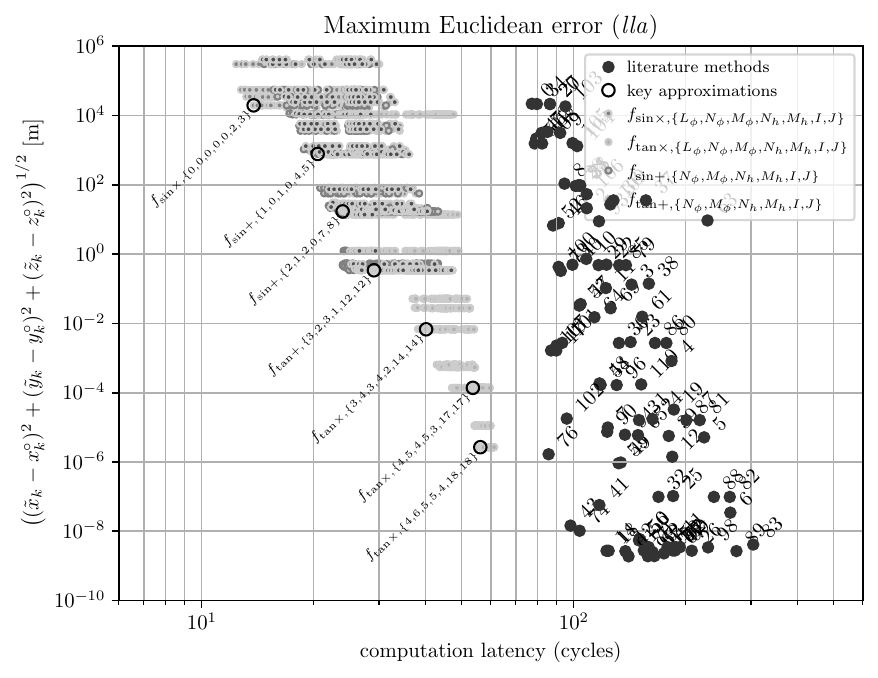}
\includegraphics[width=0.49\textwidth]{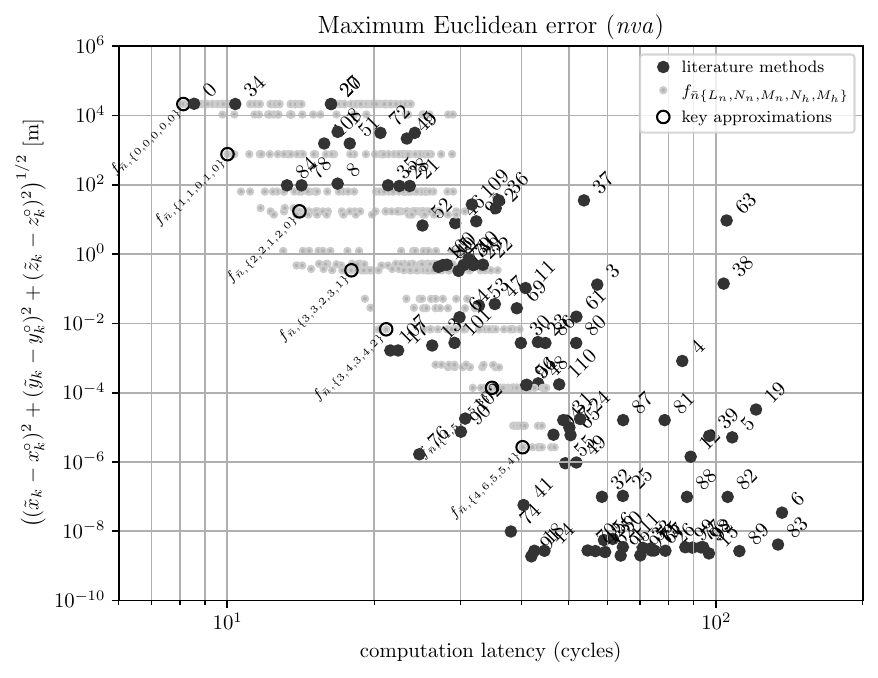}
\includegraphics[width=0.49\textwidth]{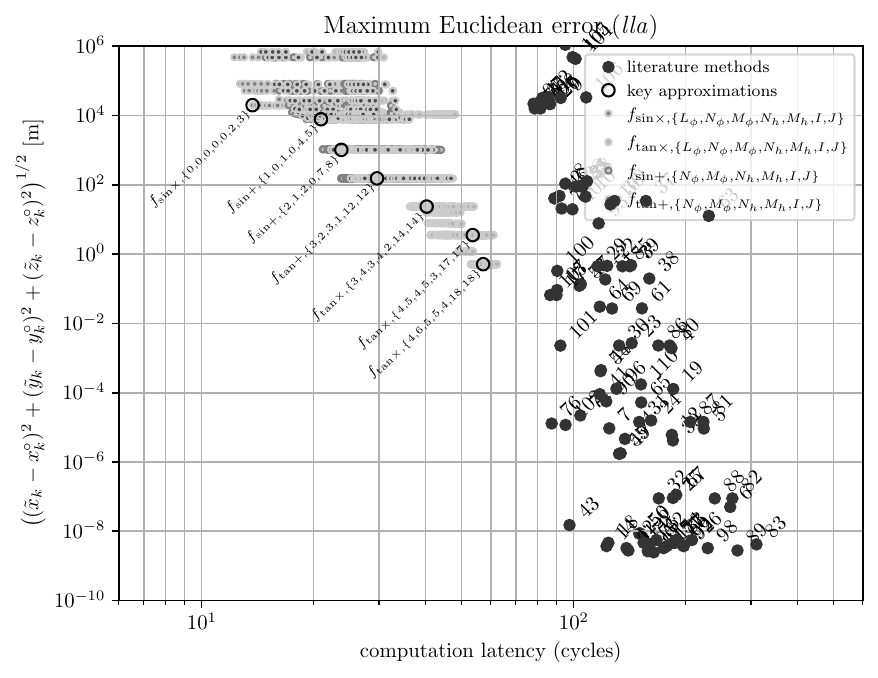}
\includegraphics[width=0.49\textwidth]{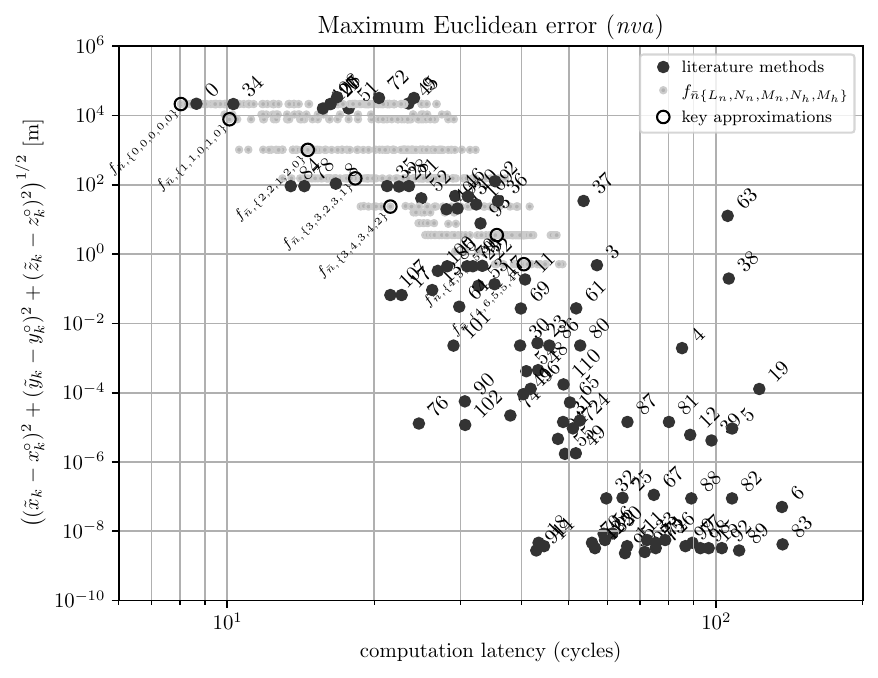}
\caption{\footnotesize Euclidean error for the altitude ranges $[-5000,32\,000]$m, $[-5000,500\,000]$m and $[-5000,5\,000\,000]$m. Comparing the Euclidean errors for the altitude range $[-5000,100\,000]$m shown in Fig.~\ref{fig:lat_and_alt_err} with those for the ranges $[-5000,32\,000]$m, $[-5000,500\,000]$m, the overall picture is the same but, for the range $[-5000,5\,000\,000]$m, the polynomial methods (with single polynomials over the complete range) becomes inefficient and perform poorly. Note that some methods have numerical problems at very large altitudes~\cite{Ward2020} and that the method by~\cite{Hmam2018} performs surprisingly well for large ranges.}\label{fig:err_alt}
\end{center}
\end{figure}

\begin{figure}[!ht]
\begin{center}
\includegraphics[width=0.49\textwidth]{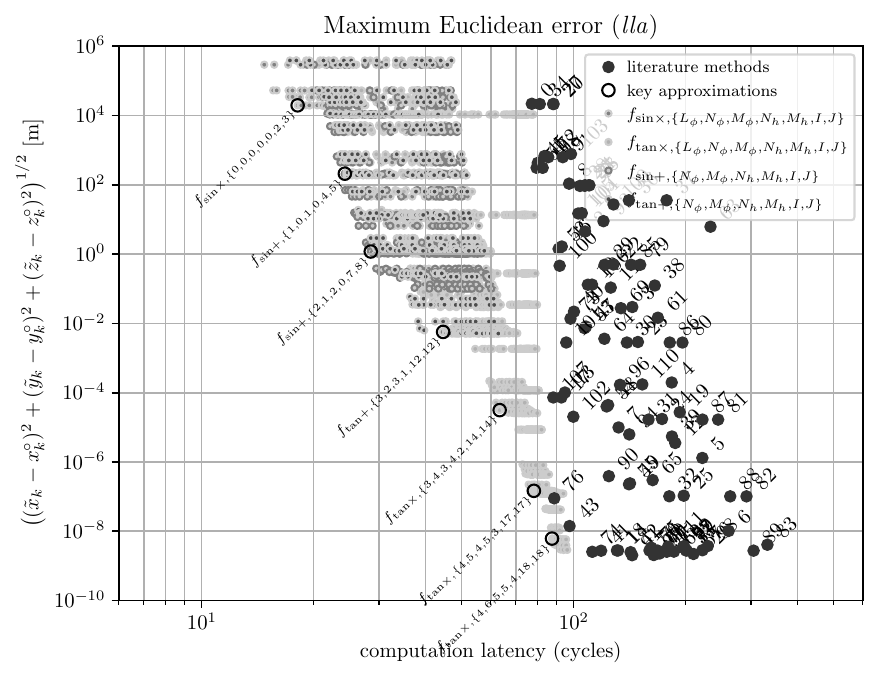}
\includegraphics[width=0.49\textwidth]{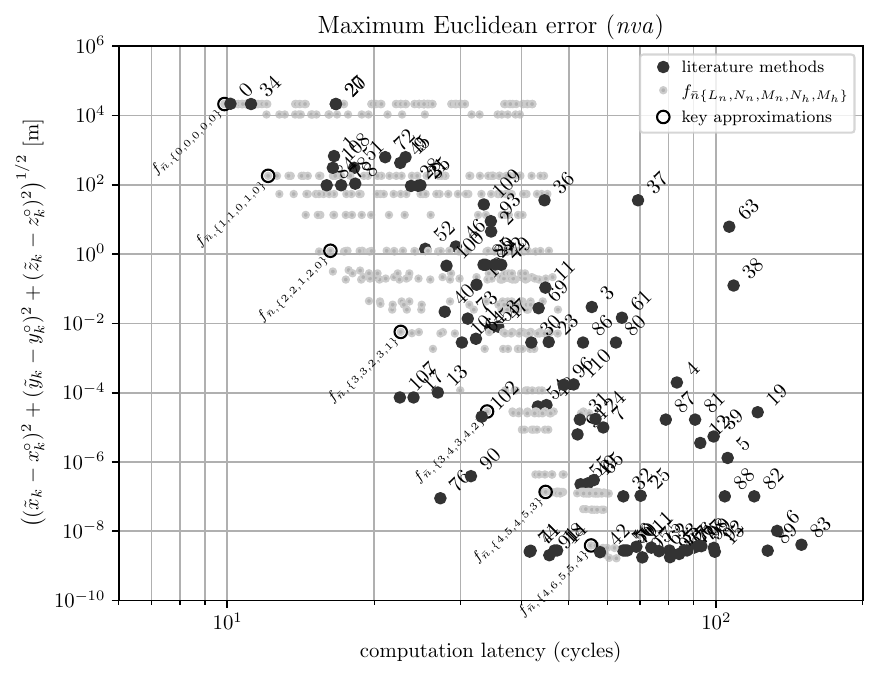}
\caption{\footnotesize Euclidean errors for the altitude range $[-5000,100\,000]$m all methods compiled without AVX support (compiler flag \texttt{-mno-avx}). The error results are essentially identical but the latency results are worse, especially for the polynomial methods, i.e. the polynomial methods can to a large extent be vectorized and the lack of SIMD operations favor other methods. However, SIMD instructions have been prevalent in CPUs for more than a decade and CPUs without SIMD might have different arithmetic performance profiles. Therefore, too certain conclusions should not be drawn from these results.}\label{fig:err_avx}
\end{center}
\end{figure}

Additional results for smaller and larger altitude ranges as well as for methods compiled without AVX support are provided in Fig.~\ref{fig:err_alt} and Fig.~\ref{fig:err_avx}, respectively.

\section{Other benchmarks and benchmark cross-validation}\label{sec:cross_validation}
The results can only partially be compared with other results found in the literature. Foremost, there are no other benchmarks considering \emph{nva} performance. Further, the only extensive modern benchmark found in the literature is~\cite{Ward2020}. There is also an early interesting benchmark~\cite{Voll1990} but it is somewhat too old and has too poor code standard to be of interest for latency measurement comparison. Worth mentioning is also the solid but narrow benchmark~\cite{Bajorek2014}. Frequently cited is the benchmark~\cite{Gerdan1999} but it is too small to be of general interest. In addition, a few more benchmarks can be found~\cite{Simkooei2002} and ~\cite{Fok2003}. However, they are of such poor quality that they can largely be ignored. A small benchmark can also be found in~\cite{Wahlberg2009} (Swedish) and a review of methods in~\cite{Burtch2006}. Further, most articles presenting transformation methods comes with some benchmark but they are generally too small to draw far reaching conclusions.

\begin{figure}[!h]
\begin{center}
\includegraphics[width=0.49\textwidth]{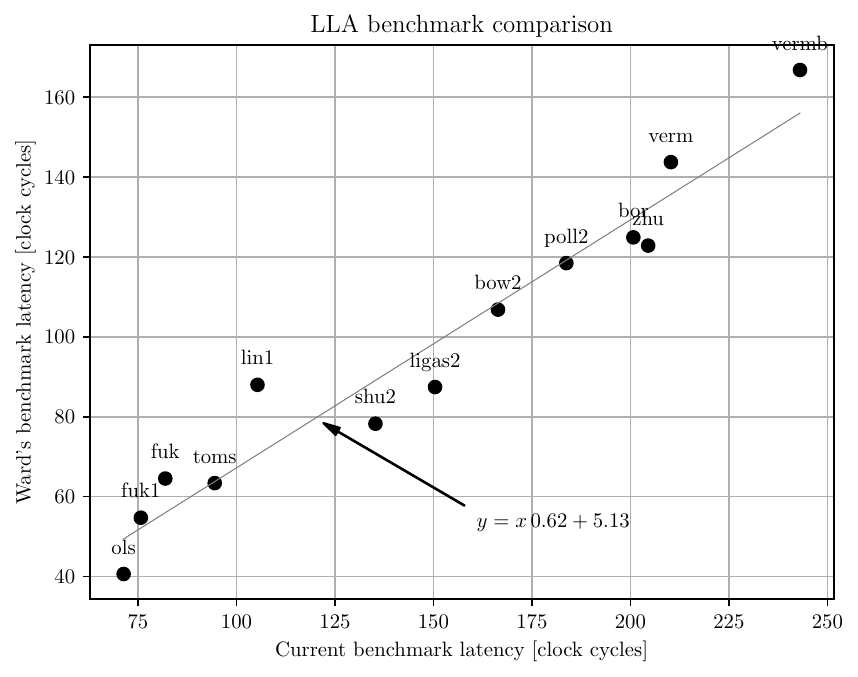}
\includegraphics[width=0.49\textwidth]{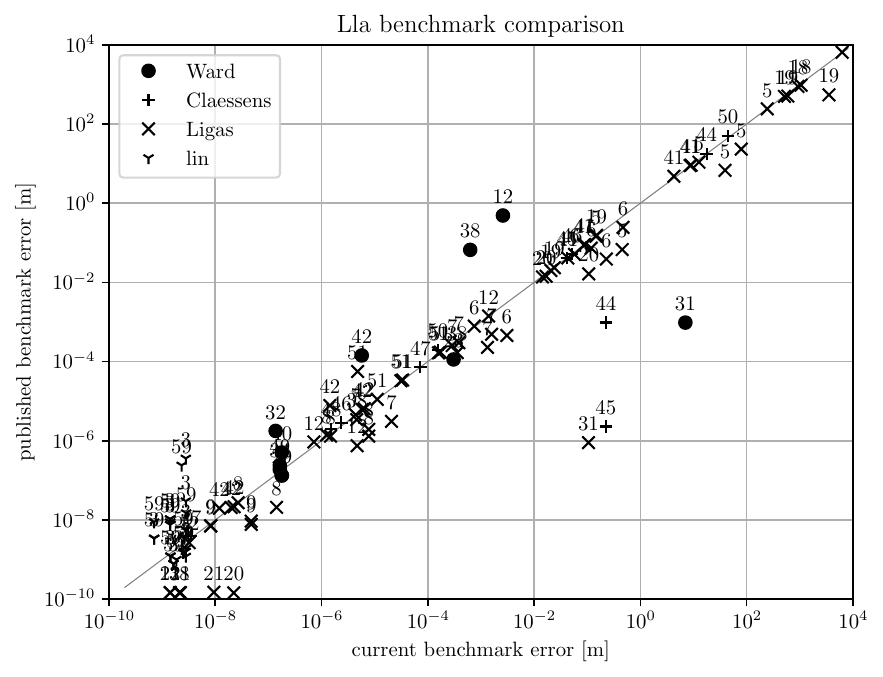}
\caption{\footnotesize Latency measurement comparison of the current benchmark and latency measurements for~\cite{Ward2020}. The latency measurements are not those of the article but rather the latest measurements retrieved from \url{https://github.com/planet36/ecef-geodetic} at 2023-12-11. Further, a latency of $\atantwo$ of 75 clock cycles has been subtracted from the latency measurements of the current benchmark to compensate for longitude not being computed. A least squares fit line is also shown.}\label{fig:bench_comp}
\end{center}
\end{figure}

Absolute timing results are scarce in the literature.
A comparison of latency measurements of transformation methods overlapping with those of~\cite{Ward2020} is provided in the first plot of Fig.~\ref{fig:bench_comp}, together with a least squares line fit.  Performance variation figures presented in~\cite{Claessens2019} indicate that differences of 5-10\% can be expected even for identical implementations. Hence, the deviation from the linear fit can be expected. Further, the latencies was measured on an Intel Core i7 1185G7 (current) and on an Intel Core i9 13950HX (Ward). From CPU single core benchmark results, the difference can be expected to be +45\% which is in line with 0.65 (+54\%). Finally, the small constant part indicate that the measurement overhead is low.
Altogether, this provides an additional confidence in the latency measurements.
Some other timing results can also be found in the literature. In~\cite{Fukushima2006}, latencies are provided for 8 methods but they are in the range $[500,5000]$ and clearly too old to be comparable.

Maximum error results are more common and results from Ward~\cite{Ward2020}, Claessens~\cite{Claessens2019}, Ligas and Banasik~\cite{Ligas2011} and Lin and Wang~\cite{Lin1995} are compared against error results produced by the current benchmark machinery in the second plot of Fig.~\ref{fig:bench_comp}. The results are a mix of Euclidean distance, latitude errors and altitude errors, all converted to distances in meter. Ideally all results should fall on the $y=x$ line included in the plot. Most results are spot on while some results deviate considerably. Specifically
\begin{enumerate}
\item[-] All altitude error results presented in~\cite{Ligas2011}, except those of ~\cite{Heiskanen1967}, appear to be off by a factor 1000. From comments in the code from~\cite{Ward2020}, others appear to have hade the same experience. The results are included in the plot corrected for this.
\item[-] Altitude error results presented in ~\cite{Ligas2011} of~\cite{Heiskanen1967} are roughly a factor of 8 off. There is no obvious explanation for this and the results have been excluded not to bloat the plot.
\item[-] Results presented in ~\cite{Ligas2011} of~\cite{Fukushima1999} appear to have one more iteration than stated. The results are plotted against the apparent correct number of iterations in the plots.
\item[-] Multiple error results for~\cite{Borkowski1987} are considerably higher. This is likely because the presented benchmarks ignore or miss the poor performance of the method very close to the poles.
\item[-] Methods by \cite{Sampson1982} adapted in~\cite{Claessens2019} have a considerably lower performance than expected. Code is provided so likely something is wrong in the original description.
%\item[-] TODO Why is ward off?
%\item[-] TODO Include values from Voll.
%\item[-] TODO INclude values from Gerdan.
\end{enumerate}
Other small variations can be explained by slight implementation differences and the sampling not capturing the largest errors. Altogether, this provides an additional confidence in the error measurements.

%Fok Matlab, exclude poles, convergence check of 0.5mm, i.e. very different accuracies are compared.
%Simkooen, Matlab
%Gerdan C++, relative time, single iteration except hirvonen, ONLY POINTS IN AUSTRALIA. Maximum error! Right order of stuff.

%How does the errors scale with the altitude range.

%The error and latency results are roughly in agreement with those of~\cite{Gerdan1999}. No absolute latency results are provided but the order of the latency results are the same, eventhough the range of our relative timing is considerably higher. On the other hand, this can be expected since the latency in ~\cite{Gerdan1999} is measured of the complete test sample set causing timing to be inclusive of system interrupts, sample retrieval, etc. The error results are only order of magnitude the same which can be expected since different coordinate ranges are considered.

\section{Some more comments about the benchmark}\label{sec:appendix_benchmark}
The intention of this paper is \emph{not} to be a review and benchmark of methods for ECEF to geodetic coordinate transformations. Hence, it is rather brief when it comes to benchmark details. Nonetheless, it contains the most extensive published benchmark of said methods which makes the benchmark valuable in itself, and due to the scarcity of benchmarks in the literature, some further comments about the benchmark are appropriate:% and benchmarking transformation methods in general are necessarily given here.
% Due to the limited size and poor quality of benchmarks in the literature, some further comments about benchmarking these types of methods are given
\begin{itemize}
\item The benchmark methods have been selected from a combination of commonly cited methods and methods with expected good performance. From~\cite{Nilsson2024}, it can clearly be seen not to be an exhaustive set. However, common methods are included. Poorly described methods without a clear implementation have been excluded. For example, the promising method~\cite{Turner2009} is lacking some coefficients. I hope I am doing better.
\item Micro-benchmarking on modern superscalar CPUs with modern compilers is challenging to say the least. To mention a few challenges:
\begin{itemize}
\item Different cores typically provide different performance. Scheduling is correlated over at least seconds. Hence, just looping over benchmark methods from first to last will not give stable results. Rather the benchmarks methods will have to be looped over multiple times and results aggregated. Here 100 loops over the benchmark methods are used.
\item Many modern CPUs have dynamic frequency scaling. This makes latency measurements sensitive in terms of time dependent on the thermal and power supply state of the CPU and make conversion to clock cycles more difficult. Further, some instructions like AVX may be more impacted than others. 
\item To avoid branch prediction, the provided data have to be unpredictably drawn from a pre-generated set. Unfortunately, to make it unpredictable, the base set may not fit in the L1 cache which will introduce infrequent L2 fetch latencies of ~10 clock cycles. So the data set will have to be large enough but not too large to introduce significant L2/L3 cache misses and the sample drawing has to be randomly generated at runtime to avoid the compiler seeing through it. Here $2^{14}=16384$ base samples with randomly shuffled indices to retrieve them have been used. %On the other hand, whether some degree of data correlation should be included is a choice that has to be made. After all, most library functions and most real world applications demonstrates some degree of data correlation.
\item Regular CPU clocks have a typical accuracies of tenth to hundreds of ns, roughly the time of a transformation. This means sets of transformations has to be measured. On the other hand, these measurement sets cannot be too large since then system interruption may corrupt a large portion of the measurements. Here measurement sets of $1000$ samples repeated $10\,000$ times (with different samples) for every benchmark methods loop have been used.
\item Modern compilers are very good at removing unused code and it may interleave multiple transformations. To avoid this, some benchmark loop logic need to be made dependent on the result. We have used the micro benchmarking framework \texttt{Nanobench}~\cite{Ankerl2022} to handle this. Coding your own benchmark harness is a challenging task. Don't do it!
\end{itemize}
\item Many iterative methods only require one or a handful of iterations to reach numerical accuracy for reasonable altitude ranges. Iterating such methods until convergence, rather than a fixed number of iterations, possibly dependent on $p$, will incur a significantly higher computational cost. Iterating until convergence and using an altitude range tuned for the convergence of a particular method is a good way to make that method appear better than all other methods. This is a common trap to fall into~\cite{Claessens2007}. Don't iterate until convergence when benchmarking!
\item Clearly, different transformation methods do demonstrate different accuracies, both in terms of absolute Euclidean error and in terms of different latitude and altitude errors, and applications do have varying accuracy requirements. Hence, the accuracy has to be taken into account when comparing the computational cost of methods. Further, since the relation between computational cost and latency is roughly logarithmic, it should be plotted in logarithmic scale. Interpreting just commonly provided tabulated values is hard!
\item Comparing \emph{lla} and \emph{nva} coordinate transformations, as noted, it can be seen that \emph{lla} methods are dominated by the evaluation of inverse trigonometric functions. This means that the core computational cost of methods are mixed by less relevant common components. As a result, \emph{nva} versions should preferably be compared.
\item Standard library trigonometric functions are typically implemented with some lookup tables. This means that they are cache sensitive. Consequently, different sample orders can give significantly different performance values. The same goes for branch-prediction, as discussed above, which to a lesser extent can affect values. As a result, one has to be very clear on what and in what order samples are provided to the methods, and simply be aware of that the results will depend on how the benchmarking is done, which is another reason to favor \emph{nva} methods.
\item Longitude has a singularity at the poles, which is another reason for favoring comparison of the n-vector versions. However, in practice, this is of less concern when using double precision. First, the likelihood of randomly sampling a problematic point is minimal (the pole would naturally be sampled with quasi-random sequences but $\atantwo(0,0)$ is well defined for most implementations). Second, the problem is inherent to longitude and can trivially be solved by simply ignoring samples too close to the poles. On the other hand, some methods have poor latitude accuracy at the poles, see e.g.~\cite{Fukushima1999}, and some methods demonstrate narrow maximum error peaks at other points, see e.g.~\cite{Osen2017}. Hence, when benchmarking with respect to maximum error, pure random or quasi-random sampling cannot be used but some points will have to be manually identified and added.
\item It may appear surprising that the accuracy for longitude vary between methods. This is probably because the C standard (since C99) allows floating point contraction, see Section~6.5 of~\cite{C11LanguageStandard}. GCC enables contraction by default and different contractions will result in different numerical results which is manifested in different maximum errors. These differences are of little practical importance but it demonstrates to what length the compiler goes to optimize the compiled methods. Consequently, benchmarking should be done in some low level language such as C/C++ and one has to be clear on the compiler and compiler options.
\end{itemize}
Few, if any, benchmarks found in the literature comes close to considering the above points. See Appendix~\ref{sec:cross_validation} for some more comments.
%There are three specific benchmarks found in the literature are \cite{Gerdan1999}, \cite{Simkooei2002} and \cite{Fok2003}. None of them comes close to considering the items above.

%Fok2003
%Methods are code in Matlab!
%Exclude poles (since two methods fail for the poles)
%Very few sample points. 0.1deg intervals.
%Convergence criteria at 0.5mm for altitude and 0.2mm for latitude

%Simkooei
%Very few points, 1deg.
%Matlab!

%Gerdan
%Only for Australia
%Reasonable results but vague on many details.

%\begin{itemize}
%\item The benchmark is the most extensive published benchmark in terms of the number of benchmarked methods, number of error measures and number and even distribution of samples.
%\item Its the first benchmark demonstrating the accuracy versus latency performance in a clear way rather than just tabulated values.
%\item Benchmark including n-vector version giving consistent results with a regular latitude-longituden versions demonstrating that these results are not artifact of errors around the poles. Further, it gives a picture of the gain of using n-vector representation 
%\end{itemize}

\section{Labelled approximations}\label{appendix:labeled_approx}
In the following Table~\ref{tab:poly} to Table~\ref{tab:poly_trig}, the polynomials of the key approximations, labeled in the result plots, are provided.

\begin{table}[ht]
\centering
\tiny
%auto-ignore
\begin{tabularx}{\textwidth}{|r@{\hspace{1mm}}X|}
\hline
$\sigma_{0}(d) $\!\!\!\!\!\!\!\!\!\!\!\!\!\!\!&\,\,\,\,\,\,\,\,\,\,\,\,\,\,\,=\,$ 9.9999716626081590\!\cdot\! 10^{-1}$\\
$\sigma_{1}(d) $\!\!\!\!\!\!\!\!\!\!\!\!\!\!\!&\,\,\,\,\,\,\,\,\,\,\,\,\,\,\,=\,$ 9.9999716626081590\!\cdot\! 10^{-1}$\\
$\sigma_{2}(d) $\!\!\!\!\!\!\!\!\!\!\!\!\!\!\!&\,\,\,\,\,\,\,\,\,\,\,\,\,\,\,=\,$ 9.9999999999933083\!\cdot\! 10^{-1} -4.9999952770986831\!\cdot\! 10^{-1}  d$\\
$\sigma_{3}(d) $\!\!\!\!\!\!\!\!\!\!\!\!\!\!\!&\,\,\,\,\,\,\,\,\,\,\,\,\,\,\,=\,$ 9.9999999999933083\!\cdot\! 10^{-1} -4.9999952770986831\!\cdot\! 10^{-1}  d$\\
$\sigma_{4}(d) $\!\!\!\!\!\!\!\!\!\!\!\!\!\!\!&\,\,\,\,\,\,\,\,\,\,\,\,\,\,\,=\,$ 1.0000000000000000 -4.9999999999989962\!\cdot\! 10^{-1}  d + 4.1666643052156936\!\cdot\! 10^{-2}  d^{2}$\\
\hline
$\tau_{0}(w, d) $\!\!\!\!\!\!\!\!\!\!\!\!\!\!\!&\,\,\,\,\,\,\,\,\,\,\,\,\,\,\,=\,$ 0.0$\\
$\tau_{1}(w, d) $\!\!\!\!\!\!\!\!\!\!\!\!\!\!\!&\,\,\,\,\,\,\,\,\,\,\,\,\,\,\,=\,$ w  9.9999905541991507\!\cdot\! 10^{-1}$\\
$\tau_{2}(w, d) $\!\!\!\!\!\!\!\!\!\!\!\!\!\!\!&\,\,\,\,\,\,\,\,\,\,\,\,\,\,\,=\,$ w  9.9999905541991507\!\cdot\! 10^{-1}$\\
$\tau_{3}(w, d) $\!\!\!\!\!\!\!\!\!\!\!\!\!\!\!&\,\,\,\,\,\,\,\,\,\,\,\,\,\,\,=\,$ w  ( 9.9999999999986617\!\cdot\! 10^{-1} -1.6666657220863013\!\cdot\! 10^{-1}  d )$\\
$\tau_{4}(w, d) $\!\!\!\!\!\!\!\!\!\!\!\!\!\!\!&\,\,\,\,\,\,\,\,\,\,\,\,\,\,\,=\,$ w  ( 9.9999999999986617\!\cdot\! 10^{-1} -1.6666657220863013\!\cdot\! 10^{-1}  d )$\\
\hline
\end{tabularx}

\vspace{2mm}
\caption{\label{tab:poly}\footnotesize Approximation polynomials of labelled ECEF to geodetic coordinate approximations as implemented with Estrin. The coefficients are computed for $h\in[-5000,100\,000]$ and $h_0=0$.
}
\end{table}

\begin{table}[ht]
\centering
\tiny
%auto-ignore
\begin{tabularx}{\textwidth}{|r@{\hspace{1mm}}X|}
\hline
$\mu_{0,0}(u, v) $\!\!\!\!\!\!\!\!\!\!\!\!\!\!\!&\,\,\,\,\,\,\,\,\,\,\,\,\,\,\,=\,$ -6367431.3222291581 + u$\\
$\mu_{1,0}(u, v) $\!\!\!\!\!\!\!\!\!\!\!\!\!\!\!&\,\,\,\,\,\,\,\,\,\,\,\,\,\,\,=\,$ -6378123.6318397466 + u + 21384.619221178389  v$\\
$\mu_{2,0}(u, v) $\!\!\!\!\!\!\!\!\!\!\!\!\!\!\!&\,\,\,\,\,\,\,\,\,\,\,\,\,\,\,=\,$ -6378136.9666263694 + u + 21491.297514157781  v -106.67829297939183  v^{2}$\\
$\mu_{3,1}(u, v) $\!\!\!\!\!\!\!\!\!\!\!\!\!\!\!&\,\,\,\,\,\,\,\,\,\,\,\,\,\,\,=\,$ -6378136.9994595172 + 0.99999999993245003  u + (21635.897371183597$\\&$ -2.2448844415581047\!\cdot\! 10^{-5}  u)  v + (-255.12110929990945 + 2.2892370728809709\!\cdot\! 10^{-5}  u$\\&$ + (3.9094878910461248 -4.4352564908397368\!\cdot\! 10^{-7}  u)  v)  v^{2}$\\
$\mu_{4,2}(u, v) $\!\!\!\!\!\!\!\!\!\!\!\!\!\!\!&\,\,\,\,\,\,\,\,\,\,\,\,\,\,\,=\,$ -6378136.9999909624 + 0.99999999999763889  u + 1.5802065572606207\!\cdot\! 10^{-19}  u^{2}$\\&$ + (21780.894277501458 -6.7654346288251749\!\cdot\! 10^{-5}  u + 3.523327586518059\!\cdot\! 10^{-12}  u^{2})  v$\\&$ + (-405.06255735505755 + 6.9626677884681548\!\cdot\! 10^{-5}  u -3.6418186823541659\!\cdot\! 10^{-12}  u^{2}$\\&$ + (9.0523504657800462 -2.0213005835652036\!\cdot\! 10^{-6}  u + 1.2163741069533121\!\cdot\! 10^{-13}  u^{2})  v)  v^{2}$\\&$ + (-0.19833376155550902 + 4.897368096169858\!\cdot\! 10^{-8}  u -3.1466289533477168\!\cdot\! 10^{-15}  u^{2})  v^{4}$\\
$\mu_{5,3}(u, v) $\!\!\!\!\!\!\!\!\!\!\!\!\!\!\!&\,\,\,\,\,\,\,\,\,\,\,\,\,\,\,=\,$ -6378136.999999837 + 0.99999999999993427  u + (9.0086894408412461\!\cdot\! 10^{-21}$\\&$ -4.1721771358431153\!\cdot\! 10^{-28}  u)  u^{2} + (21926.846232683445 -0.0001359137163728073  u$\\&$ + (1.4164341421699455\!\cdot\! 10^{-11} -5.529324196314687\!\cdot\! 10^{-19}  u)  u^{2})  v + (-557.90712954128696$\\&$ + 0.00014110897616511272  u + (-1.4785186246301743\!\cdot\! 10^{-11} + 5.7903314015836112\!\cdot\! 10^{-19}  u)  u^{2}$\\&$ + (16.200215643865732 -5.3621764869014012\!\cdot\! 10^{-6}  u + (6.4221743608384831\!\cdot\! 10^{-13}$\\&$ -2.7041524201192779\!\cdot\! 10^{-20}  u)  u^{2})  v)  v^{2} + (-0.46569261099818271 + 1.7163926304246266\!\cdot\! 10^{-7}  u$\\&$ + (-2.2001911719684914\!\cdot\! 10^{-14} + 9.6932580539601789\!\cdot\! 10^{-22}  u)  u^{2} + (0.01212864332278238$\\&$ -4.7225675478058751\!\cdot\! 10^{-9}  u + (6.2930011258784072\!\cdot\! 10^{-16} -2.8522125189628692\!\cdot\! 10^{-23}  u)  u^{2})  v)  v^{4}$\\
$\mu_{5,4}(u, v) $\!\!\!\!\!\!\!\!\!\!\!\!\!\!\!&\,\,\,\,\,\,\,\,\,\,\,\,\,\,\,=\,$ -6378136.9999995893 + 0.99999999999977973  u + (4.5142727491369904\!\cdot\! 10^{-20}$\\&$ -4.1725151386930229\!\cdot\! 10^{-27}  u)  u^{2} + 1.4635120955670929\!\cdot\! 10^{-34}  u^{4} + (22073.775734502189$\\&$ -0.00022753572483565161  u + (3.5589046523527045\!\cdot\! 10^{-11} -2.7795229385686731\!\cdot\! 10^{-18}  u)  u^{2}$\\&$ + 8.6774096768703165\!\cdot\! 10^{-26}  u^{4})  v + (-713.77257463984938 + 0.00023830325013161956  u$\\&$ + (-3.7512900428756296\!\cdot\! 10^{-11} + 2.9410407690609742\!\cdot\! 10^{-18}  u)  u^{2} -9.2051543802841124\!\cdot\! 10^{-26}  u^{4}$\\&$ + (25.535665117187818 -1.1183565979926647\!\cdot\! 10^{-5}  u + (2.0034813315984979\!\cdot\! 10^{-12}$\\&$ -1.6851285921853015\!\cdot\! 10^{-19}  u)  u^{2} + 5.5133913606023886\!\cdot\! 10^{-27}  u^{4})  v)  v^{2} + (-0.87962551332934436$\\&$ + 4.2975946940615239\!\cdot\! 10^{-7}  u + (-8.2360392773041159\!\cdot\! 10^{-14} + 7.2421780503870395\!\cdot\! 10^{-21}  u)  u^{2}$\\&$ -2.4446464398246533\!\cdot\! 10^{-28}  u^{4} + (0.026555348121184361 -1.3718782035050124\!\cdot\! 10^{-8}  u$\\&$ + (2.7329656904472876\!\cdot\! 10^{-15} -2.4714925716231548\!\cdot\! 10^{-22}  u)  u^{2} + 8.5203151372608724\!\cdot\! 10^{-30}  u^{4})  v)  v^{4}$\\
\hline
$\omega_{0,0}(u, v) $\!\!\!\!\!\!\!\!\!\!\!\!\!\!\!&\,\,\,\,\,\,\,\,\,\,\,\,\,\,\,=\,$ 0.0$\\
$\omega_{1,0}(u, v) $\!\!\!\!\!\!\!\!\!\!\!\!\!\!\!&\,\,\,\,\,\,\,\,\,\,\,\,\,\,\,=\,$ 0.0066677813753770136$\\
$\omega_{2,1}(u, v) $\!\!\!\!\!\!\!\!\!\!\!\!\!\!\!&\,\,\,\,\,\,\,\,\,\,\,\,\,\,\,=\,$ 0.013446184736230014 -1.0515236264437181\!\cdot\! 10^{-9}  u + (-0.00022196483792195034$\\&$ + 2.4231357903357331\!\cdot\! 10^{-11}  u)  v$\\
$\omega_{3,2}(u, v) $\!\!\!\!\!\!\!\!\!\!\!\!\!\!\!&\,\,\,\,\,\,\,\,\,\,\,\,\,\,\,=\,$ 0.02023800884030541 -3.1689947923483929\!\cdot\! 10^{-9}  u + 1.6503606600562022\!\cdot\! 10^{-16}  u^{2}$\\&$ + (-0.00047856006021247579 + 1.0339367946779417\!\cdot\! 10^{-10}  u -6.1276159848868637\!\cdot\! 10^{-18}  u^{2})  v$\\&$ + (1.2255186842036269\!\cdot\! 10^{-5} -2.9789880393313049\!\cdot\! 10^{-12}  u + 1.8956055600526669\!\cdot\! 10^{-19}  u^{2})  v^{2}$\\
$\omega_{4,3}(u, v) $\!\!\!\!\!\!\!\!\!\!\!\!\!\!\!&\,\,\,\,\,\,\,\,\,\,\,\,\,\,\,=\,$ 0.027074537946419158 -6.3663290353839142\!\cdot\! 10^{-9}  u + (6.6347135933514977\!\cdot\! 10^{-16}$\\&$ -2.5899885739079588\!\cdot\! 10^{-23}  u)  u^{2} + (-0.00082614540139308665 + 2.6591180000141378\!\cdot\! 10^{-10}  u$\\&$ + (-3.1458081288862306\!\cdot\! 10^{-17} + 1.3160531764922644\!\cdot\! 10^{-24}  u)  u^{2})  v + (2.7326402303242933\!\cdot\! 10^{-5}$\\&$ -9.9255846855924077\!\cdot\! 10^{-12}  u + (1.2611748377621174\!\cdot\! 10^{-18} -5.524714330136761\!\cdot\! 10^{-26}  u)  u^{2}$\\&$ + (-8.3406214314548869\!\cdot\! 10^{-7} + 3.2207409051958092\!\cdot\! 10^{-13}  u + (-4.2673520748401045\!\cdot\! 10^{-20}$\\&$ + 1.9262444612240382\!\cdot\! 10^{-27}  u)  u^{2})  v)  v^{2}$\\
$\omega_{5,4}(u, v) $\!\!\!\!\!\!\!\!\!\!\!\!\!\!\!&\,\,\,\,\,\,\,\,\,\,\,\,\,\,\,=\,$ 0.033956858269743512 -1.0657992633478044\!\cdot\! 10^{-8}  u + (1.6670252730549704\!\cdot\! 10^{-15}$\\&$ -1.3019553810283226\!\cdot\! 10^{-22}  u)  u^{2} + 4.0645825278533089\!\cdot\! 10^{-30}  u^{4} + (-0.0012704734728483523$\\&$ + 5.4298359021842561\!\cdot\! 10^{-10}  u + (-9.6247756886895796\!\cdot\! 10^{-17} + 8.0493882503509578\!\cdot\! 10^{-24}  u)  u^{2}$\\&$ -2.6240930039793388\!\cdot\! 10^{-31}  u^{4})  v + (5.0407946329043319\!\cdot\! 10^{-5} -2.4311236288225156\!\cdot\! 10^{-11}  u$\\&$ + (4.6237683172662731\!\cdot\! 10^{-18} -4.046026381146591\!\cdot\! 10^{-25}  u)  u^{2} + 1.3611682445895033\!\cdot\! 10^{-32}  u^{4}$\\&$ + (-1.9142270212250502\!\cdot\! 10^{-6} + 9.8416784098310913\!\cdot\! 10^{-13}  u + (-1.9547420630659842\!\cdot\! 10^{-19}$\\&$ + 1.7642687942146882\!\cdot\! 10^{-26}  u)  u^{2} -6.074031842567329\!\cdot\! 10^{-34}  u^{4})  v)  v^{2} + (6.3806316097197935\!\cdot\! 10^{-8}$\\&$ -3.4050334493087929\!\cdot\! 10^{-14}  u + (6.9509879641250328\!\cdot\! 10^{-21} -6.4057707012819846\!\cdot\! 10^{-28}  u)  u^{2}$\\&$ + 2.2416300605419168\!\cdot\! 10^{-35}  u^{4})  v^{4}$\\
$\omega_{6,5}(u, v) $\!\!\!\!\!\!\!\!\!\!\!\!\!\!\!&\,\,\,\,\,\,\,\,\,\,\,\,\,\,\,=\,$ 0.040885295839059019 -1.6058512798514048\!\cdot\! 10^{-8}  u + (3.3508256373764818\!\cdot\! 10^{-15}$\\&$ -3.9268407514331456\!\cdot\! 10^{-22}  u)  u^{2} + (2.452407337947089\!\cdot\! 10^{-29} -6.3787271347333624\!\cdot\! 10^{-37}  u)  u^{4}$\\&$ + (-0.0018135837610434001 + 9.6632292183719524\!\cdot\! 10^{-10}  u + (-2.2823864407320369\!\cdot\! 10^{-16}$\\&$ + 2.8625533791616239\!\cdot\! 10^{-23}  u)  u^{2} + (-1.8662042533586911\!\cdot\! 10^{-30} + 5.0002116900156677\!\cdot\! 10^{-38}  u)  u^{4})  v$\\&$ + (8.3959005641911052\!\cdot\! 10^{-5} -5.0462939089357817\!\cdot\! 10^{-11}  u + (1.2777379582642763\!\cdot\! 10^{-17}$\\&$ -1.6756621695205854\!\cdot\! 10^{-24}  u)  u^{2} + (1.1268286476356885\!\cdot\! 10^{-31} -3.0887598999489422\!\cdot\! 10^{-39}  u)  u^{4}$\\&$ + (-3.6754339736361957\!\cdot\! 10^{-6} + 2.3558477370497502\!\cdot\! 10^{-12}  u + (-6.2286650409724249\!\cdot\! 10^{-19}$\\&$ + 8.4234421419378366\!\cdot\! 10^{-26}  u)  u^{2} + (-5.7956335293000243\!\cdot\! 10^{-33} + 1.6169761725178128\!\cdot\! 10^{-40}  u)  u^{4})  v)  v^{2}$\\&$ + (1.5042738265816568\!\cdot\! 10^{-7} -1.0030850101437109\!\cdot\! 10^{-13}  u + (2.7300583918580043\!\cdot\! 10^{-20}$\\&$ -3.7741601218646458\!\cdot\! 10^{-27}  u)  u^{2} + (2.6418278049580857\!\cdot\! 10^{-34} -7.473104708913223\!\cdot\! 10^{-42}  u)  u^{4}$\\&$ + (-5.2615868937848073\!\cdot\! 10^{-9} + 3.5969760005163239\!\cdot\! 10^{-15}  u + (-9.9797774889619664\!\cdot\! 10^{-22}$\\&$ + 1.4008066097662379\!\cdot\! 10^{-28}  u)  u^{2} + (-9.926763165672736\!\cdot\! 10^{-36} + 2.8366680915535596\!\cdot\! 10^{-43}  u)  u^{4})  v)  v^{4}$\\
\hline
\end{tabularx}

\vspace{2mm}
\caption{\label{tab:poly2}\footnotesize Approximation polynomials of labeled ECEF to geodetic coordinate approximations as implemented with Estrin. The coefficients are computed for $h\in[-5000,100\,000]$ and $h_0=0$.
}
\end{table}

\begin{table}[ht]
\centering
\tiny
%auto-ignore
\begin{tabularx}{\textwidth}{|r@{\hspace{1mm}}X|}
\hline
$\xi_{3}(x) $\!\!\!\!\!\!\!\!\!\!\!\!\!\!\!&\,\,\,\,\,\,\,\,\,\,\,\,\,\,\,=\,$ x  (9.9535795470534027\!\cdot\! 10^{1} -2.8869023791774234\!\cdot\! 10^{1}  x^{2} + 7.9339041370619872\!\cdot\! 10^{2}  x^{4})$\\
$\xi_{5}(x) $\!\!\!\!\!\!\!\!\!\!\!\!\!\!\!&\,\,\,\,\,\,\,\,\,\,\,\,\,\,\,=\,$ x  (9.9986632904710004\!\cdot\! 10^{1} -3.3030478126799267\!\cdot\! 10^{1}  x^{2} + (1.8015928178377459\!\cdot\! 10^{1}$\\&$ -8.5156335459610334\!\cdot\! 10^{2}  x^{2})  x^{4} + 2.0845107810858079\!\cdot\! 10^{2}  x^{8})$\\
$\xi_{8}(x) $\!\!\!\!\!\!\!\!\!\!\!\!\!\!\!&\,\,\,\,\,\,\,\,\,\,\,\,\,\,\,=\,$ x  (9.9999933557895379\!\cdot\! 10^{1} -3.3329860785866898\!\cdot\! 10^{1}  x^{2} + (1.9946565664948636\!\cdot\! 10^{1}$\\&$ -1.3908629610769453\!\cdot\! 10^{1}  x^{2})  x^{4} + (9.6421974777848821\!\cdot\! 10^{2} -5.5912328807716874\!\cdot\! 10^{2}  x^{2}$\\&$ + (2.1862959296299865\!\cdot\! 10^{2} -4.0545676075818409\!\cdot\! 10^{3}  x^{2})  x^{4})  x^{8})$\\
$\xi_{12}(x) $\!\!\!\!\!\!\!\!\!\!\!\!\!\!\!&\,\,\,\,\,\,\,\,\,\,\,\,\,\,\,=\,$ x  (9.9999999943022189\!\cdot\! 10^{1} -3.3333327040718965\!\cdot\! 10^{1}  x^{2} + (1.9999793539915971\!\cdot\! 10^{1}$\\&$ -1.4282551389892477\!\cdot\! 10^{1}  x^{2})  x^{4} + (1.1083671099533618\!\cdot\! 10^{1} -8.9411184337301530\!\cdot\! 10^{2}  x^{2}$\\&$ + (7.1430746289771899\!\cdot\! 10^{2} -5.2514556999956749\!\cdot\! 10^{2}  x^{2})  x^{4})  x^{8} + (3.2232566609514774\!\cdot\! 10^{2}$\\&$ -1.4721216068835910\!\cdot\! 10^{2}  x^{2} + (4.2936462844677191\!\cdot\! 10^{3} -5.8769992093623553\!\cdot\! 10^{4}  x^{2})  x^{4})  x^{16})$\\
$\xi_{14}(x) $\!\!\!\!\!\!\!\!\!\!\!\!\!\!\!&\,\,\,\,\,\,\,\,\,\,\,\,\,\,\,=\,$ x  (9.9999999998328136\!\cdot\! 10^{1} -3.3333333086677668\!\cdot\! 10^{1}  x^{2} + (1.9999989164783083\!\cdot\! 10^{1}$\\&$ -1.4285491033770566\!\cdot\! 10^{1}  x^{2})  x^{4} + (1.1108488389521914\!\cdot\! 10^{1} -9.0713495063830683\!\cdot\! 10^{2}  x^{2}$\\&$ + (7.5933101852628750\!\cdot\! 10^{2} -6.3109521418777435\!\cdot\! 10^{2}  x^{2})  x^{4})  x^{8} + (4.9445357766039380\!\cdot\! 10^{2}$\\&$ -3.3981220148545598\!\cdot\! 10^{2}  x^{2} + (1.8820607710988513\!\cdot\! 10^{2} -7.6115686823698953\!\cdot\! 10^{3}  x^{2})  x^{4}$\\&$ + (1.9542989485087399\!\cdot\! 10^{3} -2.3593188960412083\!\cdot\! 10^{4}  x^{2})  x^{8})  x^{16})$\\
$\xi_{17}(x) $\!\!\!\!\!\!\!\!\!\!\!\!\!\!\!&\,\,\,\,\,\,\,\,\,\,\,\,\,\,\,=\,$ x  (9.9999999999991586\!\cdot\! 10^{1} -3.3333333331539649\!\cdot\! 10^{1}  x^{2} + (1.9999999885901307\!\cdot\! 10^{1}$\\&$ -1.4285710866597597\!\cdot\! 10^{1}  x^{2})  x^{4} + (1.1111052308663949\!\cdot\! 10^{1} -9.0902614649300906\!\cdot\! 10^{2}  x^{2}$\\&$ + (7.6874146803011537\!\cdot\! 10^{2} -6.6400996804419749\!\cdot\! 10^{2}  x^{2})  x^{4})  x^{8} + (5.7751981894433748\!\cdot\! 10^{2}$\\&$ -4.9340121701547388\!\cdot\! 10^{2}  x^{2} + (3.9762413378979828\!\cdot\! 10^{2} -2.8629853703447561\!\cdot\! 10^{2}  x^{2})  x^{4}$\\&$ + (1.7310365496208254\!\cdot\! 10^{2} -8.2096187905995513\!\cdot\! 10^{3}  x^{2} + (2.8093457170694520\!\cdot\! 10^{3}$\\&$ -6.0940031581488599\!\cdot\! 10^{4}  x^{2})  x^{4})  x^{8})  x^{16} + 6.2436108681934464\!\cdot\! 10^{5}  x^{32})$\\
$\xi_{18}(x) $\!\!\!\!\!\!\!\!\!\!\!\!\!\!\!&\,\,\,\,\,\,\,\,\,\,\,\,\,\,\,=\,$ x  (9.9999999999998558\!\cdot\! 10^{1} -3.3333333332990472\!\cdot\! 10^{1}  x^{2} + (1.9999999975665970\!\cdot\! 10^{1}$\\&$ -1.4285713471275124\!\cdot\! 10^{1}  x^{2})  x^{4} + (1.1111095442381502\!\cdot\! 10^{1} -9.0907156394328137\!\cdot\! 10^{2}  x^{2}$\\&$ + (7.6906649156526998\!\cdot\! 10^{2} -6.6566100086412207\!\cdot\! 10^{2}  x^{2})  x^{4})  x^{8} + (5.8364603332581473\!\cdot\! 10^{2}$\\&$ -5.1030977240766378\!\cdot\! 10^{2}  x^{2} + (4.3267805093829844\!\cdot\! 10^{2} -3.4100063401437874\!\cdot\! 10^{2}  x^{2})  x^{4}$\\&$ + (2.3697852281270090\!\cdot\! 10^{2} -1.3702546920513891\!\cdot\! 10^{2}  x^{2} + (6.1837955998122361\!\cdot\! 10^{3}$\\&$ -2.0101395373502504\!\cdot\! 10^{3}  x^{2})  x^{4})  x^{8})  x^{16} + (4.1436296328711983\!\cdot\! 10^{4} -4.0407586855442829\!\cdot\! 10^{5}  x^{2})  x^{32})$\\
\hline
$\chi_{2}(x) $\!\!\!\!\!\!\!\!\!\!\!\!\!\!\!&\,\,\,\,\,\,\,\,\,\,\,\,\,\,\,=\,$ 1.5702116976458603 -2.0212058400388336\!\cdot\! 10^{1}  x + 4.6707077880155363\!\cdot\! 10^{2}  x^{2}$\\
$\chi_{4}(x) $\!\!\!\!\!\!\!\!\!\!\!\!\!\!\!&\,\,\,\,\,\,\,\,\,\,\,\,\,\,\,=\,$ 1.5707878616372565 -2.1412466400217967\!\cdot\! 10^{1}  x + (8.4666692360295800\!\cdot\! 10^{2}$\\&$ -3.5756539685967685\!\cdot\! 10^{2}  x)  x^{2} + 8.6486772213302613\!\cdot\! 10^{3}  x^{4}$\\
$\chi_{7}(x) $\!\!\!\!\!\!\!\!\!\!\!\!\!\!\!&\,\,\,\,\,\,\,\,\,\,\,\,\,\,\,=\,$ 1.5707963049952714 -2.1459880383418343\!\cdot\! 10^{1}  x + (8.8979049893930584\!\cdot\! 10^{2}$\\&$ -5.0174715212327148\!\cdot\! 10^{2}  x)  x^{2} + (3.0893053197591073\!\cdot\! 10^{2} -1.7089810809721586\!\cdot\! 10^{2}  x$\\&$ + (6.6712932596193130\!\cdot\! 10^{3} -1.2628309167103800\!\cdot\! 10^{3}  x)  x^{2})  x^{4}$\\
$\chi_{12}(x) $\!\!\!\!\!\!\!\!\!\!\!\!\!\!\!&\,\,\,\,\,\,\,\,\,\,\,\,\,\,\,=\,$ 1.5707963267933006 -2.1460183603240527\!\cdot\! 10^{1}  x + (8.9048588758505052\!\cdot\! 10^{2}$\\&$ -5.0792024609074237\!\cdot\! 10^{2}  x)  x^{2} + (3.3671622816568150\!\cdot\! 10^{2} -2.4303153862137010\!\cdot\! 10^{2}  x$\\&$ + (1.8329796462243492\!\cdot\! 10^{2} -1.3751038185856593\!\cdot\! 10^{2}  x)  x^{2})  x^{4} + (9.5272136428286426\!\cdot\! 10^{3}$\\&$ -5.5320203601637119\!\cdot\! 10^{3}  x + (2.4008869075915833\!\cdot\! 10^{3} -6.6857373783667137\!\cdot\! 10^{4}  x)  x^{2}$\\&$ + 8.7773781127018793\!\cdot\! 10^{5}  x^{4})  x^{8}$\\
$\chi_{14}(x) $\!\!\!\!\!\!\!\!\!\!\!\!\!\!\!&\,\,\,\,\,\,\,\,\,\,\,\,\,\,\,=\,$ 1.5707963267948586 -2.1460183658456538\!\cdot\! 10^{1}  x + (8.9048621136695723\!\cdot\! 10^{2}$\\&$ -5.0792770716881518\!\cdot\! 10^{2}  x)  x^{2} + (3.3680571282738509\!\cdot\! 10^{2} -2.4367246446961202\!\cdot\! 10^{2}  x$\\&$ + (1.8626098028204800\!\cdot\! 10^{2} -1.4676637467188599\!\cdot\! 10^{2}  x)  x^{2})  x^{4} + (1.1531109639172660\!\cdot\! 10^{2}$\\&$ -8.5675083245346239\!\cdot\! 10^{3}  x + (5.6003920887711281\!\cdot\! 10^{3} -2.9601736597984054\!\cdot\! 10^{3}  x)  x^{2}$\\&$ + (1.1466969584531041\!\cdot\! 10^{3} -2.8304614323065713\!\cdot\! 10^{4}  x + 3.2965787399023498\!\cdot\! 10^{5}  x^{2})  x^{4})  x^{8}$\\
$\chi_{17}(x) $\!\!\!\!\!\!\!\!\!\!\!\!\!\!\!&\,\,\,\,\,\,\,\,\,\,\,\,\,\,\,=\,$ 1.5707963267948965 -2.1460183660245246\!\cdot\! 10^{1}  x + (8.9048622536943916\!\cdot\! 10^{2}$\\&$ -5.0792814046672612\!\cdot\! 10^{2}  x)  x^{2} + (3.3681275570828086\!\cdot\! 10^{2} -2.4374172925348297\!\cdot\! 10^{2}  x$\\&$ + (1.8670862830280291\!\cdot\! 10^{2} -1.4876918953706647\!\cdot\! 10^{2}  x)  x^{2})  x^{4} + (1.2172747383436669\!\cdot\! 10^{2}$\\&$ -1.0071291531222582\!\cdot\! 10^{2}  x + (8.2100842886296591\!\cdot\! 10^{3} -6.3278993837482987\!\cdot\! 10^{3}  x)  x^{2}$\\&$ + (4.3671279910859448\!\cdot\! 10^{3} -2.5369588366614161\!\cdot\! 10^{3}  x + (1.1590307122361844\!\cdot\! 10^{3}$\\&$ -3.8309521446820528\!\cdot\! 10^{4}  x)  x^{2})  x^{4})  x^{8} + (8.0476117495016973\!\cdot\! 10^{5} -8.0043584568231787\!\cdot\! 10^{6}  x)  x^{16}$\\
$\chi_{18}(x) $\!\!\!\!\!\!\!\!\!\!\!\!\!\!\!&\,\,\,\,\,\,\,\,\,\,\,\,\,\,\,=\,$ 1.5707963267948966 -2.1460183660253421\!\cdot\! 10^{1}  x + (8.9048622545903351\!\cdot\! 10^{2}$\\&$ -5.0792814434974340\!\cdot\! 10^{2}  x)  x^{2} + (3.3681284430401487\!\cdot\! 10^{2} -2.4374295730838710\!\cdot\! 10^{2}  x$\\&$ + (1.8671988159799984\!\cdot\! 10^{2} -1.4884115235275848\!\cdot\! 10^{2}  x)  x^{2})  x^{4} + (1.2206039463975348\!\cdot\! 10^{2}$\\&$ -1.0185410804411290\!\cdot\! 10^{2}  x + (8.5042859077692046\!\cdot\! 10^{3} -6.9026076875376219\!\cdot\! 10^{3}  x)  x^{2}$\\&$ + (5.2181956820067259\!\cdot\! 10^{3} -3.4855347796590377\!\cdot\! 10^{3}  x + (1.9414549204238104\!\cdot\! 10^{3}$\\&$ -8.4603434241738747\!\cdot\! 10^{4}  x)  x^{2})  x^{4})  x^{8} + (2.6620763505161977\!\cdot\! 10^{4} -5.3242183459453063\!\cdot\! 10^{5}  x$\\&$ + 5.0486339748452772\!\cdot\! 10^{6}  x^{2})  x^{16}$\\
\hline
\end{tabularx}

\vspace{2mm}
\caption{\label{tab:poly_trig}\footnotesize Inverse trigonometric function approximation polynomials of labeled ECEF to geodetic coordinate (lla) approximations as implemented with Estrin.
}
\end{table}

\end{document}